\documentclass[aps,prd,twocolumn,nofootinbib,superscriptaddress,floatfix]{revtex4-2}

\usepackage{graphicx}      % 插入图片
\usepackage{amssymb}       % 额外数学符号
\usepackage{amsmath}       % 数学公式增强
\usepackage{latexsym}      % 更多符号
\usepackage{cancel}        % 数学中的删除线
\usepackage[normalem]{ulem} % 下划线等
\usepackage{url}           % URL 排版
\usepackage{soul}          % 文字高亮等
\usepackage{verbatim}      % 多行注释
\usepackage{multirow}      % 表格合并单元格
\usepackage{mathrsfs}      % 花体字母
\usepackage{float}         % 浮动体控制
\usepackage[dvipsnames]{xcolor} % 彩色（已包含 color 功能）
\usepackage{mathtools}     % amsmath 扩展
\usepackage{slashed}       % 费曼斜线
\usepackage{physics}       % 物理符号
\usepackage{epstopdf}      % EPS 转 PDF
\usepackage{subfigure}     % 并排子图
\usepackage{bbold}         % 黑体数字
\usepackage{wasysym}       % 更多符号
\usepackage{feynmp}        % 费曼图

\usepackage[colorlinks=true, citecolor=blue, linkcolor=blue, urlcolor=blue]{hyperref}

\newcommand{\beq}{\begin{eqnarray}}
\newcommand{\eeq}{\end{eqnarray}}

\makeatletter
\newcommand{\myBig}{\bBigg@{1.75}}
\makeatother

\usepackage{orcidlink}
\usepackage{etoolbox}

\begin{document}
\title{Stability analysis and double critical phenomenon in the Einstein-Maxwell-scalar theory} 

%EMs理论中的稳定性分析以及超临界现象

\author{Zi-Qiang Zhao\orcidlink{0009-0009-7859-3655}}%\email{zhaoziqiang@stumail.neu.edu.cn}
\affiliation{Liaoning Key Laboratory of Cosmology and Astrophysics, College of Sciences, Northeastern University, Shenyang 110819, China}
\author{Mei-Ling Yan\orcidlink{0009-0000-8877-5262}}%\email{yanmeiling@mails.neu.edu.cn}
\affiliation{Liaoning Key Laboratory of Cosmology and Astrophysics, College of Sciences, Northeastern University, Shenyang 110819, China}
\author{Zhang-Yu Nie\orcidlink{0000-0001-7064-247X}}\email{niezy@kust.edu.cn}
\affiliation{Center for Gravitation and Astrophysics, Kunming University of Science and Technology, Kunming 650500, China}
\author{Jing-Fei Zhang\orcidlink{0000-0002-3512-2804}}
\affiliation{Liaoning Key Laboratory of Cosmology and Astrophysics, College of Sciences, Northeastern University, Shenyang 110819, China}
\author{Xin Zhang\orcidlink{0000-0002-6029-1933}}\email{zhangxin@neu.edu.cn}
\affiliation{Liaoning Key Laboratory of Cosmology and Astrophysics, College of Sciences, Northeastern University, Shenyang 110819, China}
\affiliation{MOE Key Laboratory of Data Analytics and Optimization for Smart Industry, Northeastern University, Shenyang 110819, China}
\affiliation{National Frontiers Science Center for Industrial Intelligence and Systems Optimization, Northeastern University, Shenyang 110819, China}

%我们研究了包含高阶自相互作用项$\lambda |\psi|^4$、$\tau|\psi|^6$以及非最小耦合$h(\psi)=\mathrm{e}^{\alpha\psi^2}$的全息超导模型中的相变行为与动力学稳定性。热力学与动力学稳定性分析表明，该系统的热力学稳定性和动力学稳定性是一致的。相图分析揭示了丰富的临界与超临界现象：固定$\lambda<0$和$\alpha$，增大$\tau$使一阶相变区域收缩至临界点并进入超临界区；变化$\alpha$时，系统可呈现无临界点,单一临界点或者是最引人注目的双重临界现象——系统随$\alpha$增大先后两次进出超临界区。这种单参数驱动的双重临界现象在全息超导模型中属首次报道，揭示了非最小耦合与高阶相互作用项之间复杂的非单调耦合效应。
\begin{abstract}
We investigate the dynamical stability and phase transition behavior in a holographic superfluid model incorporating higher-order self-interaction terms $\lambda |\psi|^4$, $\tau|\psi|^6$, and a non-minimal coupling $h(\psi)=\mathrm{e}^{\alpha|\psi|^2}$. Thermodynamic and dynamical stability analyzes show that the thermodynamic stability and dynamical stability of the system are consistent. Phase diagram analysis reveals rich critical and supercritical phenomena. For fixed $\lambda<0$ and $\alpha$, increasing $\tau$ shrinks the first-order phase transition region to a critical point and then enters the supercritical region. When varying $\alpha$, the system can exhibit no critical point or, most notably, a double critical phenomenon in which, as $\alpha$ increases, the system first enters the supercritical region and then re-enters the first-order phase transition region. This double critical phenomenon driven by a single parameter is reported for the first time in holographic superfluid models, revealing a complex nonmonotonic coupling effect between the non-minimal coupling and higher-order interaction terms.
\end{abstract}
\maketitle
\section{Introduction}
%在全息对偶的框架下，引力体系的经典动力学与强耦合量子场论中的平衡态及非平衡性质之间建立了一座桥梁。其中，全息超导模型作为全息对偶在凝聚态物理中的应用，为我们理解超导/超流相变机制提供了全新的视角。标准的全息超导模型通常考虑一个在反德西特（AdS）黑洞背景下，与麦克斯韦场耦合的复标量场。在低温下，标量场会凝聚，自发破缺U(1)对称性，从而形成超导相，这一过程对应着二阶相变。
Within the framework of holographic duality \cite{Maldacena:1997re}, a bridge is established between the classical dynamics of gravitational systems and the equilibrium and nonequilibrium properties of strongly coupled quantum field theories. As an application of holographic duality in condensed matter physics, the holographic superfluid model provides a novel perspective for understanding the mechanism of superfluid and superfluid phase transitions \cite{Hartnoll:2008vx,Hartnoll:2008kx,Herzog:2010vz}. The standard holographic superfluid model typically considers a complex scalar field coupled to a Maxwell field in an Anti-de Sitter (AdS) black hole background. At low temperatures, the scalar field condenses and spontaneously breaks the $U(1)$ symmetry, giving rise to the superconducting phase, a process that corresponds to a second-order phase transition.

%为了更深入地研究强耦合场的物理特性，研究者们在标准模型的基础上引入了各种修正，例如不同的引力背景模型、p波序参量和D波序。有些情况下，研究者们还会考虑度时空背景的反作用以及各种序参量之间的共存和竞争。这些修正项可以丰富系统的相结构，诱导出一阶相变、零阶相变甚至更复杂的临界行为。特别是，文献[]中已经发现,在单一的s波模型中，当标量场自相互作用包含负的$\lambda |\psi|^4$项时，系统会出现零阶相变，但该相变在热力学和动力学上均不稳定，预示着其难以在物理上实现。而进一步引入更高阶的$\tau|\psi|^6$项，则可以稳定系统，并产生由一阶和二阶相变组合而成的“COW”相变（即一阶相变和二阶相变的混合）或者是单一的一阶相变（normal解和超流解的一阶相变），展现出更为丰富的相图。尽管已有诸多进展，但对这类包含多重相互作用（$\lambda |\psi|^4$和$\tau|\psi|^6$）和非最小耦合（$h(\psi)=\mathrm{e}^{\alpha\psi^2}$）的Einstein-Maxwell标量理论，其完整的相结构和动力学稳定性图景仍有待揭示。特别是，不同参数对系统的影响往往是耦合且非单调的，这可能导致一些全新的、未被发现的临界现象。
To further investigate the physical properties of strongly coupled fields, various modifications have been introduced into the standard model, such as different gravitational backgrounds \cite{Qiao:2020hkx,Pan:2021jii,Zhang:2023uuq,Ghorai:2021uby,Cai:2009hn}, $p$-wave \cite{Gubser:2008wv,Cai:2013aca}, and $d$-wave \cite{Chen:2010mk,Kim:2013oba} order parameters. In some cases, the backreaction of the spacetime background \cite{Hartnoll:2008kx,Cai:2013aca,Pan:2021jii,Cai:2013wma,Wang:2016jov,Wang:2019vaq,Zhao:2025tqq,Zhao:2025vtr,Zhao:2025odj,Zhang:2025hkb,Zhang:2025tsa}, as well as the coexistence and competition among different order parameters \cite{Basu:2010fa,Musso:2013ija,Nie:2013sda,Donos:2013woa,Li:2017wbi,Nie:2015zia,Nie:2014qma,Amado:2013lia,Zhang:2021vwp,Zhang:2025tsa}, is also considered. These modifications can enrich the phase structure of the system by inducing first-order phase transitions, zeroth-order phase transitions, or even more complex critical behavior. In particular, it has been found in Ref. \cite{Zhao:2022jvs} that in a single $s$-wave model, the inclusion of a negative $\lambda |\psi|^4$ term in the scalar self-interaction leads to a zeroth-order phase transition. However, this phase transition is thermodynamically and dynamically unstable, suggesting that it is unlikely to be realized physically. The further introduction of a higher-order $\tau |\psi|^6$ term can stabilize the system and give rise to a Cave-of-Wind (COW) phase transition, a combination of first-order and second-order phase transitions, or a single first-order phase transition (a first-order phase transition between the normal and superfluid solutions), thereby exhibiting a richer phase diagram. This approach of realizing multiple phase transition models through the addition of higher-order terms provides a useful template for studying nonequilibrium evolution in first-order phase transitions within holographic systems \cite{Chen:2022tfy,Zhao:2023ffs}. In addition, holographic duality has also been extensively applied in the study of nonequilibrium dynamical evolution \cite{Xia:2020cjl,delCampo:2021rak,Li:2021iph,Li:2021dwp,Xia:2021xap,Zeng:2022hut,Yang:2025bsw,Xia:2026yrj,Xia:2023pom,Su:2023vqa,Zhao:2023ffs,An:2024ebg,Yang:2024hom,Xia:2024ton,Zeng:2024rwn,Janik:2015iry,Janik:2016btb,Janik:2017ykj,Attems:2017ezz,Attems:2019yqn,Bellantuono:2019wbn,Attems:2020qkg,Chen:2022tfy,Ning:2023edr}.

%此外，全息理论在宇宙学里面也有非常广泛的引用，也就是全息暗能量\cite{Zhang:2005hs,Zhang:2009un,Fu:2011ab,Li:2012spm,Zhang:2015rha,Li:2024qus}

%这种通过添加高阶项实现多种相变模型的方法，为我们研究全息系统中的一阶相变中的非平衡演化提供了一个非常有用的模板。除此之外，全息对偶理论在非平衡动态演化方面也有着相当多的运用。

Despite these advances, the complete phase structure and dynamical stability analysis of the Einstein–Maxwell–scalar theory with multiple interactions ($\lambda |\psi|^4$ and $\tau|\psi|^6$) and non-minimal coupling ($h(\psi)=\mathrm{e}^{\alpha|\psi|^2}$) remain to be uncovered. In particular, the influence of different parameters is often coupled and nonmonotonic, which may lead to entirely new and unexplored critical phenomenon. It is worth noting that critical and supercritical phenomenon are important not only in condensed matter systems but also in black hole physics. Moreover, recent studies suggest that the supercritical phase is not a single uniform phase. Distinct supercritical subphases can still be identified using different approaches \cite{Yoon_2018,PhysRevLett.111.145901,Bolmatov2013,Prescher_2017,Bolmatov_2015,Fomin_2018,Fomin2015,PhysRevE.85.031203,2023PhRvR...5a3149H,Jiang_2025,Xu_2005,Ruppeiner_2012,PhysRevLett.112.135701,PhysRevE.95.052120,Gallo2014,Zhao:2025ecg,Xu:2025jrk,Li:2025tdd,Wang:2025ctk,Li:2025lrq,Anand:2025rzh,Guo:2026xlk,Zhao:2026mkx}.
%在这篇文章中，我们主要研究非线性耦合项与非最小耦合系数对相结构的调控作用，这使得我们可以在一个简单的全息超导模型中实现更加丰富的相结构，并且使得我们可以更方便的去研究临界和超临界现象。值得注意的是，临界和超临界现象不仅在凝聚态系统中非常重要，在黑洞系统中同样如此。并且前沿研究表明，超临界相并非单一的纯相，它仍然可以通过不同的方法区分出不同的超临界亚相。

%这个工作主要研究了非最小耦合参数和高阶非线性项对相变类型的调控作用。本文的结构安排如下：第二部分详细阐述全息模型框架，包括作用量、运动方程、边界条件以及自由能和准正规模的计算方法。第三部分分别对二级相变、零级相变和COW相变进行热力学与动力学稳定性分析。第四部分绘制相图，系统展示$\tau$和$\alpha$对相结构的影响，并重点论述三种不同临界行为以及双重临界现象的发现。第五部分总结全文并对未来工作进行展望。
In this paper, we focus on the role of nonlinear interaction terms and the non-minimal coupling coefficient in tuning the phase structure. This allows us to realize a richer phase diagram within a simple holographic superfluid model and provides a convenient platform for investigating critical and supercritical phenomenon. The remainder of this paper is organized as follows. Sect.~\ref{sec2} elaborates on the holographic model framework, including the action, equations of motion, boundary conditions, and the methods for computing the free energy and quasinormal modes. Sect.~\ref{sec3} presents the thermodynamic and dynamical stability analyzes for the second-order, zeroth-order, and COW phase transitions, respectively. Sect.~\ref{sec4} constructs the phase diagrams, systematically illustrating the effects of $\tau$ and $\alpha$ on the phase structure. Sect.~\ref{sec5} concludes the paper and provides an outlook for future work.

\section{Holographic setup}\label{sec2}
In this manuscript, we consider the non-minimal coupling Einstein–Maxwell–scalar theory with the addition of two higher-order nonlinear terms $\lambda |\psi|^4$ and $\tau|\psi|^6$. The Lagrangian of our model takes the following form
\begin{align}
\mathcal{L}_{m}=&-\frac{1}{4}h(\psi)F_{\mu\nu}F^{\mu\nu}-D_{\mu}\psi^{\ast} D^{\mu}\psi-m^{2}\psi^*\psi\nonumber\\
&-\lambda(\psi^{\ast}\psi)^{2}-\tau(\psi^{\ast}\psi)^{3},\label{Lag}
\end{align}
in which $h(\psi)=\mathrm{e}^{\alpha \psi^{\ast}\psi}$. $D_{\mu}\psi=\nabla_{\mu}\psi-\mathrm{i} A_\mu\psi$ is the standard covariant derivative term of the charged scalar field $\psi$ and $F_{\mu\nu}=\nabla_{\mu}A_{\nu}-\nabla_{\nu}A_{\mu}$ is the Maxwell field strength. We adopt an ansatz of the form
\begin{align}\label{ansatz}
\psi=\psi(r)~, \quad A_{t}=\phi(r)~,
\end{align}
% %并且黑洞度规线元如下
% We take the following ansatz
and the metric of the black hole is given by
\begin{align}
\mathrm{d}s^{2}=-f(r)\mathrm{d}t^{2}+\frac{1}{f(r)}\mathrm{d}r^{2}+r^{2}\mathrm{d}x^{2}+r^{2}\mathrm{d}y^{2},
\end{align}
where the emblackening function $f(r)$ is given by
\begin{align}
f(r)=\frac{r^2}{L^2}-\frac{r_h^3}{r}~,
\end{align}
where $r_h$ is the radius of the black hole event horizon, and its Hawking temperature is
\begin{align}
T= \frac{3 r_h}{4\pi L^2}.
\end{align}
The equations of motion are
\begin{align}
\psi  \left(\frac{\phi ^2}{f^2}+\frac{\text{$\alpha $} \mathrm{e}^{\text{$\alpha $} \psi ^2}
   \phi '^2}{2 f}+\frac{2}{f L^2}\right)+\psi '
   \left(\frac{f'}{f}+\frac{2}{r}\right)&\nonumber\\
   -\frac{2 \lambda  \psi ^3}{f}-\frac{3 \tau  \psi
   ^5}{f}+\psi ''&=0~,\\
   2 \text{$\alpha $} \psi  \phi ' \psi '-\frac{2 \phi  \psi ^2 \mathrm{e}^{-\text{$\alpha $} \psi
   ^2}}{f}+\frac{2 \phi '}{r}+\phi ''&=0~.
\end{align}
To solve the above equations, the boundary conditions need to be specified. The boundary conditions at the AdS boundary are as follows
\begin{align}
\phi(r)=\mu-\frac{\rho}{r}+...~,\qquad \psi=\frac{{\psi^{(1)}}}{r}+\frac{{\psi^{(2)}}}{r^2}...~.
\end{align}
Here, $\mu$ and $\rho$ denote the chemical potential and the charge density, respectively. In this work, we consider the canonical ensemble, and thus we choose fixed charge density as the boundary condition. Furthermore, we adopt the source-free quantization scheme, which implies $\psi^{(1)} = 0$, and the non-vanishing vacuum expectation value is given by $\langle O_2 \rangle = \psi^{(2)}$.
The asymptotic expansion near the horizon is given by
\begin{align}
&\phi(r)=\phi_{1}(r-r_{h})+\mathcal{O}((r-r_{h})^{2})~,\\
&\psi(r)=\psi_{0}+\psi_{1}(r-r_{h})+\mathcal{O}(r-r_{h})~.
\end{align}
In this work, due to the introduction of higher-order nonlinear terms, various types of phase transitions can be realized. Therefore, we introduce the free energy to determine the phase transition type. In this manuscript, since we work in the probe limit, the free energy does not include contributions from the spacetime and can thus be obtained from the scalar field. The free energy takes the following form
\begin{align}
	G=\frac{V_2}{T}\bigg(\frac{\mu\rho}{2L^2}+\int_{r_h}^{\infty}\bigg(\frac{ r^2 \phi^2 \psi^2}{f}-r^2 \lambda\psi^4-2r^2\tau\psi^6\nonumber\\
    +\frac{1}{2}\mathrm{e}^{\alpha\psi^2}r^2\alpha\psi^2\phi'^2\bigg)\mathrm{d}r\bigg)~.
\end{align}
Here, $V_2$ is just the volume of the spatial boundary manifold. In the rest of this manuscript, we take $r_h=1$, $m^2=-2$ and $L=1$.

In addition, to analyze the dynamical stability of a system, we also need to compute its quasinormal modes. We adopt perturbations of the following form
\begin{align}
\delta \psi=\tilde{\sigma}(r,t,x)+\mathrm{i}\tilde{\eta}(r,t,x)~,~\delta A_\mu=\tilde{a}_\mu(r,t,x)~.
\end{align}
The specific form of the perturbation is $\mathrm{e}^{-\mathrm{i}(\omega t - kx)}$. Substituting this perturbation into the equations of motion and expanding to linear order, we obtain the final perturbation equations
\begin{align}
    \mathrm{i} a_t \psi  \omega +\frac{\mathrm{i} a_t f k \psi }{r^2}-f^2 \eta ''+\eta '
   (-\frac{2 f^2}{r}-f f')&\nonumber\\
   +\eta  (-\frac{1}{2} \alpha  f \mathrm{e}^{\alpha  \psi ^2}
   \phi '^2+\frac{f k^2}{r^2}+2 f \lambda  \psi ^2+3 f \tau  \psi ^4&\nonumber\\
   -2 f-\omega
   ^2-\phi ^2)+2 \mathrm{i} \sigma  \omega  \phi=0&~,\\
   %_____________________________________________
   \alpha  f \psi  \mathrm{e}^{\alpha  \psi ^2} a_t' \phi '\nonumber+2 a_t \psi  \phi +f^2 \sigma
   ''+\sigma ' (\frac{2 f^2}{r}+f f')\nonumber\\
   +\sigma  (\alpha ^2 f \psi ^2 \mathrm{e}^{\alpha 
   \psi ^2} \phi '^2+\frac{1}{2} \alpha  f \mathrm{e}^{\alpha  \psi ^2} \phi
   '^2-\frac{f k^2}{r^2}\nonumber\\-6 f \lambda  \psi ^2-15 f \tau  \psi ^4+2 f+\omega ^2+\phi
   ^2)+2 \mathrm{i} \eta  \omega  \phi=0&~,\\
    %_____________________________________________
   f
   a_t''+a_t' (2 \alpha  f \psi  \psi '+\frac{2 f}{r})-2 \mathrm{i} \eta  \psi  \omega  \mathrm{e}^{-\alpha  \psi ^2}\nonumber\\
   +a_t (-2
   \psi ^2 \mathrm{e}^{-\alpha  \psi ^2}-\frac{k^2}{r^2})-\frac{a_x k \omega }{r^2}\nonumber\\
   +2 \alpha  f \psi  \sigma ' \phi '+2 \alpha  f \sigma  \psi 
   \phi ''+2 \alpha  f \sigma  \psi ' \phi '\nonumber\\
   +\frac{4 \alpha  f \sigma  \psi  \phi '}{r}-4 \sigma  \psi  \phi  \mathrm{e}^{-\alpha  \psi ^2}+4 \alpha
   ^2 f \sigma  \psi ^2 \psi ' \phi '=0&~,\\
%_____________________________________________
   \frac{a_t k \omega }{f}+f a_x''+a_x' f'+2 \alpha  f \psi  a_x' \psi '\nonumber\\
   -2
   a_x \psi ^2 \mathrm{e}^{-\alpha  \psi ^2}+\frac{a_x \omega ^2}{f}+2 \mathrm{i} \eta  k \psi 
   \mathrm{e}^{-\alpha  \psi ^2}=0&~,
\end{align}
and a constraint equation
\begin{align}
    \mathrm{i} r^2 \omega  a_t'+\mathrm{i} f k a_x'-2 f r^2 \psi  \mathrm{e}^{-\alpha  \psi ^2} \eta '&\nonumber\\
    +2 f \eta  r^2
   \mathrm{e}^{-\alpha  \psi ^2} \psi '+2 \mathrm{i} \alpha  r^2 \sigma  \psi  \omega  \phi '=0&~.
\end{align}
To solve for the quasinormal modes, boundary conditions also need to be imposed. We adopt ingoing boundary conditions
\begin{align}
&\eta(r)=(r-1)^\xi(\eta^{(0)}+\eta^{(1)}(r-1)+...)~,\\
&\sigma(r)=(r-1)^\xi(\sigma^{(0)}+\sigma^{(1)}(r-1)+...)~,\\
&a_t(r)=(r-1)^{\xi+1}(a^{(0)}_t+a^{(1)}_t(r-1)+...)~,\\
&a_x(r)=(r-1)^{\xi}(a^{(0)}_x+a^{(1)}_x(r-1)+...)~,
\end{align}
with $\xi=-\mathrm{i}\omega/3$. Among the above boundary conditions, there are three independent solutions characterized by the horizon coefficients. To fully determine the quasinormal mode frequencies of the system, we also need a set of linearly independent solutions, which can be obtained via gauge transformations
\begin{align}
\eta^{IV}=\mathrm{i}\beta\psi~,\sigma^{IV}=0~,a^{IV}_t=\beta\omega~,a^{IV}_x=-\beta k~,
\end{align}
here, $\beta$ is an arbitrary constant.
When solving the complete equations of motion, each set of independent variables $\{ \eta^{(0)}, \sigma^{(0)}, a_x^{(0)} \}$ determines a set of solutions. Together with the solutions obtained from gauge transformations, this yields the matrix required for solving the quasinormal modes. For the quasinormal modes frequencies, we require
\begin{align}
0 = \frac{1}{\beta} \det
\begin{pmatrix}
\eta^{I} & \eta^{II} & \eta^{{III}} & \eta^{{IV}} \\
\sigma^{{I}} & \sigma^{{II}} & \sigma^{{III}} & \sigma^{ {IV}} \\
a_{t}^{ {I}} & a_{t}^{ {II}} & a_{t}^{ {III}} & a_{t}^{ {IV}} \\
a_{x}^{ {I}} & a_{x}^{ {II}} & a_{x}^{ {III}} & a_{x}^{ {IV}}
\end{pmatrix}.
\end{align}
%在这篇文章中，我们采用谱方法来求解似正规模的频率。在求解出似正规模的频率之后，系统的稳定性通过似正规模的虚部是否小于0来判断。如果最低的mode虚部大于0，那么此时系统就是不稳定的，反之，如果系统没有虚部大于0的mode，那么系统就是稳定的。
In this work, we employ the spectral method to solve for the quasinormal mode frequencies. After obtaining the quasinormal mode frequencies, the stability of the system is determined by whether the imaginary part is negative. If the imaginary part of the lowest mode is positive, the system is unstable. Conversely, if no mode has a positive imaginary part, the system is stable.

\section{Stability analysis}\label{sec3}
%对于一个定义良好的系统，他的热力学稳定性和动力学应该是一致的。热力学稳定性由自由能决定，动力学稳定性则由系统的微扰，也就是最低一级的QNMs决定。但是他们都来自于landscape分析中的最稳定极小值点。比如说，如果系统存在不稳定的自由能分支，那么它的landscape一定处于局域极大值点，在这一情况下，如果给这个局域极大值点施加一个微扰，那么处于鞍点的值会滚落到任意一个极小值点上。如果打开空间方向，这一过程就对应与泡泡的产生。在这一章节的剩下部分，我们将分别讨论不同相变的稳定性。
For a well-defined system, its thermodynamic stability and dynamical stability should be consistent. Thermodynamic stability is determined by the free energy, while dynamical stability is determined by the perturbations of the system, namely the lowest-lying quasinormal modes. However, both arise from the most stable local minimum in the landscape analysis \cite{Zhao:2022jvs}. For instance, if an unstable branch of the free energy exists, its landscape must be located at a local maximum. In this case, if a perturbation is applied to this local maximum, the system will roll down to one of the local minima. When spatial directions are turned on, this process corresponds to bubble formation \cite{Janik:2015iry,Janik:2017ykj,Attems:2019yqn,Bellantuono:2019wbn,Li:2020ayr,Chen:2022tfy,Zhao:2023ffs,Chen:2024pyy,Zhao:2026eav,Jin:2026lzp,Zhao:2026jqp,Zhao:2026als}. In the remainder of this section, we will discuss the stability of different phase transitions separately.

\subsection{Second-order phase transition}
%我们在图1中给出了凝聚和自由能，在过了临界点之后，normal解通过二阶相变转变为超流解。从自由能上来看，它表现为超流解具有更低的自由能。
We start from the simplest case of a second-order phase transition. In our model, when $\lambda = \tau = 0$, it reduces to the simplest second-order phase transition model.
\begin{figure}
\centering
\includegraphics[width=1\columnwidth]{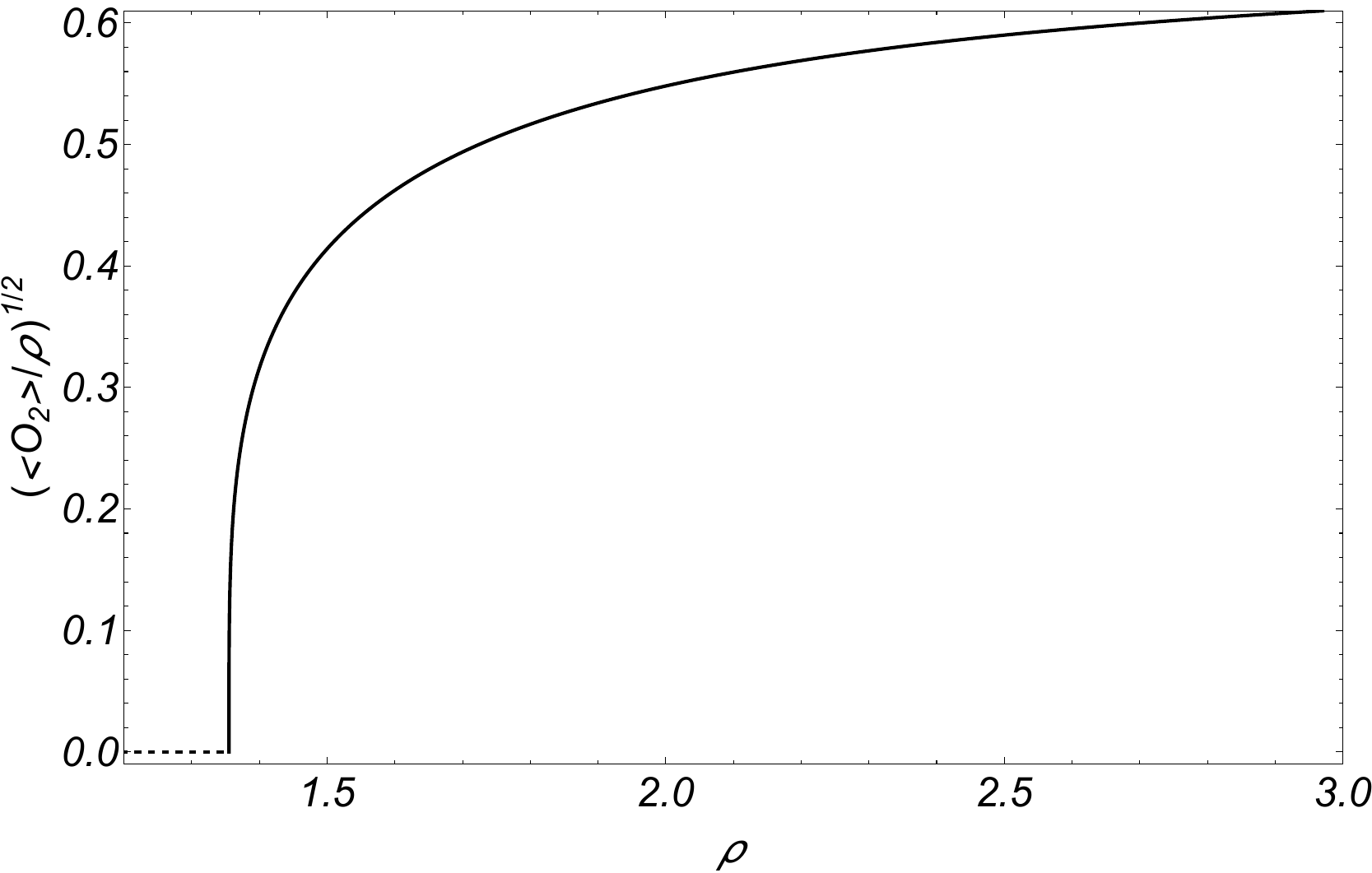}
\includegraphics[width=1\columnwidth]{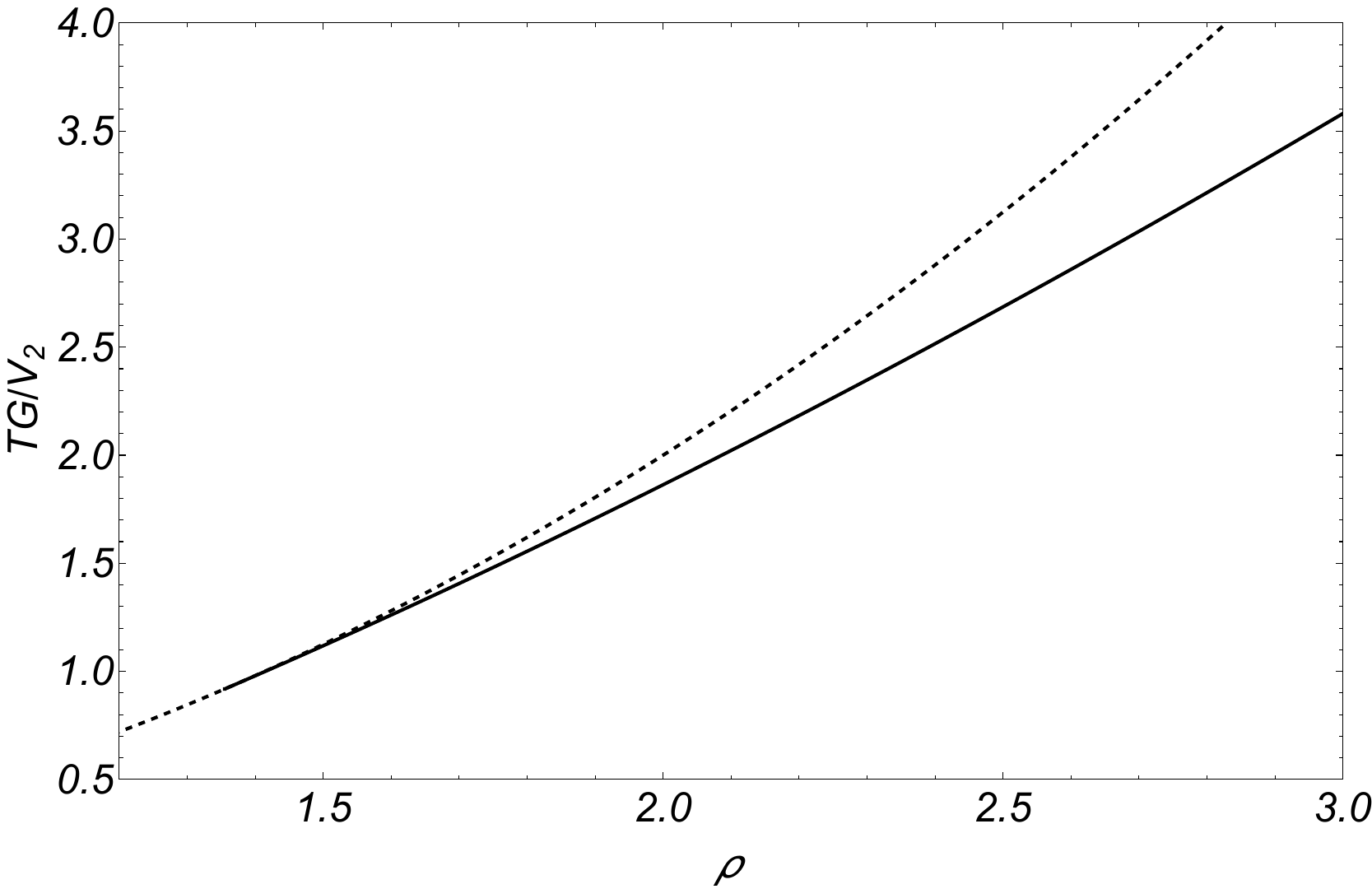}
\caption{The condensate and free energy for $\lambda=0$ and $\tau=0$ with $\alpha=5$. The dashed lines correspond to the normal solution, and the solid lines correspond to the superfluid solution.}\label{co2nd}
\end{figure}
In Fig.~\ref{co2nd}, we present the condensate and the free energy. Beyond the critical point, the normal solution transforms into the superfluid solution via a second-order phase transition. As reflected in the free energy, the superfluid solution exhibits a lower free energy.

%对于一个系统，它的动力学稳定性应该和热力学稳定性是一致的。这种动力学稳定性表现在QNMs上就是系统没有虚部为正的模式。我们在图2中给出了二阶相变的似正规模随着凝聚值的变化情况。在凝聚值比较小的时候，黑色圆圈表示的最低一级模式会向下运动，蓝色三角形表示的mode会向上运动，随后在$(O_2/rho)^1/2=0.428$处发生跨越，此时最低级的mode就变成了蓝色三角形表示的mode。随着凝聚值的继续增大，黑色圆圈表示的mode会与另外一个mode发生碰撞，转变成一对实部不为零的左右对称的mode，也就是图2中绿色正方形表示的mode

%其中，蓝色三角形和黑色圆圈表示实部为零的mode，绿色方块表示实部不为零的mode。随着凝聚的增大，黑色圆圈表示的最低一级的mode会在$(\langle O_2 \rangle/\rho)^{1/2} = 0.428$的位置发生一次crossover，此时最低一级的mode会变成蓝色三角形表示的mode。
\begin{figure}
\centering
\includegraphics[width=0.98\columnwidth]{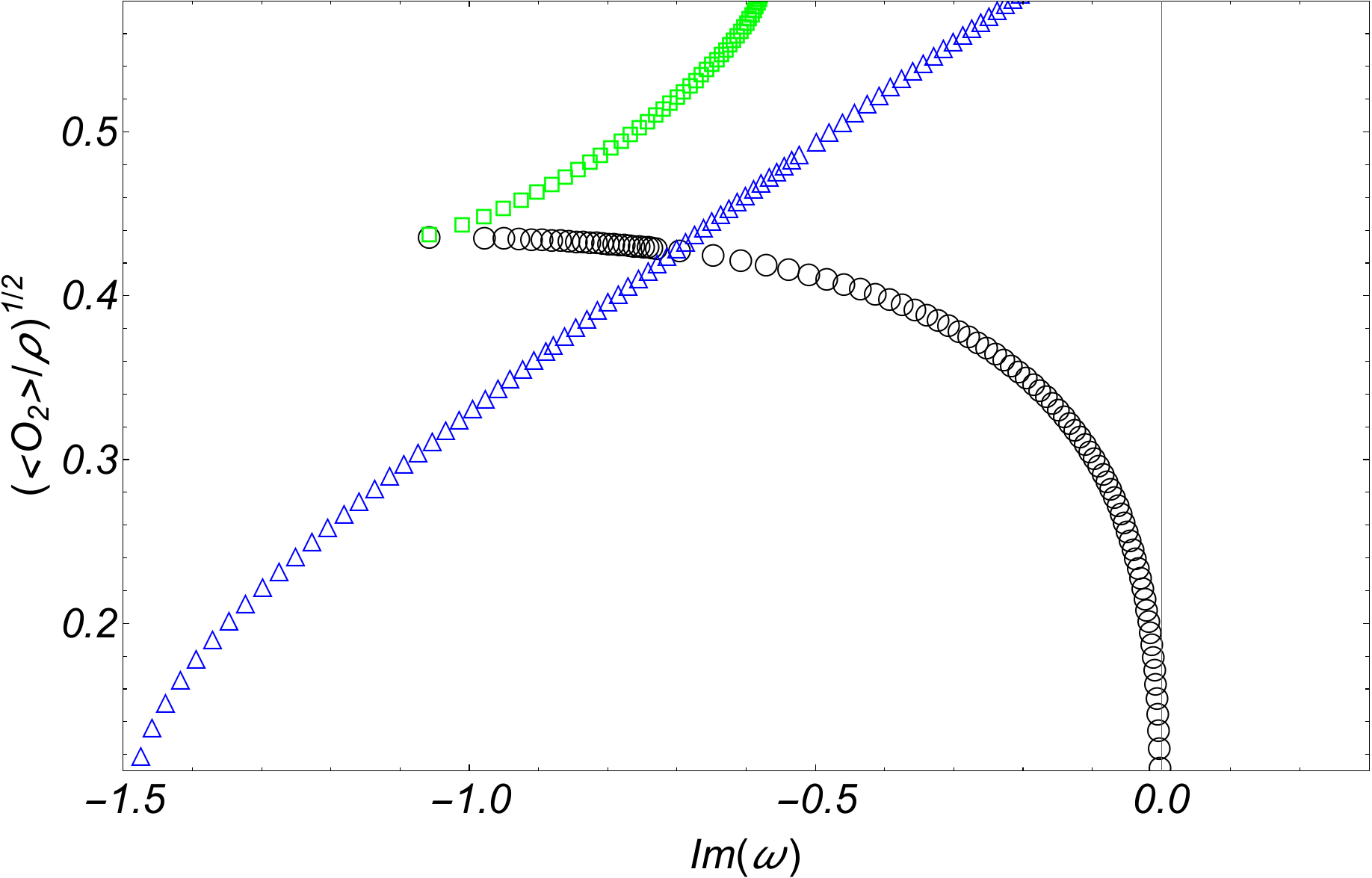}
\includegraphics[width=1\columnwidth]{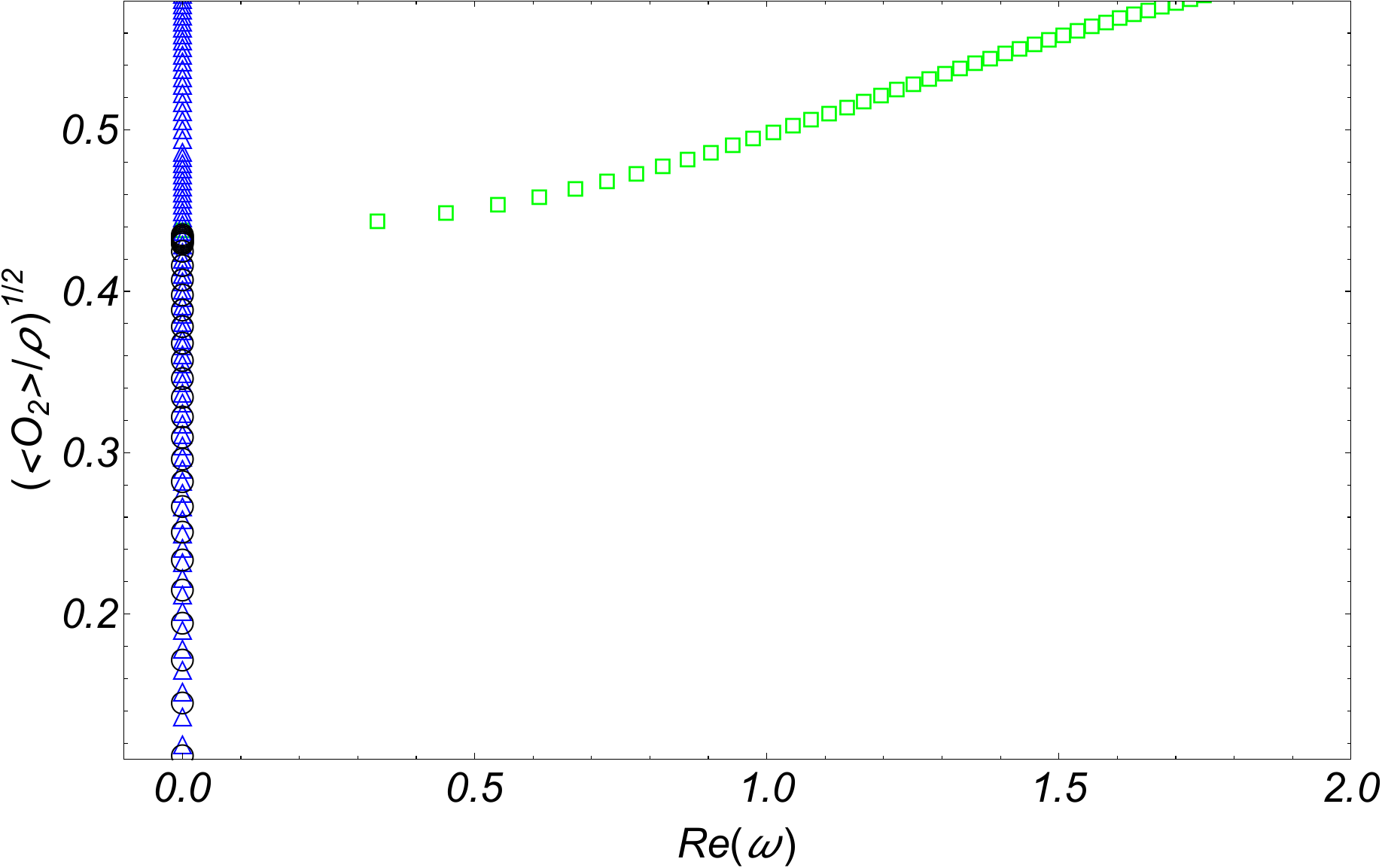}
\caption{The quasinormal modes for $\lambda=0$ and $\tau=0$ with $\alpha=5$. The blue triangles and black circles represent modes with vanishing real part, whereas the green squares represent modes with nonvanishing real part. As the condensate increases, the lowest-lying mode represented by the black circles undergoes a crossover at $(\langle O_2 \rangle/\rho)^{1/2} = 0.428$, where it becomes the mode represented by the blue triangles.}\label{QNMs2nd}
\end{figure}
For a given system, its dynamical stability should be consistent with its thermodynamic stability. This dynamical stability is reflected in the quasinormal modes by the absence of modes with positive imaginary parts. In Fig.~\ref{QNMs2nd}, we present the quasinormal modes for the second-order phase transition as a function of the condensate. When the condensate is relatively small, the lowest mode represented by black circles moves downward, while the mode represented by blue triangles moves upward. Subsequently, an avoided crossing occurs at $(\langle O_2 \rangle/\rho)^{1/2} = 0.428$, after which the lowest mode becomes the one represented by blue triangles. As the condensate further increases, the mode represented by black circles collides with another mode and transforms into a pair of modes with opposite nonzero real parts, i.e., the modes represented by green squares in Fig.~\ref{QNMs2nd}. Although the modes in the second-order phase transition undergo various changes, the system remains stable overall, which is consistent with the results obtained from the free energy.
%尽管二阶相变的模式会发生多种多样的模式上的改变，但系统总体总是稳定的，这一点正好和自由能的结果一致。

\subsection{Zeroth-order phase transition}\label{zerothorderPT}
%接下来我们将讨论一下零阶相变的稳定性。文章[]中，作者已经通过热力学和动力学的方法详细分析过零阶相变的稳定性，结果表明零阶相变是一个不稳定的模型，理论上不应该发生。在这个工作中，我们也计算了一下零阶相变的热力学和动力学稳定性。在图3中，我们给出了零阶相变的凝聚图和自由能。从自由能上可以看出来，此时系统还存在一支自由能更高的解，也就是图3中的红色实线。在图4中，我们给出了0阶相变的似正规模的计算结果。可以看到，当系统的凝聚值处于自由能更高的解，也就是红色分支的时候，系统的似正规模也从虚部小于零变成了虚部大于零，也就是图4中红色正方形表示的模式。这意味着从动力学上来讲，此时系统也是不稳定的
Next, we will discuss the stability of the zeroth-order phase transition. In Ref. \cite{Zhao:2022jvs}, the authors have analyzed the stability of the zeroth-order phase transition in detail using thermodynamic and dynamical methods, showing that the zeroth-order phase transition corresponds to an unstable model and should not occur in theory. In this work, we also compute the thermodynamic and dynamical stability of the zeroth-order phase transition. 

\begin{figure}
\centering
\includegraphics[width=0.98\columnwidth]{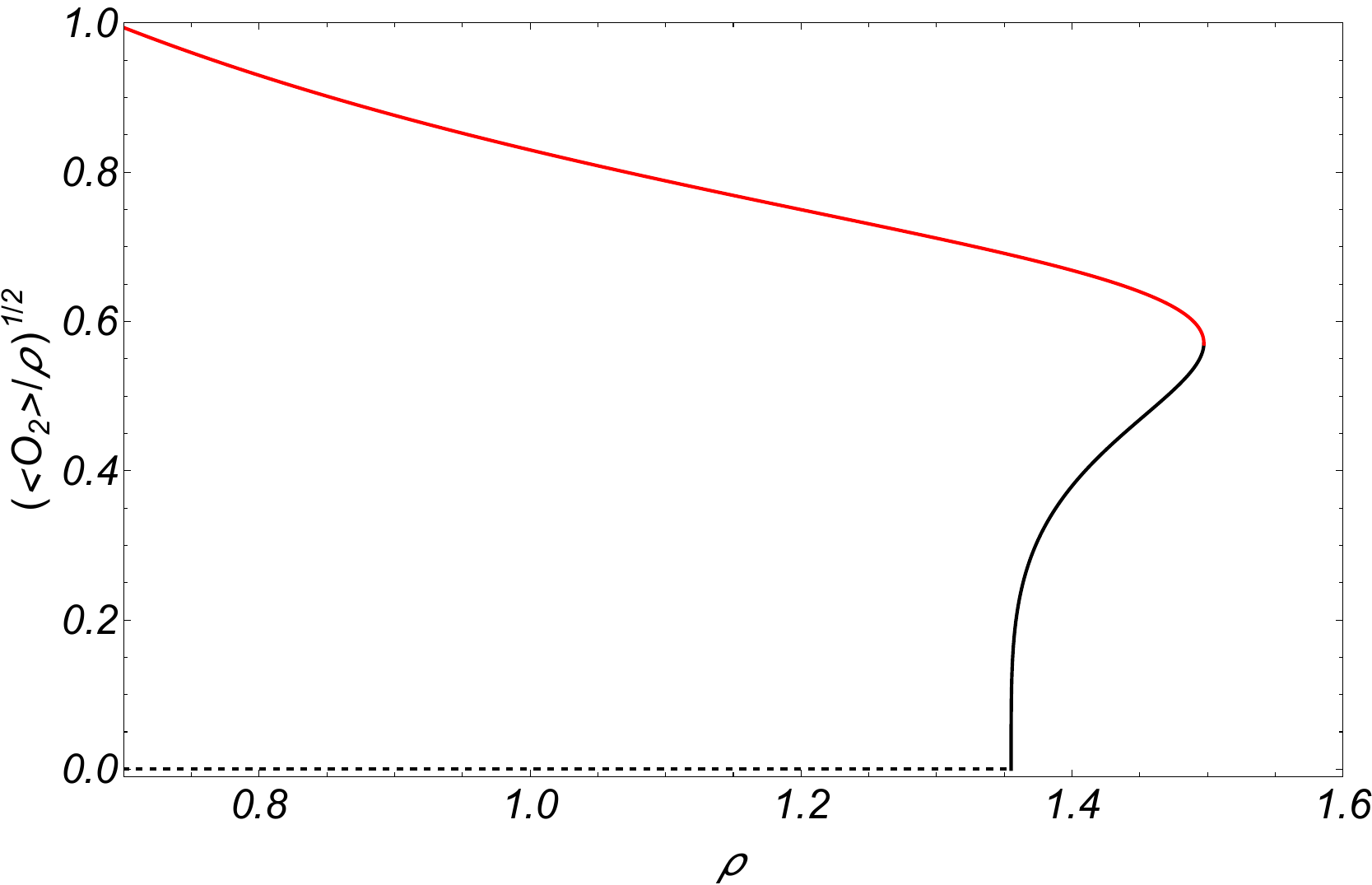}
\includegraphics[width=1\columnwidth]{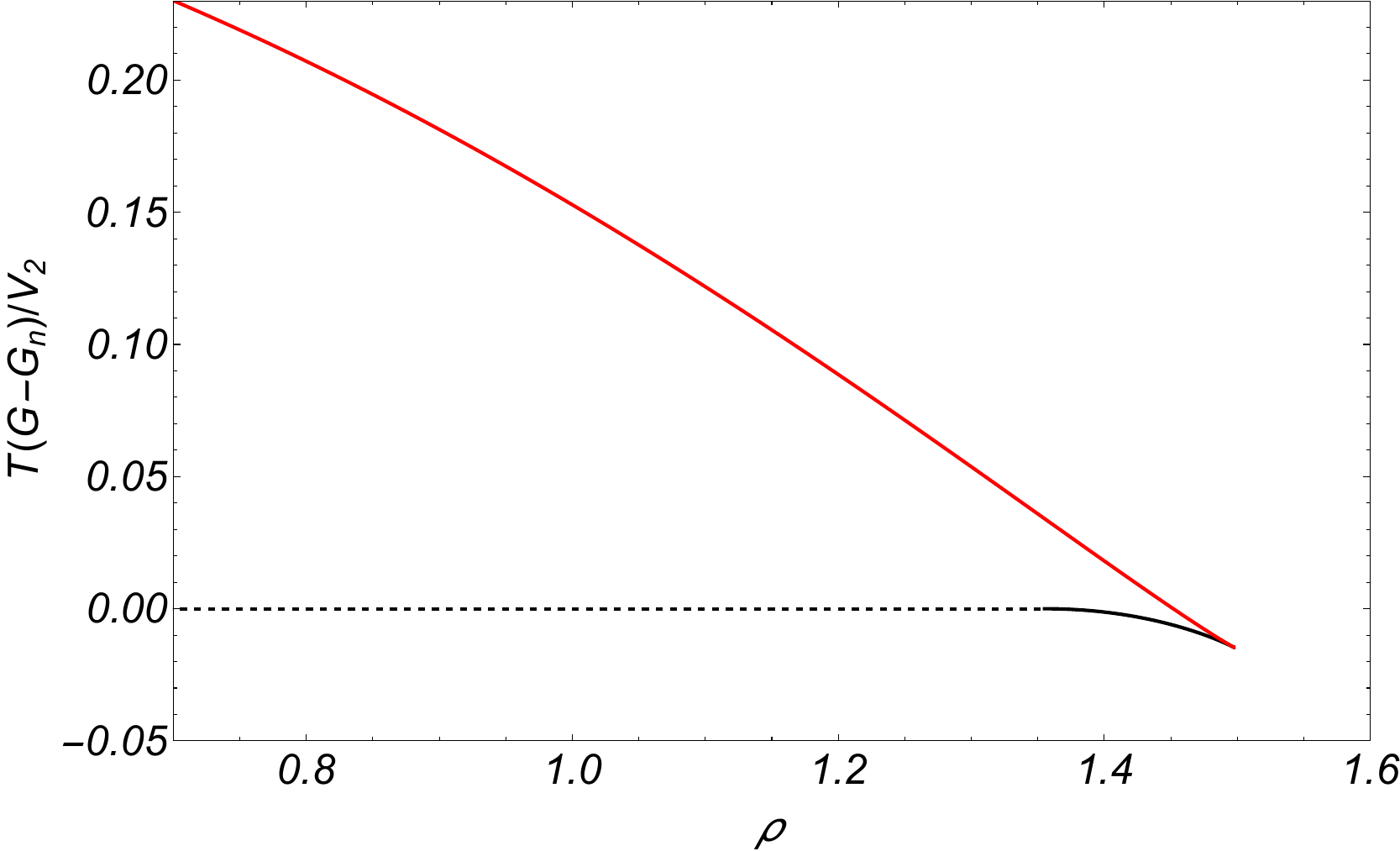}
\caption{The condensate and free energy for $\lambda=-4$ and $\tau=0$ with $\alpha=5$. The dashed lines correspond to the normal solution, and the solid lines correspond to the superfluid solution.}\label{c0th}
\end{figure}

\begin{figure}
\centering
\includegraphics[width=0.98\columnwidth]{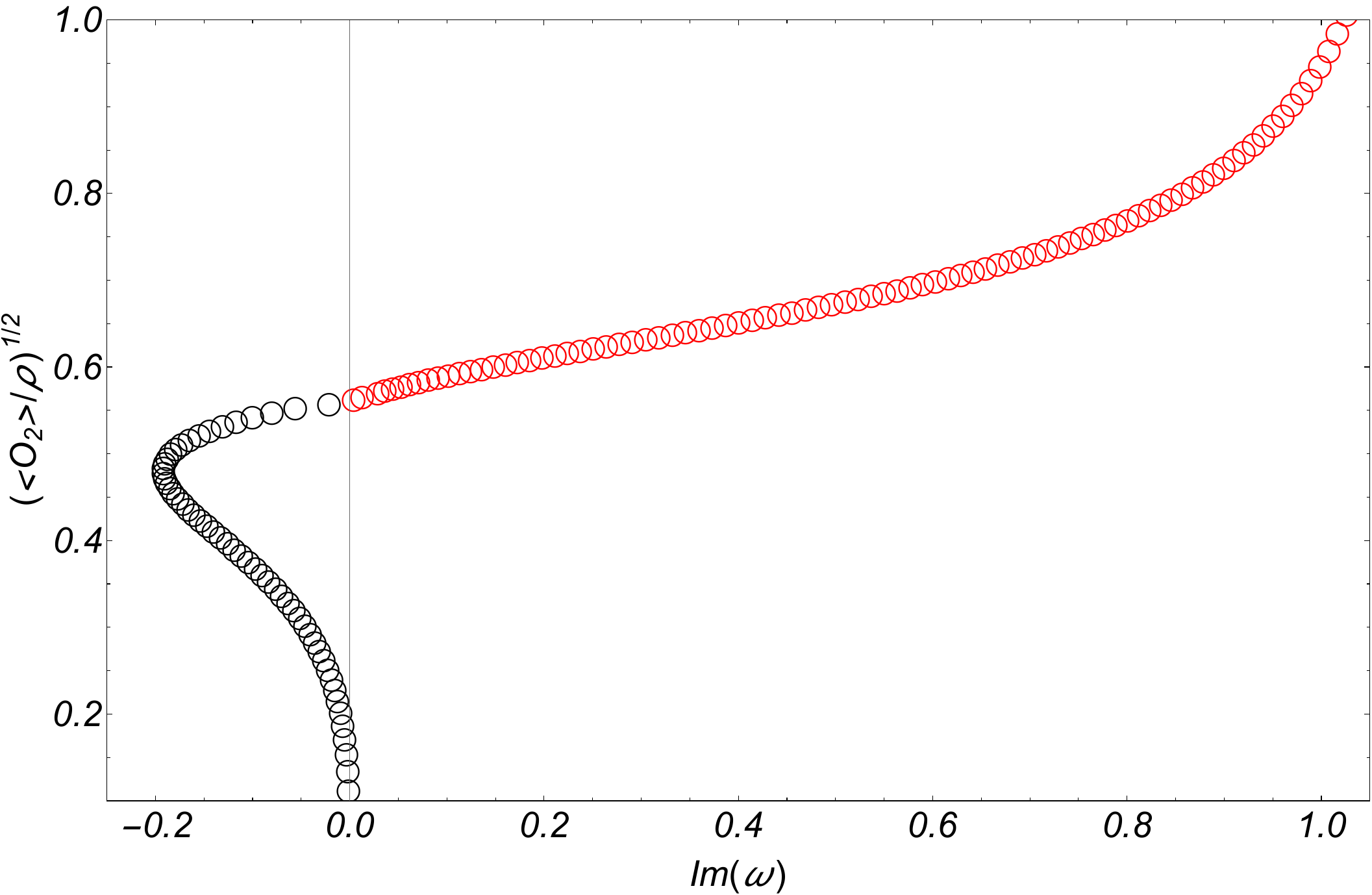}
\caption{The quasinormal modes for $\lambda=-4$ and $\tau=0$ with $\alpha=5$. Black circles correspond to modes on the stable branch in Fig.~\ref{c0th}, while red circles correspond to modes on the unstable branch. The point where the quasinormal modes changes from the stable to the unstable branch corresponds to the point where the condensate changes from black solid line to red solid line in Fig.~\ref{c0th}. All these modes have vanishing real parts.}\label{QNMs0th}
\end{figure}
%黑色圆圈对应于图c0th中稳定分支的modes，红色则对应于不稳定分支的modes，似正规模从稳定转变到不稳定的点对应于图c0th中凝聚图从黑色转变到红色的点。这里面所有modes都是实部为零的modes。

In Fig.~\ref{c0th}, we present the condensate and the free energy for the zeroth-order phase transition. From the free energy, it can be seen that there also exists a solution with higher free energy, i.e., the red solid line in Fig.~\ref{c0th}. In Fig.~\ref{QNMs0th}, we present the quasinormal mode results for the zeroth-order phase transition. It can be observed that when the system lies on the branch with higher free energy, i.e., the red branch, the imaginary part of the quasinormal modes changes from negative to positive, corresponding to the mode represented by the red circles in Fig.~\ref{QNMs0th}. This indicates that, from a dynamical perspective, the system is also unstable.

\subsection{COW phase transition}
%当我们打开tau参数的影响时，情况会更加复杂。如果只有一个负的lambda参数，对应的就是section\ref{zerothorderPT}描述的零阶相变的情况。但是如果此时我们考虑更高阶项的影响，系统在凝聚值足够大的时候还会出现一段自由能更低的解，此时系统将从零阶相变转变成cow相变。我们在图5中给出了COW相变的凝聚图和自由能曲线。值得注意的是，cow相变是一阶相变和二阶相变的组合，系统先经过二阶相变转变成超流解1(图5中黑色实线)，然后通过一阶相变转变为超流解2(图5中蓝色实线)。从自由能上来看，红色实线代表的区域是不稳定的区域。

\begin{figure}
\centering
\includegraphics[width=0.98\columnwidth]{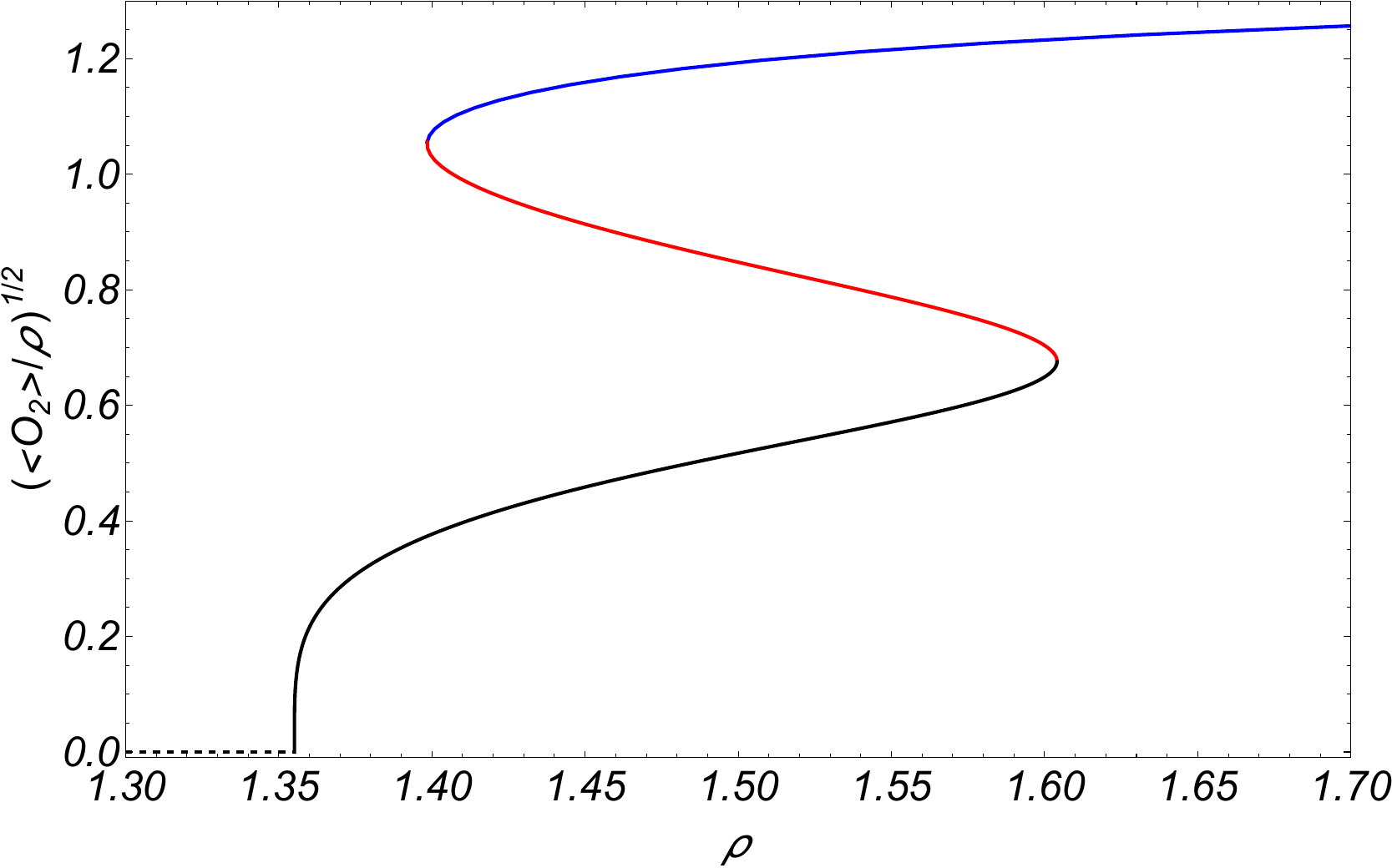}
\includegraphics[width=1\columnwidth]{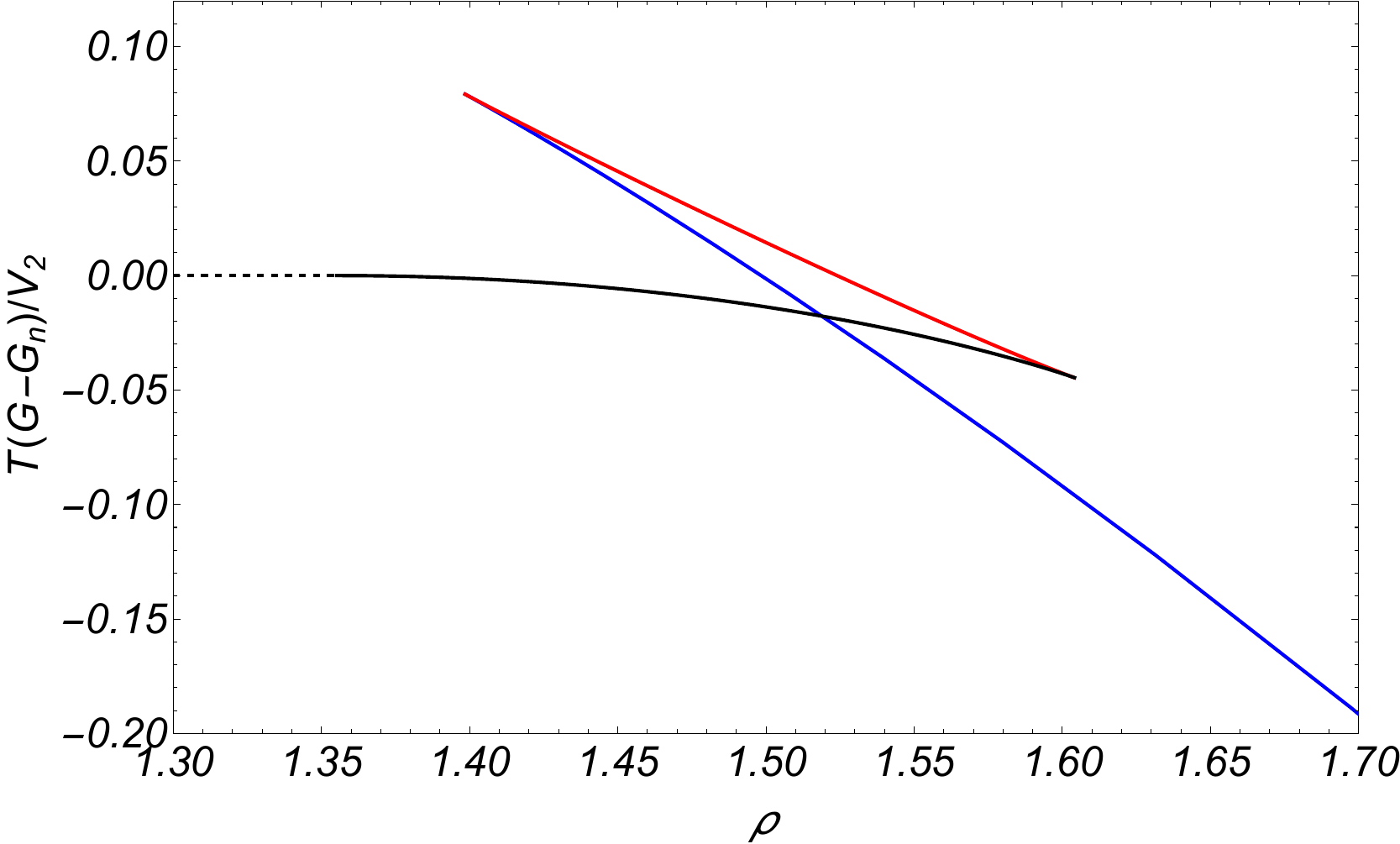}
\caption{The condensate and free energy for $\lambda=-4$ and $\tau=2.68$ with $\alpha=5$. The dashed lines correspond to the normal solution, and the solid lines correspond to the superfluid solution.}\label{cCOW}
\end{figure}

%为了验证这一点，我们同样计算了cow相变的似正规模。我们在图6中给出了COW相变的似正规模的计算结果，可以看到，与图5中自由能的不稳定区间一致。

\begin{figure}[!t]
\centering
\includegraphics[width=0.98\columnwidth]{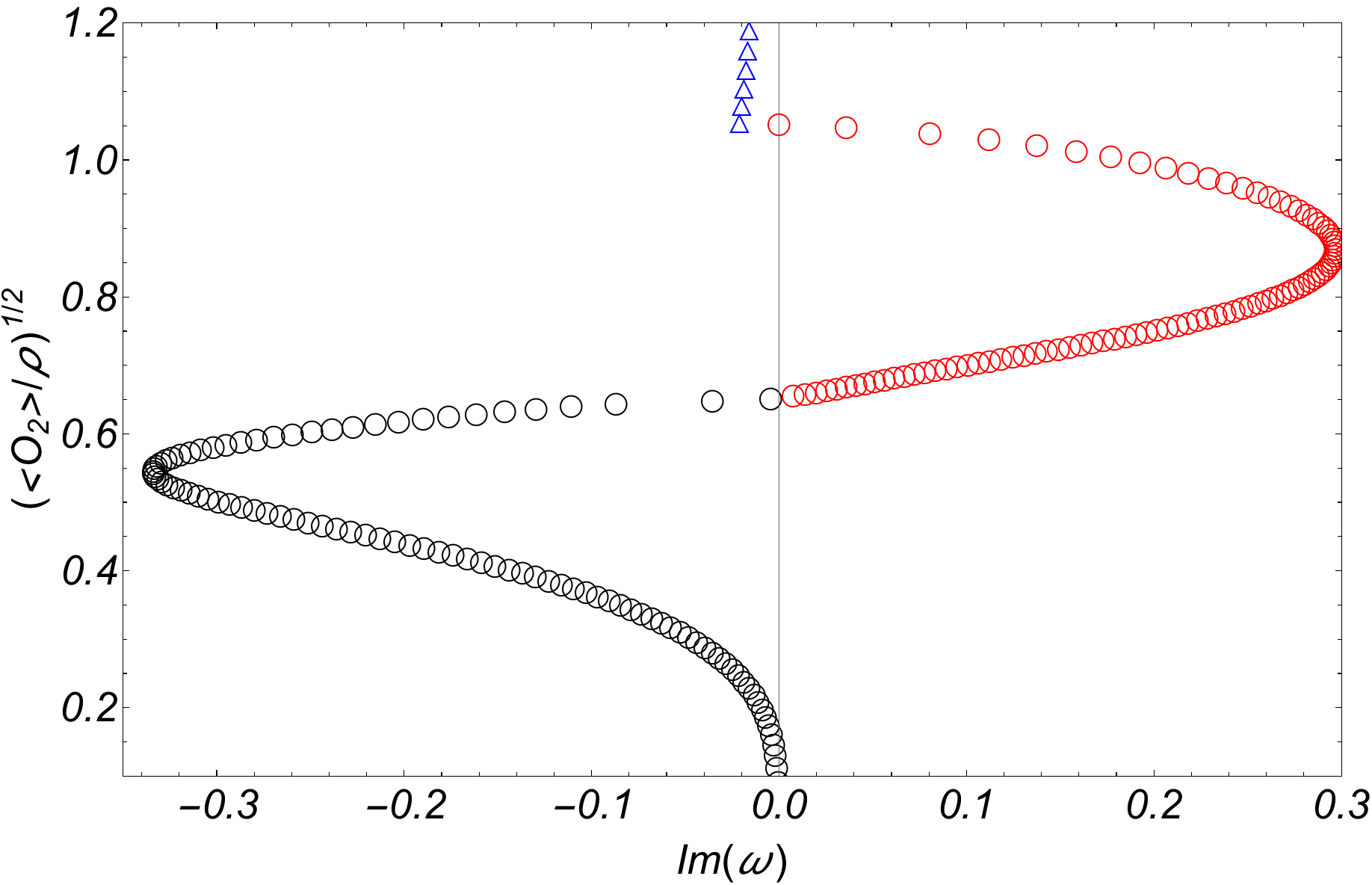}
\caption{The quasinormal modes for $\lambda=-4$ and $\tau=2.68$ with $\alpha=5$. Black circles and blue triangles correspond to modes on the stable branches in Fig.~\ref{cCOW}, while red circles correspond to modes on the unstable branch. The point where the quasinormal modes changes from stable to unstable corresponds to the point where the condensate changes from the black solid line to the red solid line in Fig.~\ref{cCOW}, and the point where it changes from unstable to stable corresponds to the point where the condensate changes from the red solid line to the blue solid line. All these modes have vanishing real parts.}\label{QNMsCOW}
\end{figure}

%黑色圆圈和蓝色三角形对应于图cCOW中稳定分支的modes，红色则对应于不稳定分支的modes，似正规模从稳定转变到不稳定的点对应于图cCOW中凝聚从黑色实线转变到红色实线的点，从不稳定转变到稳定的点则对应于图cCOW中凝聚图从红色实线转变到蓝色实线的点。这里面所有modes都是实部为零的modes。

When the effect of the term $\tau|\psi|^6$ is turned on, the situation becomes more complicated. If only a negative $\lambda$ parameter is present, the scenario corresponds to the zeroth-order phase transition described in Section \ref{zerothorderPT}. However, when the influence of higher-order term $\tau|\psi|^6$ is considered, a branch of solutions with lower free energy emerges at sufficiently large condensate values, and the system transitions from a zeroth-order phase transition to a COW phase transition. In Fig.~\ref{cCOW}, we present the condensate and the free energy curve for the COW phase transition. Notably, the COW phase transition is a combination of first-order and second-order phase transitions: the system first undergoes a second-order phase transition to superfluid solution 1 (the black solid line in Fig.~\ref{cCOW}), and then transforms into superfluid solution 2 (the blue solid line in Fig.~\ref{cCOW}) via a first-order phase transition. Briefly, COW is a complex version of a first-order phase transition. It not only contains a second-order phase transition from the normal solution to the superfluid solution, but also a first-order phase transition between superfluid solution 1 and superfluid solution 2. From the perspective of free energy, the region represented by the red solid line is unstable. To verify this, we also compute the quasinormal modes of the COW phase transition. In Fig.~\ref{QNMsCOW}, we present the quasinormal modes results for the COW phase transition. It can be seen that they are consistent with the unstable region of the free energy in Fig.~\ref{cCOW}.

%简单来说，cow是一个复杂版本的一阶相变，它不仅包含了从normal解到超流解的二阶相变，还包含了超流解1于超流解2之间的一阶相变。

%值得注意的是，cow相变情况下的QNMs的变化规律同样非常复杂。图6中蓝色三角形处发生了一次突变，这种情况和图2中黑色圆圈与蓝色三角形相互穿越的情况是一致的。因为我们关心的是系统的稳定性，而代表系统稳定的主要mode是最低的模式，所以我们并没有展示蓝色三角形表示的mode的完整演化路径，但是这也是一个非常有趣的问题。
It is worth noting that the behavior of the quasinormal mods in the COW phase transition is also quite complex. A sudden change occurs at the blue triangle in Fig.~\ref{QNMsCOW}, which is analogous to the crossing between the black circles and the blue triangles in Fig.~\ref{QNMs2nd}. Since our focus is on the stability of the system and the dominant mode governing stability is the lowest one, we do not present the complete evolution of the mode represented by the blue triangle. Nevertheless, this remains an interesting issue.

\section{Phase diagram}\label{sec4}
\begin{figure}[!t]
\centering
\includegraphics[width=0.98\columnwidth]{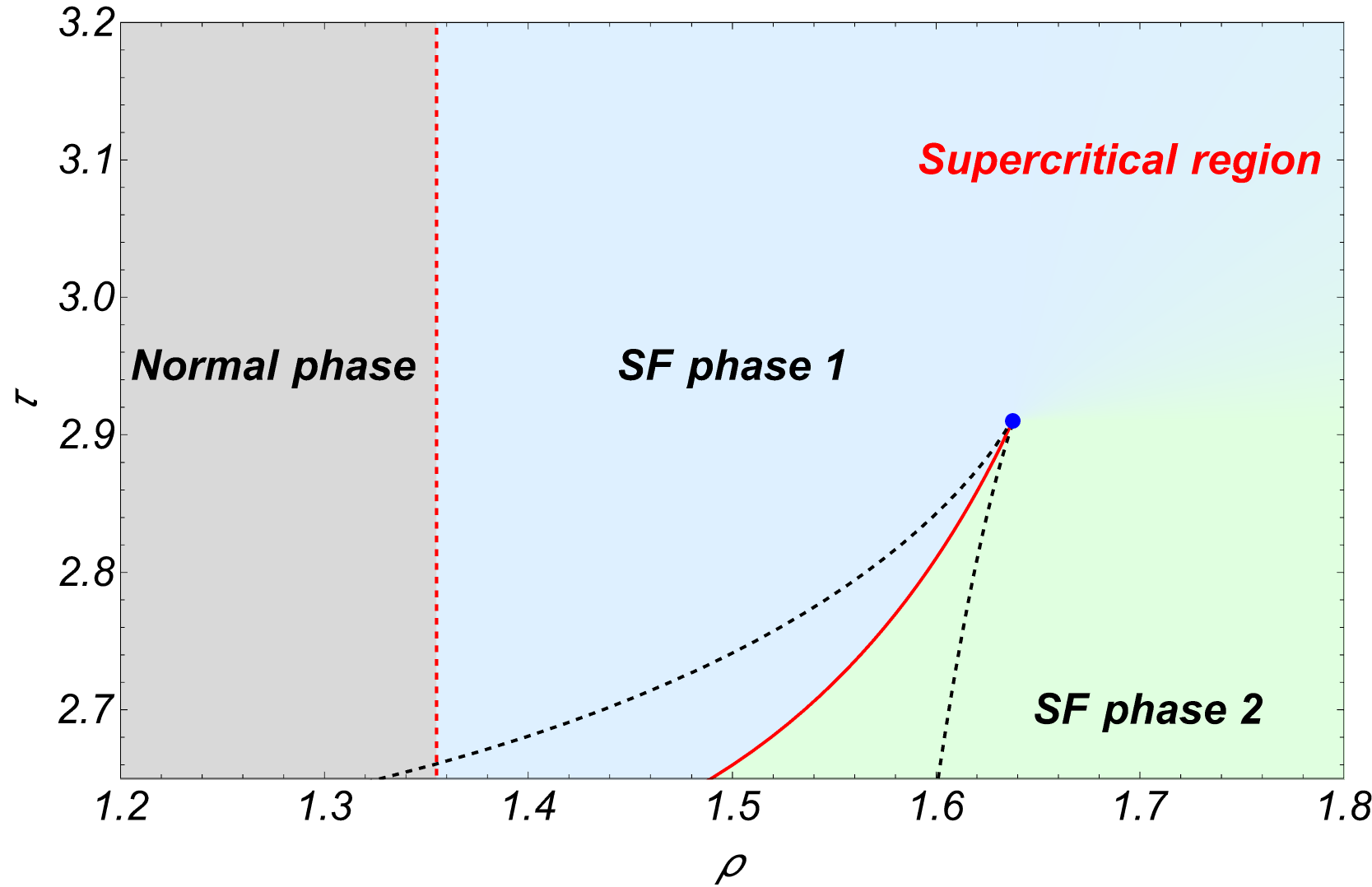}
\caption{The phase diagram for $\lambda = -4$ and $\alpha = 5$. The red dashed line denotes the critical point of the second-order phase transition, and the red solid line denotes the phase transition point of the first-order phase transition. The black dashed line represents the spinodal region of the first-order phase transition. The blue point indicates the critical point of the first-order phase transition. SF denotes the superfluid phase.}\label{TauphaseDiagram}
\end{figure}

Having discussed the stability of different types of phase transitions, we now turn to the phase diagram of this model.
In this model, due to the presence of the higher-order nonlinear terms $\lambda|\psi|^4$ and $\tau|\psi|^6$, the phase structure becomes very complex. Within this complex phase structure, a significant phenomenon can be realized, namely the critical phenomenon, where the region of the first-order phase transition shrinks, eventually reaching a critical point and then entering the supercritical region. In Fig.~\ref{TauphaseDiagram}, we present the phase diagram for fixed parameters $\lambda$ and $\alpha$  while varying $\tau$. It can be seen that the spinodal region, indicated by the dashed line, gradually shrinks as the parameter $\tau$ increases, eventually reaching the critical point marked by the blue dot. As $\tau$ increases further, the system enters the supercritical region. As the parameter $\tau$ increases, the spinodal region of the first-order phase transition gradually shrinks and reaches a critical point. If $\tau$ continues to increase, the system then enters the supercritical region. Entering the supercritical region implies the disappearance of the first-order phase transition, meaning that traditional thermodynamic methods for distinguishing between two different phases become invalid. Of course, frontier research shows that one can distinguish different supercritical subphases via the Widom line \cite{Xu_2005,Ruppeiner_2012,PhysRevLett.112.135701,PhysRevE.95.052120,Gallo2014}, the Frenkel line \cite{Yoon_2018,PhysRevLett.111.145901,Bolmatov2013,Prescher_2017,Bolmatov_2015,Fomin_2018,Fomin2015,PhysRevE.85.031203,2023PhRvR...5a3149H,Jiang_2025}, or nonequilibrium dynamical evolution \cite{Zhao:2026jqp}.

%随着tua参数的增加，一阶相变的spinodal区域会逐渐减小，并且达到一个临界点。如果此时继续增加tau，那么系统就会进入超临界区域。系统进入超临界区域意味着一阶相变的消失，也就是说传统的通过热力学区分两种不同相的方法失效了。当然，前沿研究表明，此时可以通过widom line，frenkel line或者是非平衡动力学演化的方式来区分不同的超临界亚相。

%当lambda=-4，alpha=5时候的相图。红色虚线表示二阶相变的临界点，红色实线表示一阶相变的相变点。黑色虚线表示一阶相变的spinodal区域。蓝色点表示一阶相变的相变点。SF表示superfluid，S-SF表示supercritical superfluid

\begin{figure}[!t]
\centering
\includegraphics[width=0.98\columnwidth]{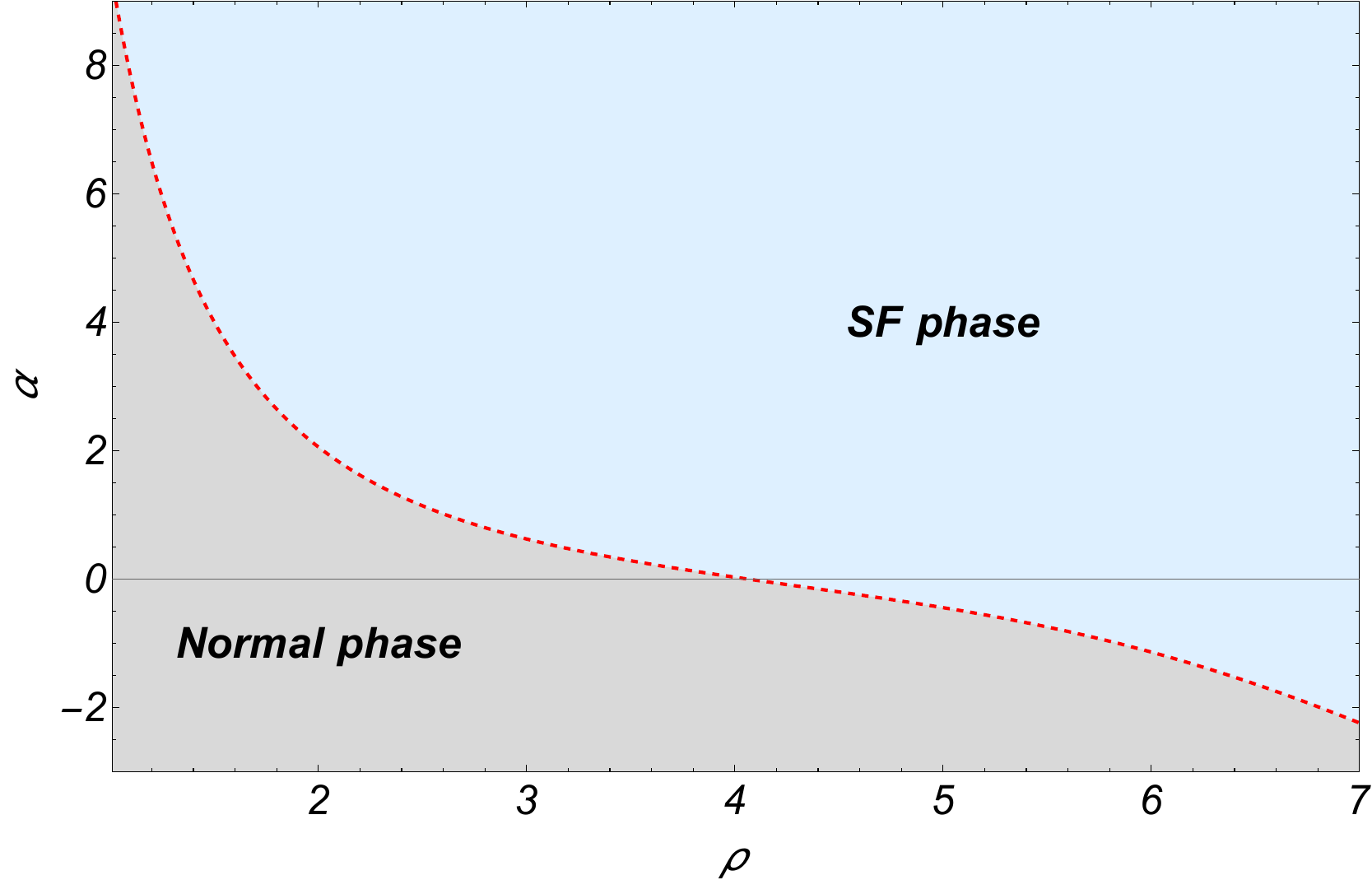}
\caption{The phase diagram for $\lambda =0$ and $\tau = 0$. The red dashed line denotes the critical point of the second-order phase transition. SF denotes the superfluid phase.}\label{criticalAlpha}
\end{figure}

%但是在我们这个模型中，因为考虑了非最小耦合，这里面还有一个活动的参数alpha，当我们放开alpha参数时，系统还会展现出更加丰富的相变行为。首先，我们计算了lambda和tau都等于0时改变alpha参数对临界点的影响，我们在图\ref{criticalAlpha}给出了计算结果。可以看到，当我们增加alpha参数时，系统的临界点rho_c会降低，减小alpha则会增大rho_c。此外，当alpha等于0时，系统会回到最标准的全息超导模型。不仅如此，alpha还可以取负值，并且继续降低alpha也会使得临界点增加。
However, in this model, due to the inclusion of non-minimal coupling, there is an additional active parameter $\alpha$. When $\alpha$ is varied, the system exhibits even richer phase transition behavior. First, we compute the effect of varying $\alpha$ on the critical point when $\lambda = \tau = 0$. The results are presented in Fig.~\ref{criticalAlpha}. It can be seen that increasing $\alpha$ lowers the critical value $\rho_c$, while decreasing $\alpha$ raises $\rho_c$. Moreover, when $\alpha = 0$, the system reduces to the standard holographic superfluid model \cite{Hartnoll:2008vx}. Notably, $\alpha$ can also take negative values, and further decreasing $\alpha$ also increases the critical point.

\begin{figure}[!t]
\centering
\includegraphics[width=0.98\columnwidth]{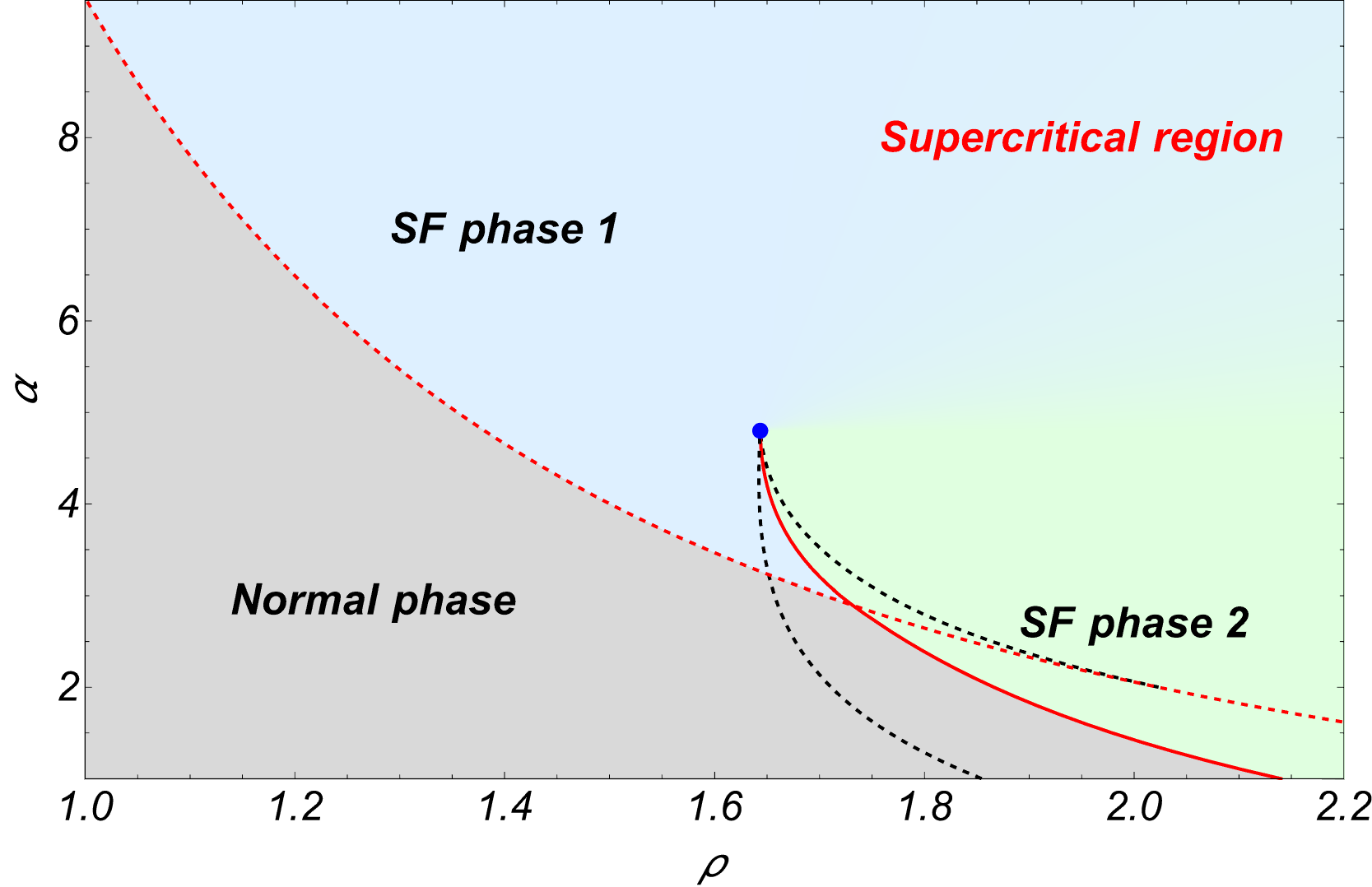}
\caption{The phase diagram for $\lambda = -4$ and $\tau = 2.93$. The red dashed line denotes the critical point of the second-order phase transition, and the red solid line denotes the phase transition point of the first-order phase transition. The black dashed line represents the spinodal region of the first-order phase transition. The blue point indicates the critical point of the first-order phase transition. SF denotes the superfluid phase.}\label{AlphaCriticalphaseDiagram}
\end{figure}

%如果此时我们同时打开lambda和tau，那么我们同样可以实现临界和超临界现象。我们在图AlphaCriticalphaseDiagram中给出了lambda=-4 τ=2.93时候的相图，可以看到，系统在增加alpha参数时同样会压缩spinodal区域，并且最终也将会进入超临界区域。与图TauphaseDiagram不同的是，改变alpha参数时系统的临界点也会同时改变，也就是图criticalAlpha中展示的趋势。这一点和反作用参数对系统的影响非常类似[]。
If we now turn on $\lambda$ and $\tau$ simultaneously, we can also realize the critical and supercritical phenomenon. In Fig.~\ref{AlphaCriticalphaseDiagram}, we present the phase diagram for $\lambda = -4$ and $\tau = 2.93$. It can be observed that increasing $\alpha$ also compresses the spinodal region, and the system eventually enters the supercritical region. Unlike the case in Fig.~\ref{TauphaseDiagram}, varying $\alpha$ also shifts the critical point, following the trend shown in Fig.~\ref{criticalAlpha}. This behavior is very similar to the effect of the back-reaction parameter on the system \cite{Zhao:2025tqq}.

\begin{figure}[!t]
\centering
\includegraphics[width=0.98\columnwidth]{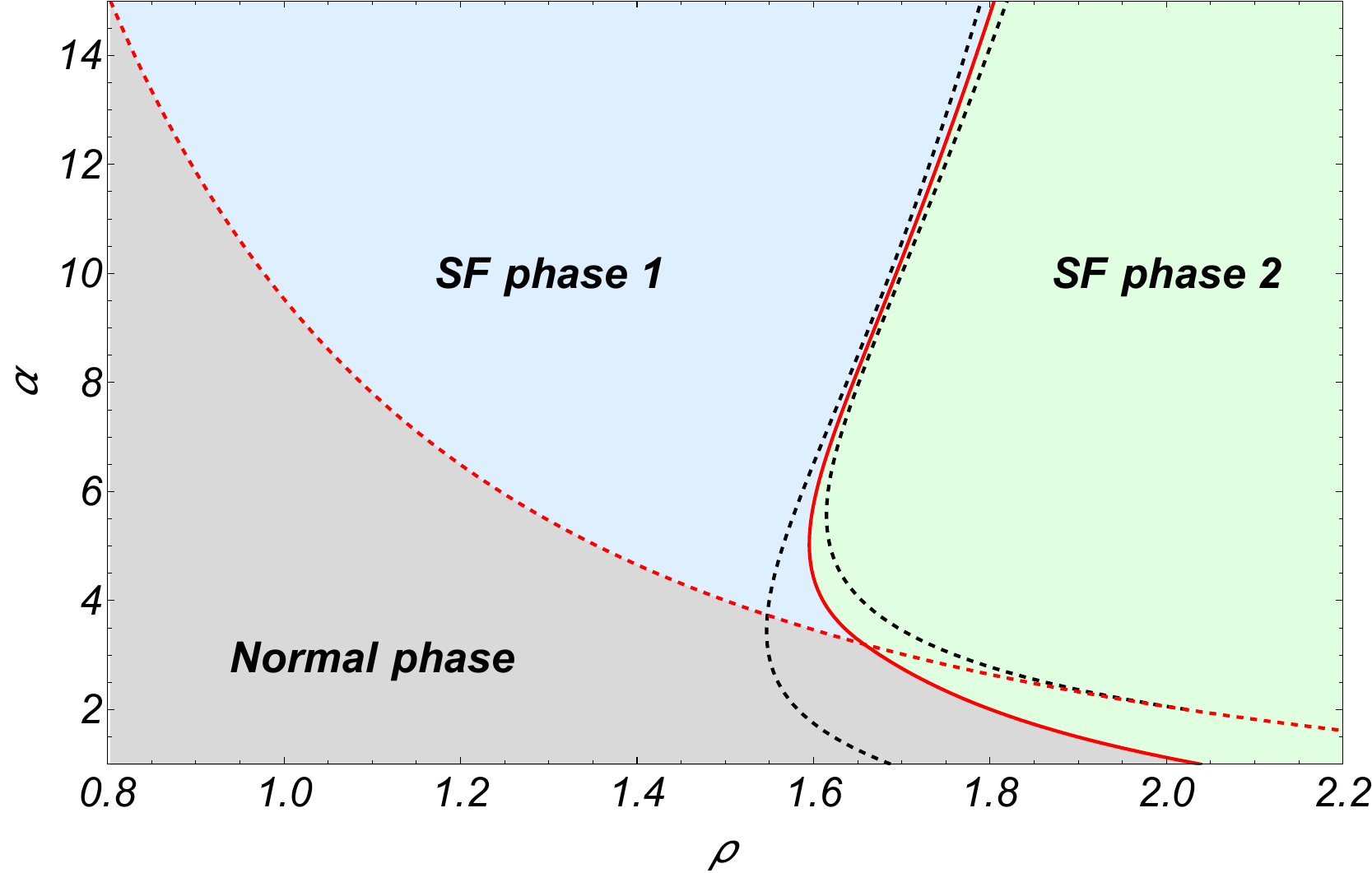}
\caption{The phase diagram for $\lambda = -4$ and $\tau = 2.8$. The red dashed line denotes the critical point of the second-order phase transition, and the red solid line denotes the phase transition point of the first-order phase transition. The black dashed line represents the spinodal region of the first-order phase transition.}\label{AlphaNoCriticalphaseDiagram}
\end{figure}

%但是，如果当我们选择一个比较小的tau参数时，改变alpha参数同样可能会出现图AlphaNoCriticalphaseDiagram中展示的结果，即系统没有临界点。尽管从零开始增加alpha时，一阶相变的spinodal区域也会收缩，但是在缩小到一定程度之后，随着alpha参数的继续增大，他又会继续扩张。
However, if a relatively small value of $\tau$ is chosen, varying $\alpha$ may lead to the scenario shown in Fig.~\ref{AlphaNoCriticalphaseDiagram}, where the system exhibits no critical point. Although increasing $\alpha$ from zero initially shrinks the spinodal region of the first-order phase transition, beyond a certain point further increasing $\alpha$ causes the spinodal region to expand again.

\begin{figure}[!t]
\centering
\includegraphics[width=0.98\columnwidth]{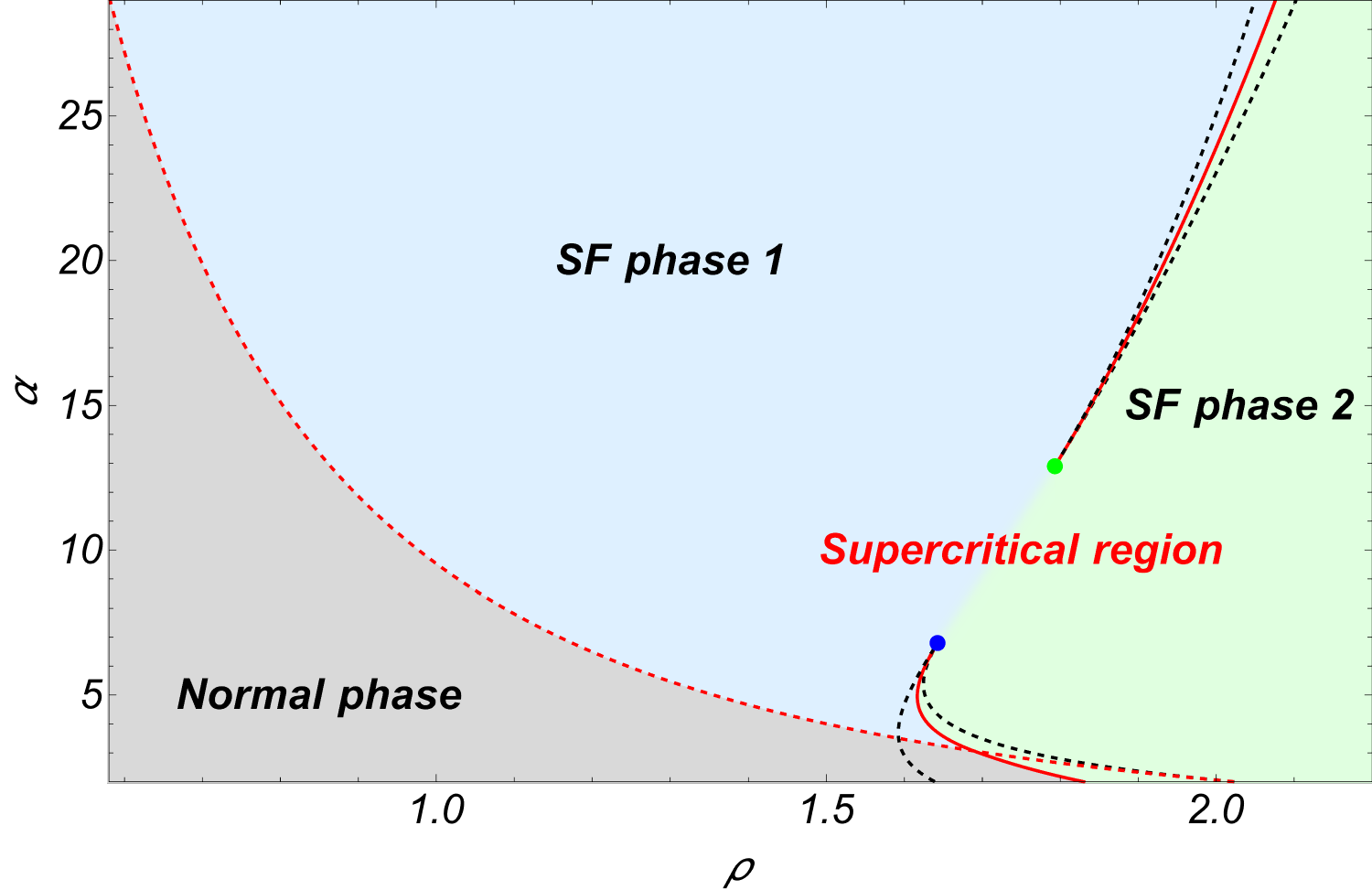}
\caption{The phase diagram for $\lambda = -4$ and $\tau = 2.85$. The red dashed line denotes the critical point of the second-order phase transition, and the red solid line denotes the phase transition point of the first-order phase transition. The black dashed line represents the spinodal region of the first-order phase transition. The blue and green points indicate the critical point of the first-order phase transition. SF denotes the superfluid phase.}\label{doubleCriticalphaseDiagram}
\end{figure}

%最后，我们来讨论一下这个工作里面最有意思的部分。图AlphaNoCriticalphaseDiagram的现象告诉我们，alpha对相结构的影响并不是单调的，这一结果表明图AlphaCriticalphaseDiagram中的临界点似乎并不是一个单一的临界点，如果我们继续增加alpha参数，可能会发现另外一个临界点。但是，由于tau参数太大，这个临界点可能离现在的临界点非常远。于是，我们选择了一个相对来说比较小的tau参数，并且在遇到第一个临界点之后，继续增加alpha，不出意外的，我们发现了第二个临界点。这一结果我们展示在了图doubleCriticalphaseDiagram中。可以看到，alpha在超过第一个临界点（蓝色点）之后，系统进入了超临界区域，但是随后会遇到第二个临界点（绿色点），继续增加alpha系统则会从超临界再次转变成一阶相变主导的区域。
Finally, we discuss the most intriguing aspect of this work. The phenomenon shown in Fig.~\ref{AlphaNoCriticalphaseDiagram} indicates that the effect of $\alpha$ on the phase structure is non-monotonic. This suggests that the critical point in Fig.~\ref{AlphaCriticalphaseDiagram} may not be a single critical point. If we continue increasing $\alpha$, another critical point might emerge. However, for a large value of $\tau$, this additional critical point could be far from the current one. To explore this, we choose a relatively small $\tau$ and, after encountering the first critical point, continue increasing $\alpha$. As expected, a second critical point is observed. This result is presented in Fig.~\ref{doubleCriticalphaseDiagram}. It can be seen that after $\alpha$ exceeds the first critical point (blue point), the system enters the supercritical region, but subsequently encounters a second critical point (green point). Further increasing $\alpha$ then drives the system from the supercritical region back to a region dominated by first-order phase transitions. Conventional phase diagrams typically contain only a single critical point. When we turn on the coupling parameter $\alpha$, however, the system exhibits two critical points. This implies that, regardless of whether $\alpha$ is increased or decreased, the system can always find a path into the supercritical region.

%传统的相图一般只包含一个临界点，在我们放开耦合参数alpha的情况下，我们会得到一个包含两个临界点的系统。这意味着，不管是增大还是减小alpha，系统总能找到一个路径进入超临界区域。

\section{Conclusions}\label{sec5}
%本文旨在系统研究由$\lambda |\psi|^4$、$\tau|\psi|^6$以及非最小耦合$\mathrm{e}^{\alpha\psi^2}$共同作用的Einstein-Maxwell标量理论中的稳定性与相变行为。在稳定性分析上，我们采用自由能和似正规模，我们发现，在自由能呈现出不稳定性的区间上，似正规模的结果同样展现出了动力学上的不稳定性，这一结果与最小耦合的结果一致。在相图分析中，我们发现了丰富的临界与超临界行为。固定$\lambda<0$和$\alpha$，增大$\tau$会使一阶相变区域（spinodal区域）逐渐收缩，最终汇聚于一个临界点，之后系统进入超临界区域。更为有趣的是，当变化非最小耦合参数$\alpha$时，系统的行为呈现出两种不同的情形。对于比较小的$\tau$，系统可能完全不出现临界点：随着$\alpha$增大，一阶相变区域先收缩后扩张，spinodal区域始终无法收缩至一点，因而系统始终停留在一阶相变主导区。最引人注目的是第二种情形，对于比较大的$\tau$值，系统在增大$\alpha$的过程中会先后出现两个临界点：系统首先从一阶相变主导区进入超临界区，随后又从超临界区重新进入一阶相变主导区。这种在单一参数驱动下的“双重临界现象”在全息超导模型中尚属首次报道。它揭示了非最小耦合参数$\alpha$与高阶相互作用项$\tau|\psi|^6$之间复杂的非单调耦合效应，表明$\alpha$不仅仅是一个简单的调控参数，其改变可以导致相结构发生反转式的重构。
This paper systematically investigated the stability and phase transition behavior in the Einstein–Maxwell–scalar theory incorporating $\lambda |\psi|^4$, $\tau|\psi|^6$, and the non-minimal coupling $\mathrm{e}^{\alpha|\psi|^2}$. For the stability analysis, we employ the free energy and quasinormal modes. We find that in regions where the free energy indicates instability, the quasinormal modes consistently exhibit dynamical instability, which is consistent with the results obtained in the case of $\alpha=0$ \cite{Zhao:2022jvs}.

In the phase diagram analysis, we observe rich critical and supercritical behaviors. For fixed $\lambda<0$ and $\alpha$, increasing $\tau$ gradually shrinks the first-order phase transition region (the spinodal region), which eventually converges to a critical point, beyond which the system enters the supercritical region. More intriguingly, varying the non-minimal coupling parameter $\alpha$ leads to two distinct scenarios. For relatively small $\tau$, the system may exhibit no critical point at all: as $\alpha$ increases, the first-order region first shrinks and then expands, and the spinodal region never collapses to a point, so the system remains in a region dominated by first-order phase transitions. The most striking case is the second scenario. For sufficiently large $\tau$, as $\alpha$ increases, the system sequentially undergoes two critical points: it first transitions from a first-order-dominated region to the supercritical region, and then reenters a first-order-dominated region. This “double critical phenomenon” driven by a single parameter has not been previously reported in holographic superfluid models. It reveals a complex nonmonotonic coupling effect between the non-minimal coupling parameter $\alpha$ and the higher-order interaction term $\tau|\psi|^6$, indicating that $\alpha$ not only serves as a simple tuning parameter but can also induce a structural reversal in the phase diagram.

%总结上来讲，这个工作为我们研究全息模型中的复杂的相结构提供了一个新的方法，通过非最小耦合，我们可以得到临界甚至是双临界现象。未来我们可以继续研究这种双临界系统的超临界跨越，这是一个非常有意思的研究方向。此外，这种存在双临界点的系统是否在动力学上也会表现出奇异的现象，也是一个非常有意思的问题。
In summary, this work provides a new approach for investigating complex phase structures in holographic models. Through the non-minimal coupling, critical and even double critical phenomenon can be realized. Future work may further explore the supercritical crossover in such double critical systems, which represents a highly interesting direction. Moreover, whether systems exhibiting double critical points also display exotic dynamical behavior remains an intriguing question.

\section*{Acknowledgement}
This work is supported by the National Natural Science Foundation of China (grant nos. 12533001, 12575049, 12473001, 12205039, 12305058, 11965013 and 12575054). ZYN is partially supported by Yunnan High-level Talent Training Support Plan Young $\&$ Elite Talents Project (grant no. YNWR-QNBJ-2018-181). This work is also supported by the National SKA Program of China (grant nos. 2022SKA0110200 and 2022SKA0110203) and the 111 Project (grant no. B16009).

\bibliography{reference}

%apsrev4-2.bst 2019-01-14 (MD) hand-edited version of apsrev4-1.bst
%Control: key (0)
%Control: author (8) initials jnrlst
%Control: editor formatted (1) identically to author
%Control: production of article title (0) allowed
%Control: page (0) single
%Control: year (1) truncated
%Control: production of eprint (0) enabled
\begin{thebibliography}{84}%
\makeatletter
\providecommand \@ifxundefined [1]{%
 \@ifx{#1\undefined}
}%
\providecommand \@ifnum [1]{%
 \ifnum #1\expandafter \@firstoftwo
 \else \expandafter \@secondoftwo
 \fi
}%
\providecommand \@ifx [1]{%
 \ifx #1\expandafter \@firstoftwo
 \else \expandafter \@secondoftwo
 \fi
}%
\providecommand \natexlab [1]{#1}%
\providecommand \enquote  [1]{``#1''}%
\providecommand \bibnamefont  [1]{#1}%
\providecommand \bibfnamefont [1]{#1}%
\providecommand \citenamefont [1]{#1}%
\providecommand \href@noop [0]{\@secondoftwo}%
\providecommand \href [0]{\begingroup \@sanitize@url \@href}%
\providecommand \@href[1]{\@@startlink{#1}\@@href}%
\providecommand \@@href[1]{\endgroup#1\@@endlink}%
\providecommand \@sanitize@url [0]{\catcode `\\12\catcode `\$12\catcode
  `\&12\catcode `\#12\catcode `\^12\catcode `\_12\catcode `\%12\relax}%
\providecommand \@@startlink[1]{}%
\providecommand \@@endlink[0]{}%
\providecommand \url  [0]{\begingroup\@sanitize@url \@url }%
\providecommand \@url [1]{\endgroup\@href {#1}{\urlprefix }}%
\providecommand \urlprefix  [0]{URL }%
\providecommand \Eprint [0]{\href }%
\providecommand \doibase [0]{https://doi.org/}%
\providecommand \selectlanguage [0]{\@gobble}%
\providecommand \bibinfo  [0]{\@secondoftwo}%
\providecommand \bibfield  [0]{\@secondoftwo}%
\providecommand \translation [1]{[#1]}%
\providecommand \BibitemOpen [0]{}%
\providecommand \bibitemStop [0]{}%
\providecommand \bibitemNoStop [0]{.\EOS\space}%
\providecommand \EOS [0]{\spacefactor3000\relax}%
\providecommand \BibitemShut  [1]{\csname bibitem#1\endcsname}%
\let\auto@bib@innerbib\@empty
%</preamble>
\bibitem [{\citenamefont {Maldacena}(1998)}]{Maldacena:1997re}%
  \BibitemOpen
  \bibfield  {author} {\bibinfo {author} {\bibfnamefont {J.~M.}\ \bibnamefont
  {Maldacena}},\ }\bibfield  {title} {\bibinfo {title} {{The Large N limit of
  superconformal field theories and supergravity}},\ }\href
  {https://doi.org/10.4310/ATMP.1998.v2.n2.a1} {\bibfield  {journal} {\bibinfo
  {journal} {Adv. Theor. Math. Phys.}\ }\textbf {\bibinfo {volume} {2}},\
  \bibinfo {pages} {231} (\bibinfo {year} {1998})},\ \Eprint
  {https://arxiv.org/abs/hep-th/9711200} {arXiv:hep-th/9711200} \BibitemShut
  {NoStop}%
\bibitem [{\citenamefont {Hartnoll}\ \emph
  {et~al.}(2008{\natexlab{a}})\citenamefont {Hartnoll}, \citenamefont
  {Herzog},\ and\ \citenamefont {Horowitz}}]{Hartnoll:2008vx}%
  \BibitemOpen
  \bibfield  {author} {\bibinfo {author} {\bibfnamefont {S.~A.}\ \bibnamefont
  {Hartnoll}}, \bibinfo {author} {\bibfnamefont {C.~P.}\ \bibnamefont
  {Herzog}},\ and\ \bibinfo {author} {\bibfnamefont {G.~T.}\ \bibnamefont
  {Horowitz}},\ }\bibfield  {title} {\bibinfo {title} {{Building a Holographic
  Superconductor}},\ }\href {https://doi.org/10.1103/PhysRevLett.101.031601}
  {\bibfield  {journal} {\bibinfo  {journal} {Phys. Rev. Lett.}\ }\textbf
  {\bibinfo {volume} {101}},\ \bibinfo {pages} {031601} (\bibinfo {year}
  {2008}{\natexlab{a}})},\ \Eprint {https://arxiv.org/abs/0803.3295}
  {arXiv:0803.3295 [hep-th]} \BibitemShut {NoStop}%
\bibitem [{\citenamefont {Hartnoll}\ \emph
  {et~al.}(2008{\natexlab{b}})\citenamefont {Hartnoll}, \citenamefont
  {Herzog},\ and\ \citenamefont {Horowitz}}]{Hartnoll:2008kx}%
  \BibitemOpen
  \bibfield  {author} {\bibinfo {author} {\bibfnamefont {S.~A.}\ \bibnamefont
  {Hartnoll}}, \bibinfo {author} {\bibfnamefont {C.~P.}\ \bibnamefont
  {Herzog}},\ and\ \bibinfo {author} {\bibfnamefont {G.~T.}\ \bibnamefont
  {Horowitz}},\ }\bibfield  {title} {\bibinfo {title} {{Holographic
  Superconductors}},\ }\href {https://doi.org/10.1088/1126-6708/2008/12/015}
  {\bibfield  {journal} {\bibinfo  {journal} {JHEP}\ }\textbf {\bibinfo
  {volume} {12}},\ \bibinfo {pages} {015}},\ \Eprint
  {https://arxiv.org/abs/0810.1563} {arXiv:0810.1563 [hep-th]} \BibitemShut
  {NoStop}%
\bibitem [{\citenamefont {Herzog}(2010)}]{Herzog:2010vz}%
  \BibitemOpen
  \bibfield  {author} {\bibinfo {author} {\bibfnamefont {C.~P.}\ \bibnamefont
  {Herzog}},\ }\bibfield  {title} {\bibinfo {title} {{An Analytic Holographic
  Superconductor}},\ }\href {https://doi.org/10.1103/PhysRevD.81.126009}
  {\bibfield  {journal} {\bibinfo  {journal} {Phys. Rev. D}\ }\textbf {\bibinfo
  {volume} {81}},\ \bibinfo {pages} {126009} (\bibinfo {year} {2010})},\
  \Eprint {https://arxiv.org/abs/1003.3278} {arXiv:1003.3278 [hep-th]}
  \BibitemShut {NoStop}%
\bibitem [{\citenamefont {Qiao}\ \emph {et~al.}(2020)\citenamefont {Qiao},
  \citenamefont {OuYang}, \citenamefont {Wang}, \citenamefont {Pan},\ and\
  \citenamefont {Jing}}]{Qiao:2020hkx}%
  \BibitemOpen
  \bibfield  {author} {\bibinfo {author} {\bibfnamefont {X.}~\bibnamefont
  {Qiao}}, \bibinfo {author} {\bibfnamefont {L.}~\bibnamefont {OuYang}},
  \bibinfo {author} {\bibfnamefont {D.}~\bibnamefont {Wang}}, \bibinfo {author}
  {\bibfnamefont {Q.}~\bibnamefont {Pan}},\ and\ \bibinfo {author}
  {\bibfnamefont {J.}~\bibnamefont {Jing}},\ }\bibfield  {title} {\bibinfo
  {title} {{Holographic superconductors in 4D Einstein-Gauss-Bonnet gravity}},\
  }\href {https://doi.org/10.1007/JHEP12(2020)192} {\bibfield  {journal}
  {\bibinfo  {journal} {JHEP}\ }\textbf {\bibinfo {volume} {12}},\ \bibinfo
  {pages} {192}},\ \Eprint {https://arxiv.org/abs/2005.01007} {arXiv:2005.01007
  [hep-th]} \BibitemShut {NoStop}%
\bibitem [{\citenamefont {Pan}\ \emph {et~al.}(2021)\citenamefont {Pan},
  \citenamefont {Qiao}, \citenamefont {Wang}, \citenamefont {Pan},
  \citenamefont {Nie},\ and\ \citenamefont {Jing}}]{Pan:2021jii}%
  \BibitemOpen
  \bibfield  {author} {\bibinfo {author} {\bibfnamefont {J.}~\bibnamefont
  {Pan}}, \bibinfo {author} {\bibfnamefont {X.}~\bibnamefont {Qiao}}, \bibinfo
  {author} {\bibfnamefont {D.}~\bibnamefont {Wang}}, \bibinfo {author}
  {\bibfnamefont {Q.}~\bibnamefont {Pan}}, \bibinfo {author} {\bibfnamefont
  {Z.-Y.}\ \bibnamefont {Nie}},\ and\ \bibinfo {author} {\bibfnamefont
  {J.}~\bibnamefont {Jing}},\ }\bibfield  {title} {\bibinfo {title}
  {{Holographic superconductors in 4D Einstein-Gauss-Bonnet gravity with
  backreactions}},\ }\href {https://doi.org/10.1016/j.physletb.2021.136755}
  {\bibfield  {journal} {\bibinfo  {journal} {Phys. Lett. B}\ }\textbf
  {\bibinfo {volume} {823}},\ \bibinfo {pages} {136755} (\bibinfo {year}
  {2021})},\ \Eprint {https://arxiv.org/abs/2109.02207} {arXiv:2109.02207
  [hep-th]} \BibitemShut {NoStop}%
\bibitem [{\citenamefont {Zhang}\ \emph {et~al.}(2024)\citenamefont {Zhang},
  \citenamefont {Nie}, \citenamefont {Zeng},\ and\ \citenamefont
  {Pan}}]{Zhang:2023uuq}%
  \BibitemOpen
  \bibfield  {author} {\bibinfo {author} {\bibfnamefont {X.-K.}\ \bibnamefont
  {Zhang}}, \bibinfo {author} {\bibfnamefont {Z.-Y.}\ \bibnamefont {Nie}},
  \bibinfo {author} {\bibfnamefont {H.}~\bibnamefont {Zeng}},\ and\ \bibinfo
  {author} {\bibfnamefont {Q.}~\bibnamefont {Pan}},\ }\bibfield  {title}
  {\bibinfo {title} {{The holographic s+p model in 4D and 5D
  Einstein-Gauss-Bonnet gravity}},\ }\href
  {https://doi.org/10.1016/j.physletb.2024.138496} {\bibfield  {journal}
  {\bibinfo  {journal} {Phys. Lett. B}\ }\textbf {\bibinfo {volume} {850}},\
  \bibinfo {pages} {138496} (\bibinfo {year} {2024})},\ \Eprint
  {https://arxiv.org/abs/2306.13308} {arXiv:2306.13308 [hep-th]} \BibitemShut
  {NoStop}%
\bibitem [{\citenamefont {Ghorai}\ and\ \citenamefont
  {Gangopadhyay}(2021)}]{Ghorai:2021uby}%
  \BibitemOpen
  \bibfield  {author} {\bibinfo {author} {\bibfnamefont {D.}~\bibnamefont
  {Ghorai}}\ and\ \bibinfo {author} {\bibfnamefont {S.}~\bibnamefont
  {Gangopadhyay}},\ }\bibfield  {title} {\bibinfo {title} {{Analytical study of
  holographic superconductor with backreaction in 4d Gauss-Bonnet gravity}},\
  }\href {https://doi.org/10.1016/j.physletb.2021.136699} {\bibfield  {journal}
  {\bibinfo  {journal} {Phys. Lett. B}\ }\textbf {\bibinfo {volume} {822}},\
  \bibinfo {pages} {136699} (\bibinfo {year} {2021})},\ \Eprint
  {https://arxiv.org/abs/2105.09423} {arXiv:2105.09423 [hep-th]} \BibitemShut
  {NoStop}%
\bibitem [{\citenamefont {Cai}\ and\ \citenamefont {Zhang}(2010)}]{Cai:2009hn}%
  \BibitemOpen
  \bibfield  {author} {\bibinfo {author} {\bibfnamefont {R.-G.}\ \bibnamefont
  {Cai}}\ and\ \bibinfo {author} {\bibfnamefont {H.-Q.}\ \bibnamefont
  {Zhang}},\ }\bibfield  {title} {\bibinfo {title} {{Holographic
  Superconductors with Ho\v{r}ava-Lifshitz Black Holes}},\ }\href
  {https://doi.org/10.1103/PhysRevD.81.066003} {\bibfield  {journal} {\bibinfo
  {journal} {Phys. Rev. D}\ }\textbf {\bibinfo {volume} {81}},\ \bibinfo
  {pages} {066003} (\bibinfo {year} {2010})},\ \Eprint
  {https://arxiv.org/abs/0911.4867} {arXiv:0911.4867 [hep-th]} \BibitemShut
  {NoStop}%
\bibitem [{\citenamefont {Gubser}\ and\ \citenamefont
  {Pufu}(2008)}]{Gubser:2008wv}%
  \BibitemOpen
  \bibfield  {author} {\bibinfo {author} {\bibfnamefont {S.~S.}\ \bibnamefont
  {Gubser}}\ and\ \bibinfo {author} {\bibfnamefont {S.~S.}\ \bibnamefont
  {Pufu}},\ }\bibfield  {title} {\bibinfo {title} {{The Gravity dual of a
  p-wave superconductor}},\ }\href
  {https://doi.org/10.1088/1126-6708/2008/11/033} {\bibfield  {journal}
  {\bibinfo  {journal} {JHEP}\ }\textbf {\bibinfo {volume} {11}},\ \bibinfo
  {pages} {033}},\ \Eprint {https://arxiv.org/abs/0805.2960} {arXiv:0805.2960
  [hep-th]} \BibitemShut {NoStop}%
\bibitem [{\citenamefont {Cai}\ \emph {et~al.}(2014)\citenamefont {Cai},
  \citenamefont {Li},\ and\ \citenamefont {Li}}]{Cai:2013aca}%
  \BibitemOpen
  \bibfield  {author} {\bibinfo {author} {\bibfnamefont {R.-G.}\ \bibnamefont
  {Cai}}, \bibinfo {author} {\bibfnamefont {L.}~\bibnamefont {Li}},\ and\
  \bibinfo {author} {\bibfnamefont {L.-F.}\ \bibnamefont {Li}},\ }\bibfield
  {title} {\bibinfo {title} {{A Holographic P-wave Superconductor Model}},\
  }\href {https://doi.org/10.1007/JHEP01(2014)032} {\bibfield  {journal}
  {\bibinfo  {journal} {JHEP}\ }\textbf {\bibinfo {volume} {01}},\ \bibinfo
  {pages} {032}},\ \Eprint {https://arxiv.org/abs/1309.4877} {arXiv:1309.4877
  [hep-th]} \BibitemShut {NoStop}%
\bibitem [{\citenamefont {Chen}\ \emph {et~al.}(2010)\citenamefont {Chen},
  \citenamefont {Kao}, \citenamefont {Maity}, \citenamefont {Wen},\ and\
  \citenamefont {Yeh}}]{Chen:2010mk}%
  \BibitemOpen
  \bibfield  {author} {\bibinfo {author} {\bibfnamefont {J.-W.}\ \bibnamefont
  {Chen}}, \bibinfo {author} {\bibfnamefont {Y.-J.}\ \bibnamefont {Kao}},
  \bibinfo {author} {\bibfnamefont {D.}~\bibnamefont {Maity}}, \bibinfo
  {author} {\bibfnamefont {W.-Y.}\ \bibnamefont {Wen}},\ and\ \bibinfo {author}
  {\bibfnamefont {C.-P.}\ \bibnamefont {Yeh}},\ }\bibfield  {title} {\bibinfo
  {title} {{Towards A Holographic Model of D-Wave Superconductors}},\ }\href
  {https://doi.org/10.1103/PhysRevD.81.106008} {\bibfield  {journal} {\bibinfo
  {journal} {Phys. Rev. D}\ }\textbf {\bibinfo {volume} {81}},\ \bibinfo
  {pages} {106008} (\bibinfo {year} {2010})},\ \Eprint
  {https://arxiv.org/abs/1003.2991} {arXiv:1003.2991 [hep-th]} \BibitemShut
  {NoStop}%
\bibitem [{\citenamefont {Kim}\ and\ \citenamefont
  {Taylor}(2013)}]{Kim:2013oba}%
  \BibitemOpen
  \bibfield  {author} {\bibinfo {author} {\bibfnamefont {K.-Y.}\ \bibnamefont
  {Kim}}\ and\ \bibinfo {author} {\bibfnamefont {M.}~\bibnamefont {Taylor}},\
  }\bibfield  {title} {\bibinfo {title} {{Holographic d-wave
  superconductors}},\ }\href {https://doi.org/10.1007/JHEP08(2013)112}
  {\bibfield  {journal} {\bibinfo  {journal} {JHEP}\ }\textbf {\bibinfo
  {volume} {08}},\ \bibinfo {pages} {112}},\ \Eprint
  {https://arxiv.org/abs/1304.6729} {arXiv:1304.6729 [hep-th]} \BibitemShut
  {NoStop}%
\bibitem [{\citenamefont {Cai}\ \emph {et~al.}(2013)\citenamefont {Cai},
  \citenamefont {Li}, \citenamefont {Li},\ and\ \citenamefont
  {Wang}}]{Cai:2013wma}%
  \BibitemOpen
  \bibfield  {author} {\bibinfo {author} {\bibfnamefont {R.-G.}\ \bibnamefont
  {Cai}}, \bibinfo {author} {\bibfnamefont {L.}~\bibnamefont {Li}}, \bibinfo
  {author} {\bibfnamefont {L.-F.}\ \bibnamefont {Li}},\ and\ \bibinfo {author}
  {\bibfnamefont {Y.-Q.}\ \bibnamefont {Wang}},\ }\bibfield  {title} {\bibinfo
  {title} {{Competition and Coexistence of Order Parameters in Holographic
  Multi-Band Superconductors}},\ }\href
  {https://doi.org/10.1007/JHEP09(2013)074} {\bibfield  {journal} {\bibinfo
  {journal} {JHEP}\ }\textbf {\bibinfo {volume} {09}},\ \bibinfo {pages}
  {074}},\ \Eprint {https://arxiv.org/abs/1307.2768} {arXiv:1307.2768 [hep-th]}
  \BibitemShut {NoStop}%
\bibitem [{\citenamefont {Wang}\ and\ \citenamefont
  {Liu}(2016)}]{Wang:2016jov}%
  \BibitemOpen
  \bibfield  {author} {\bibinfo {author} {\bibfnamefont {Y.-Q.}\ \bibnamefont
  {Wang}}\ and\ \bibinfo {author} {\bibfnamefont {S.}~\bibnamefont {Liu}},\
  }\bibfield  {title} {\bibinfo {title} {{Holographic s-wave and p-wave
  Josephson junction with backreaction}},\ }\href
  {https://doi.org/10.1007/JHEP11(2016)127} {\bibfield  {journal} {\bibinfo
  {journal} {JHEP}\ }\textbf {\bibinfo {volume} {11}},\ \bibinfo {pages}
  {127}},\ \Eprint {https://arxiv.org/abs/1608.06364} {arXiv:1608.06364
  [hep-th]} \BibitemShut {NoStop}%
\bibitem [{\citenamefont {Wang}\ \emph {et~al.}(2021)\citenamefont {Wang},
  \citenamefont {Li}, \citenamefont {Liu},\ and\ \citenamefont
  {Zhong}}]{Wang:2019vaq}%
  \BibitemOpen
  \bibfield  {author} {\bibinfo {author} {\bibfnamefont {Y.-Q.}\ \bibnamefont
  {Wang}}, \bibinfo {author} {\bibfnamefont {H.-B.}\ \bibnamefont {Li}},
  \bibinfo {author} {\bibfnamefont {Y.-X.}\ \bibnamefont {Liu}},\ and\ \bibinfo
  {author} {\bibfnamefont {Y.}~\bibnamefont {Zhong}},\ }\bibfield  {title}
  {\bibinfo {title} {{Excited states of holographic superconductors with
  backreaction}},\ }\href {https://doi.org/10.1140/epjc/s10052-021-09365-5}
  {\bibfield  {journal} {\bibinfo  {journal} {Eur. Phys. J. C}\ }\textbf
  {\bibinfo {volume} {81}},\ \bibinfo {pages} {628} (\bibinfo {year} {2021})},\
  \Eprint {https://arxiv.org/abs/1911.04475} {arXiv:1911.04475 [hep-th]}
  \BibitemShut {NoStop}%
\bibitem [{\citenamefont {Zhao}\ \emph
  {et~al.}(2025{\natexlab{a}})\citenamefont {Zhao}, \citenamefont {Nie},
  \citenamefont {Zhang},\ and\ \citenamefont {Zhang}}]{Zhao:2025tqq}%
  \BibitemOpen
  \bibfield  {author} {\bibinfo {author} {\bibfnamefont {Z.-Q.}\ \bibnamefont
  {Zhao}}, \bibinfo {author} {\bibfnamefont {Z.-Y.}\ \bibnamefont {Nie}},
  \bibinfo {author} {\bibfnamefont {J.-F.}\ \bibnamefont {Zhang}},\ and\
  \bibinfo {author} {\bibfnamefont {X.}~\bibnamefont {Zhang}},\ }\bibfield
  {title} {\bibinfo {title} {{Phase transitions in a holographic superfluid
  model with non-linear terms beyond the probe limit}},\ }\href
  {https://doi.org/10.1140/epjc/s10052-025-14712-x} {\bibfield  {journal}
  {\bibinfo  {journal} {Eur. Phys. J. C}\ }\textbf {\bibinfo {volume} {85}},\
  \bibinfo {pages} {1064} (\bibinfo {year} {2025}{\natexlab{a}})},\ \Eprint
  {https://arxiv.org/abs/2506.17274} {arXiv:2506.17274 [hep-ph]} \BibitemShut
  {NoStop}%
\bibitem [{\citenamefont {Zhao}\ \emph
  {et~al.}(2026{\natexlab{a}})\citenamefont {Zhao}, \citenamefont {Nie},
  \citenamefont {Wei}, \citenamefont {Zhang},\ and\ \citenamefont
  {Zhang}}]{Zhao:2025vtr}%
  \BibitemOpen
  \bibfield  {author} {\bibinfo {author} {\bibfnamefont {Z.-Q.}\ \bibnamefont
  {Zhao}}, \bibinfo {author} {\bibfnamefont {Z.-Y.}\ \bibnamefont {Nie}},
  \bibinfo {author} {\bibfnamefont {S.-W.}\ \bibnamefont {Wei}}, \bibinfo
  {author} {\bibfnamefont {J.-F.}\ \bibnamefont {Zhang}},\ and\ \bibinfo
  {author} {\bibfnamefont {X.}~\bibnamefont {Zhang}},\ }\bibfield  {title}
  {\bibinfo {title} {{Phase transitions in scalarized topological AdS black
  holes}},\ }\href {https://doi.org/10.1103/m8vj-hgk2} {\bibfield  {journal}
  {\bibinfo  {journal} {Phys. Rev. D}\ }\textbf {\bibinfo {volume} {114}},\
  \bibinfo {pages} {024020} (\bibinfo {year} {2026}{\natexlab{a}})},\ \Eprint
  {https://arxiv.org/abs/2511.18074} {arXiv:2511.18074 [gr-qc]} \BibitemShut
  {NoStop}%
\bibitem [{\citenamefont {Zhao}\ \emph
  {et~al.}(2026{\natexlab{b}})\citenamefont {Zhao}, \citenamefont {Nie},
  \citenamefont {Zhang}, \citenamefont {An}, \citenamefont {Zhang},\ and\
  \citenamefont {Zhang}}]{Zhao:2025odj}%
  \BibitemOpen
  \bibfield  {author} {\bibinfo {author} {\bibfnamefont {Z.-Q.}\ \bibnamefont
  {Zhao}}, \bibinfo {author} {\bibfnamefont {Z.-Y.}\ \bibnamefont {Nie}},
  \bibinfo {author} {\bibfnamefont {X.-K.}\ \bibnamefont {Zhang}}, \bibinfo
  {author} {\bibfnamefont {Y.-S.}\ \bibnamefont {An}}, \bibinfo {author}
  {\bibfnamefont {J.-F.}\ \bibnamefont {Zhang}},\ and\ \bibinfo {author}
  {\bibfnamefont {X.}~\bibnamefont {Zhang}},\ }\bibfield  {title} {\bibinfo
  {title} {{Interior structure of black holes with nonlinear terms}},\ }\href
  {https://doi.org/10.1140/epjc/s10052-026-15625-z} {\bibfield  {journal}
  {\bibinfo  {journal} {Eur. Phys. J. C}\ }\textbf {\bibinfo {volume} {86}},\
  \bibinfo {pages} {447} (\bibinfo {year} {2026}{\natexlab{b}})},\ \Eprint
  {https://arxiv.org/abs/2512.24893} {arXiv:2512.24893 [gr-qc]} \BibitemShut
  {NoStop}%
\bibitem [{\citenamefont {Zhang}\ \emph {et~al.}(2025)\citenamefont {Zhang},
  \citenamefont {Zhao}, \citenamefont {Nie}, \citenamefont {Hu},\ and\
  \citenamefont {An}}]{Zhang:2025hkb}%
  \BibitemOpen
  \bibfield  {author} {\bibinfo {author} {\bibfnamefont {X.-K.}\ \bibnamefont
  {Zhang}}, \bibinfo {author} {\bibfnamefont {X.}~\bibnamefont {Zhao}},
  \bibinfo {author} {\bibfnamefont {Z.-Y.}\ \bibnamefont {Nie}}, \bibinfo
  {author} {\bibfnamefont {Y.-P.}\ \bibnamefont {Hu}},\ and\ \bibinfo {author}
  {\bibfnamefont {Y.-S.}\ \bibnamefont {An}},\ }\bibfield  {title} {\bibinfo
  {title} {{Diving into a holographic multi-band superconductor}},\ }\href
  {https://doi.org/10.1016/j.physletb.2025.139684} {\bibfield  {journal}
  {\bibinfo  {journal} {Phys. Lett. B}\ }\textbf {\bibinfo {volume} {868}},\
  \bibinfo {pages} {139684} (\bibinfo {year} {2025})},\ \Eprint
  {https://arxiv.org/abs/2411.07693} {arXiv:2411.07693 [hep-th]} \BibitemShut
  {NoStop}%
\bibitem [{\citenamefont {Zhang}\ \emph {et~al.}(2026)\citenamefont {Zhang},
  \citenamefont {Zhao}, \citenamefont {Nie}, \citenamefont {Hu},\ and\
  \citenamefont {An}}]{Zhang:2025tsa}%
  \BibitemOpen
  \bibfield  {author} {\bibinfo {author} {\bibfnamefont {X.-K.}\ \bibnamefont
  {Zhang}}, \bibinfo {author} {\bibfnamefont {X.}~\bibnamefont {Zhao}},
  \bibinfo {author} {\bibfnamefont {Z.-Y.}\ \bibnamefont {Nie}}, \bibinfo
  {author} {\bibfnamefont {Y.-P.}\ \bibnamefont {Hu}},\ and\ \bibinfo {author}
  {\bibfnamefont {Y.-S.}\ \bibnamefont {An}},\ }\bibfield  {title} {\bibinfo
  {title} {{Interior structure of the holographic s+p superconductor and
  chaotic-stable transition near the black hole singularity}},\ }\href
  {https://doi.org/10.1016/j.physletb.2025.140110} {\bibfield  {journal}
  {\bibinfo  {journal} {Phys. Lett. B}\ }\textbf {\bibinfo {volume} {872}},\
  \bibinfo {pages} {140110} (\bibinfo {year} {2026})},\ \Eprint
  {https://arxiv.org/abs/2506.19419} {arXiv:2506.19419 [hep-th]} \BibitemShut
  {NoStop}%
\bibitem [{\citenamefont {Basu}\ \emph {et~al.}(2010)\citenamefont {Basu},
  \citenamefont {He}, \citenamefont {Mukherjee}, \citenamefont {Rozali},\ and\
  \citenamefont {Shieh}}]{Basu:2010fa}%
  \BibitemOpen
  \bibfield  {author} {\bibinfo {author} {\bibfnamefont {P.}~\bibnamefont
  {Basu}}, \bibinfo {author} {\bibfnamefont {J.}~\bibnamefont {He}}, \bibinfo
  {author} {\bibfnamefont {A.}~\bibnamefont {Mukherjee}}, \bibinfo {author}
  {\bibfnamefont {M.}~\bibnamefont {Rozali}},\ and\ \bibinfo {author}
  {\bibfnamefont {H.-H.}\ \bibnamefont {Shieh}},\ }\bibfield  {title} {\bibinfo
  {title} {{Competing Holographic Orders}},\ }\href
  {https://doi.org/10.1007/JHEP10(2010)092} {\bibfield  {journal} {\bibinfo
  {journal} {JHEP}\ }\textbf {\bibinfo {volume} {10}},\ \bibinfo {pages}
  {092}},\ \Eprint {https://arxiv.org/abs/1007.3480} {arXiv:1007.3480 [hep-th]}
  \BibitemShut {NoStop}%
\bibitem [{\citenamefont {Musso}(2013)}]{Musso:2013ija}%
  \BibitemOpen
  \bibfield  {author} {\bibinfo {author} {\bibfnamefont {D.}~\bibnamefont
  {Musso}},\ }\bibfield  {title} {\bibinfo {title} {Competition/enhancement of
  two probe order parameters in the unbalanced holographic superconductor},\
  }\href {https://doi.org/10.1007/JHEP06(2013)083} {\bibfield  {journal}
  {\bibinfo  {journal} {JHEP}\ }\textbf {\bibinfo {volume} {06}},\ \bibinfo
  {pages} {083}},\ \Eprint {https://arxiv.org/abs/1302.7205} {arXiv:1302.7205
  [hep-th]} \BibitemShut {NoStop}%
\bibitem [{\citenamefont {Nie}\ \emph {et~al.}(2013)\citenamefont {Nie},
  \citenamefont {Cai}, \citenamefont {Gao},\ and\ \citenamefont
  {Zeng}}]{Nie:2013sda}%
  \BibitemOpen
  \bibfield  {author} {\bibinfo {author} {\bibfnamefont {Z.-Y.}\ \bibnamefont
  {Nie}}, \bibinfo {author} {\bibfnamefont {R.-G.}\ \bibnamefont {Cai}},
  \bibinfo {author} {\bibfnamefont {X.}~\bibnamefont {Gao}},\ and\ \bibinfo
  {author} {\bibfnamefont {H.}~\bibnamefont {Zeng}},\ }\bibfield  {title}
  {\bibinfo {title} {{Competition between the s-wave and p-wave
  superconductivity phases in a holographic model}},\ }\href
  {https://doi.org/10.1007/JHEP11(2013)087} {\bibfield  {journal} {\bibinfo
  {journal} {JHEP}\ }\textbf {\bibinfo {volume} {11}},\ \bibinfo {pages}
  {087}},\ \Eprint {https://arxiv.org/abs/1309.2204} {arXiv:1309.2204 [hep-th]}
  \BibitemShut {NoStop}%
\bibitem [{\citenamefont {Donos}\ \emph {et~al.}(2014)\citenamefont {Donos},
  \citenamefont {Gauntlett},\ and\ \citenamefont {Pantelidou}}]{Donos:2013woa}%
  \BibitemOpen
  \bibfield  {author} {\bibinfo {author} {\bibfnamefont {A.}~\bibnamefont
  {Donos}}, \bibinfo {author} {\bibfnamefont {J.~P.}\ \bibnamefont
  {Gauntlett}},\ and\ \bibinfo {author} {\bibfnamefont {C.}~\bibnamefont
  {Pantelidou}},\ }\bibfield  {title} {\bibinfo {title} {{Competing p-wave
  orders}},\ }\href {https://doi.org/10.1088/0264-9381/31/5/055007} {\bibfield
  {journal} {\bibinfo  {journal} {Class. Quant. Grav.}\ }\textbf {\bibinfo
  {volume} {31}},\ \bibinfo {pages} {055007} (\bibinfo {year} {2014})},\
  \Eprint {https://arxiv.org/abs/1310.5741} {arXiv:1310.5741 [hep-th]}
  \BibitemShut {NoStop}%
\bibitem [{\citenamefont {Li}\ \emph {et~al.}(2018)\citenamefont {Li},
  \citenamefont {Fu},\ and\ \citenamefont {Nie}}]{Li:2017wbi}%
  \BibitemOpen
  \bibfield  {author} {\bibinfo {author} {\bibfnamefont {Z.-H.}\ \bibnamefont
  {Li}}, \bibinfo {author} {\bibfnamefont {Y.-C.}\ \bibnamefont {Fu}},\ and\
  \bibinfo {author} {\bibfnamefont {Z.-Y.}\ \bibnamefont {Nie}},\ }\bibfield
  {title} {\bibinfo {title} {{Competing s-wave orders from
  Einstein\textendash{}Gauss\textendash{}Bonnet gravity}},\ }\href
  {https://doi.org/10.1016/j.physletb.2017.11.031} {\bibfield  {journal}
  {\bibinfo  {journal} {Phys. Lett. B}\ }\textbf {\bibinfo {volume} {776}},\
  \bibinfo {pages} {115} (\bibinfo {year} {2018})},\ \Eprint
  {https://arxiv.org/abs/1706.07893} {arXiv:1706.07893 [hep-th]} \BibitemShut
  {NoStop}%
\bibitem [{\citenamefont {Nie}\ and\ \citenamefont {Zeng}(2015)}]{Nie:2015zia}%
  \BibitemOpen
  \bibfield  {author} {\bibinfo {author} {\bibfnamefont {Z.-Y.}\ \bibnamefont
  {Nie}}\ and\ \bibinfo {author} {\bibfnamefont {H.}~\bibnamefont {Zeng}},\
  }\bibfield  {title} {\bibinfo {title} {{P-T phase diagram of a holographic
  s+p model from Gauss-Bonnet gravity}},\ }\href
  {https://doi.org/10.1007/JHEP10(2015)047} {\bibfield  {journal} {\bibinfo
  {journal} {JHEP}\ }\textbf {\bibinfo {volume} {10}},\ \bibinfo {pages}
  {047}},\ \Eprint {https://arxiv.org/abs/1505.02289} {arXiv:1505.02289
  [hep-th]} \BibitemShut {NoStop}%
\bibitem [{\citenamefont {Nie}\ \emph {et~al.}(2015)\citenamefont {Nie},
  \citenamefont {Cai}, \citenamefont {Gao}, \citenamefont {Li},\ and\
  \citenamefont {Zeng}}]{Nie:2014qma}%
  \BibitemOpen
  \bibfield  {author} {\bibinfo {author} {\bibfnamefont {Z.-Y.}\ \bibnamefont
  {Nie}}, \bibinfo {author} {\bibfnamefont {R.-G.}\ \bibnamefont {Cai}},
  \bibinfo {author} {\bibfnamefont {X.}~\bibnamefont {Gao}}, \bibinfo {author}
  {\bibfnamefont {L.}~\bibnamefont {Li}},\ and\ \bibinfo {author}
  {\bibfnamefont {H.}~\bibnamefont {Zeng}},\ }\bibfield  {title} {\bibinfo
  {title} {{Phase transitions in a holographic s $+$ p model with
  back-reaction}},\ }\href {https://doi.org/10.1140/epjc/s10052-015-3773-2}
  {\bibfield  {journal} {\bibinfo  {journal} {Eur. Phys. J. C}\ }\textbf
  {\bibinfo {volume} {75}},\ \bibinfo {pages} {559} (\bibinfo {year} {2015})},\
  \Eprint {https://arxiv.org/abs/1501.00004} {arXiv:1501.00004 [hep-th]}
  \BibitemShut {NoStop}%
\bibitem [{\citenamefont {Amado}\ \emph {et~al.}(2014)\citenamefont {Amado},
  \citenamefont {Arean}, \citenamefont {Jimenez-Alba}, \citenamefont {Melgar},\
  and\ \citenamefont {Salazar~Landea}}]{Amado:2013lia}%
  \BibitemOpen
  \bibfield  {author} {\bibinfo {author} {\bibfnamefont {I.}~\bibnamefont
  {Amado}}, \bibinfo {author} {\bibfnamefont {D.}~\bibnamefont {Arean}},
  \bibinfo {author} {\bibfnamefont {A.}~\bibnamefont {Jimenez-Alba}}, \bibinfo
  {author} {\bibfnamefont {L.}~\bibnamefont {Melgar}},\ and\ \bibinfo {author}
  {\bibfnamefont {I.}~\bibnamefont {Salazar~Landea}},\ }\bibfield  {title}
  {\bibinfo {title} {{Holographic s+p Superconductors}},\ }\href
  {https://doi.org/10.1103/PhysRevD.89.026009} {\bibfield  {journal} {\bibinfo
  {journal} {Phys. Rev. D}\ }\textbf {\bibinfo {volume} {89}},\ \bibinfo
  {pages} {026009} (\bibinfo {year} {2014})},\ \Eprint
  {https://arxiv.org/abs/1309.5086} {arXiv:1309.5086 [hep-th]} \BibitemShut
  {NoStop}%
\bibitem [{\citenamefont {Zhang}\ \emph {et~al.}(2022)\citenamefont {Zhang},
  \citenamefont {Xia}, \citenamefont {Nie},\ and\ \citenamefont
  {Zeng}}]{Zhang:2021vwp}%
  \BibitemOpen
  \bibfield  {author} {\bibinfo {author} {\bibfnamefont {X.-K.}\ \bibnamefont
  {Zhang}}, \bibinfo {author} {\bibfnamefont {C.-Y.}\ \bibnamefont {Xia}},
  \bibinfo {author} {\bibfnamefont {Z.-Y.}\ \bibnamefont {Nie}},\ and\ \bibinfo
  {author} {\bibfnamefont {H.}~\bibnamefont {Zeng}},\ }\bibfield  {title}
  {\bibinfo {title} {{Holographic multicondensate with nonlinear terms}},\
  }\href {https://doi.org/10.1103/PhysRevD.105.046016} {\bibfield  {journal}
  {\bibinfo  {journal} {Phys. Rev. D}\ }\textbf {\bibinfo {volume} {105}},\
  \bibinfo {pages} {046016} (\bibinfo {year} {2022})},\ \Eprint
  {https://arxiv.org/abs/2105.14294} {arXiv:2105.14294 [hep-th]} \BibitemShut
  {NoStop}%
\bibitem [{\citenamefont {Zhao}\ \emph {et~al.}(2023)\citenamefont {Zhao},
  \citenamefont {Zhang},\ and\ \citenamefont {Nie}}]{Zhao:2022jvs}%
  \BibitemOpen
  \bibfield  {author} {\bibinfo {author} {\bibfnamefont {Z.-Q.}\ \bibnamefont
  {Zhao}}, \bibinfo {author} {\bibfnamefont {X.-K.}\ \bibnamefont {Zhang}},\
  and\ \bibinfo {author} {\bibfnamefont {Z.-Y.}\ \bibnamefont {Nie}},\
  }\bibfield  {title} {\bibinfo {title} {{Dynamical stability from quasi normal
  modes in 2nd, 1st and 0th order holographic superfluid phase transitions}},\
  }\href {https://doi.org/10.1007/JHEP02(2023)023} {\bibfield  {journal}
  {\bibinfo  {journal} {JHEP}\ }\textbf {\bibinfo {volume} {02}},\ \bibinfo
  {pages} {023}},\ \Eprint {https://arxiv.org/abs/2211.14762} {arXiv:2211.14762
  [hep-th]} \BibitemShut {NoStop}%
\bibitem [{\citenamefont {Chen}\ \emph {et~al.}(2023)\citenamefont {Chen},
  \citenamefont {Liu}, \citenamefont {Tian}, \citenamefont {Wu},\ and\
  \citenamefont {Zhang}}]{Chen:2022tfy}%
  \BibitemOpen
  \bibfield  {author} {\bibinfo {author} {\bibfnamefont {Q.}~\bibnamefont
  {Chen}}, \bibinfo {author} {\bibfnamefont {Y.}~\bibnamefont {Liu}}, \bibinfo
  {author} {\bibfnamefont {Y.}~\bibnamefont {Tian}}, \bibinfo {author}
  {\bibfnamefont {X.}~\bibnamefont {Wu}},\ and\ \bibinfo {author}
  {\bibfnamefont {H.}~\bibnamefont {Zhang}},\ }\bibfield  {title} {\bibinfo
  {title} {{Quench dynamics in holographic first-order phase transition}},\
  }\href {https://doi.org/10.1103/PhysRevD.108.106017} {\bibfield  {journal}
  {\bibinfo  {journal} {Phys. Rev. D}\ }\textbf {\bibinfo {volume} {108}},\
  \bibinfo {pages} {106017} (\bibinfo {year} {2023})},\ \Eprint
  {https://arxiv.org/abs/2211.11291} {arXiv:2211.11291 [hep-th]} \BibitemShut
  {NoStop}%
\bibitem [{\citenamefont {Zhao}\ \emph {et~al.}(2024)\citenamefont {Zhao},
  \citenamefont {Nie}, \citenamefont {Zhao}, \citenamefont {Zeng},
  \citenamefont {Tian},\ and\ \citenamefont {Baggioli}}]{Zhao:2023ffs}%
  \BibitemOpen
  \bibfield  {author} {\bibinfo {author} {\bibfnamefont {X.}~\bibnamefont
  {Zhao}}, \bibinfo {author} {\bibfnamefont {Z.-Y.}\ \bibnamefont {Nie}},
  \bibinfo {author} {\bibfnamefont {Z.-Q.}\ \bibnamefont {Zhao}}, \bibinfo
  {author} {\bibfnamefont {H.-B.}\ \bibnamefont {Zeng}}, \bibinfo {author}
  {\bibfnamefont {Y.}~\bibnamefont {Tian}},\ and\ \bibinfo {author}
  {\bibfnamefont {M.}~\bibnamefont {Baggioli}},\ }\bibfield  {title} {\bibinfo
  {title} {{Dynamical evolution of spinodal decomposition in holographic
  superfluids}},\ }\href {https://doi.org/10.1007/JHEP02(2024)184} {\bibfield
  {journal} {\bibinfo  {journal} {JHEP}\ }\textbf {\bibinfo {volume} {02}},\
  \bibinfo {pages} {184}},\ \Eprint {https://arxiv.org/abs/2311.08277}
  {arXiv:2311.08277 [hep-th]} \BibitemShut {NoStop}%
\bibitem [{\citenamefont {Xia}\ and\ \citenamefont {Zeng}(2020)}]{Xia:2020cjl}%
  \BibitemOpen
  \bibfield  {author} {\bibinfo {author} {\bibfnamefont {C.-Y.}\ \bibnamefont
  {Xia}}\ and\ \bibinfo {author} {\bibfnamefont {H.-B.}\ \bibnamefont {Zeng}},\
  }\bibfield  {title} {\bibinfo {title} {{Winding up a finite size holographic
  superconducting ring beyond Kibble-Zurek mechanism}},\ }\href
  {https://doi.org/10.1103/PhysRevD.102.126005} {\bibfield  {journal} {\bibinfo
   {journal} {Phys. Rev. D}\ }\textbf {\bibinfo {volume} {102}},\ \bibinfo
  {pages} {126005} (\bibinfo {year} {2020})},\ \Eprint
  {https://arxiv.org/abs/2009.00435} {arXiv:2009.00435 [hep-th]} \BibitemShut
  {NoStop}%
\bibitem [{\citenamefont {del Campo}\ \emph {et~al.}(2021)\citenamefont {del
  Campo}, \citenamefont {G{\'o}mez-Ruiz}, \citenamefont {Li}, \citenamefont
  {Xia}, \citenamefont {Zeng},\ and\ \citenamefont {Zhang}}]{delCampo:2021rak}%
  \BibitemOpen
  \bibfield  {author} {\bibinfo {author} {\bibfnamefont {A.}~\bibnamefont {del
  Campo}}, \bibinfo {author} {\bibfnamefont {F.~J.}\ \bibnamefont
  {G{\'o}mez-Ruiz}}, \bibinfo {author} {\bibfnamefont {Z.-H.}\ \bibnamefont
  {Li}}, \bibinfo {author} {\bibfnamefont {C.-Y.}\ \bibnamefont {Xia}},
  \bibinfo {author} {\bibfnamefont {H.-B.}\ \bibnamefont {Zeng}},\ and\
  \bibinfo {author} {\bibfnamefont {H.-Q.}\ \bibnamefont {Zhang}},\ }\bibfield
  {title} {\bibinfo {title} {{Universal statistics of vortices in a newborn
  holographic superconductor: beyond the Kibble-Zurek mechanism}},\ }\href
  {https://doi.org/10.1007/JHEP06(2021)061} {\bibfield  {journal} {\bibinfo
  {journal} {JHEP}\ }\textbf {\bibinfo {volume} {06}},\ \bibinfo {pages}
  {061}},\ \Eprint {https://arxiv.org/abs/2101.02171} {arXiv:2101.02171
  [cond-mat.stat-mech]} \BibitemShut {NoStop}%
\bibitem [{\citenamefont {Li}\ \emph {et~al.}(2021{\natexlab{a}})\citenamefont
  {Li}, \citenamefont {Zeng},\ and\ \citenamefont {Zhang}}]{Li:2021iph}%
  \BibitemOpen
  \bibfield  {author} {\bibinfo {author} {\bibfnamefont {Z.-H.}\ \bibnamefont
  {Li}}, \bibinfo {author} {\bibfnamefont {H.-B.}\ \bibnamefont {Zeng}},\ and\
  \bibinfo {author} {\bibfnamefont {H.-Q.}\ \bibnamefont {Zhang}},\ }\bibfield
  {title} {\bibinfo {title} {{Topological Defects Formation with Momentum
  Dissipation}},\ }\href {https://doi.org/10.1007/JHEP04(2021)295} {\bibfield
  {journal} {\bibinfo  {journal} {JHEP}\ }\textbf {\bibinfo {volume} {04}},\
  \bibinfo {pages} {295}},\ \Eprint {https://arxiv.org/abs/2101.08405}
  {arXiv:2101.08405 [hep-th]} \BibitemShut {NoStop}%
\bibitem [{\citenamefont {Li}\ \emph {et~al.}(2021{\natexlab{b}})\citenamefont
  {Li}, \citenamefont {Xia}, \citenamefont {Zeng},\ and\ \citenamefont
  {Zhang}}]{Li:2021dwp}%
  \BibitemOpen
  \bibfield  {author} {\bibinfo {author} {\bibfnamefont {Z.-H.}\ \bibnamefont
  {Li}}, \bibinfo {author} {\bibfnamefont {C.-Y.}\ \bibnamefont {Xia}},
  \bibinfo {author} {\bibfnamefont {H.-B.}\ \bibnamefont {Zeng}},\ and\
  \bibinfo {author} {\bibfnamefont {H.-Q.}\ \bibnamefont {Zhang}},\ }\bibfield
  {title} {\bibinfo {title} {{Holographic topological defects and local gauge
  symmetry: clusters of strongly coupled equal-sign vortices}},\ }\href
  {https://doi.org/10.1007/JHEP10(2021)124} {\bibfield  {journal} {\bibinfo
  {journal} {JHEP}\ }\textbf {\bibinfo {volume} {10}},\ \bibinfo {pages}
  {124}},\ \Eprint {https://arxiv.org/abs/2103.01485} {arXiv:2103.01485
  [hep-th]} \BibitemShut {NoStop}%
\bibitem [{\citenamefont {Xia}\ and\ \citenamefont {Zeng}(2021)}]{Xia:2021xap}%
  \BibitemOpen
  \bibfield  {author} {\bibinfo {author} {\bibfnamefont {C.-Y.}\ \bibnamefont
  {Xia}}\ and\ \bibinfo {author} {\bibfnamefont {H.-B.}\ \bibnamefont {Zeng}},\
  }\bibfield  {title} {\bibinfo {title} {{Kibble Zurek mechanism in rapidly
  quenched phase transition dynamics}},\ }\href@noop {} {\  (\bibinfo {year}
  {2021})},\ \Eprint {https://arxiv.org/abs/2110.07969} {arXiv:2110.07969
  [cond-mat.stat-mech]} \BibitemShut {NoStop}%
\bibitem [{\citenamefont {Zeng}\ \emph {et~al.}(2023)\citenamefont {Zeng},
  \citenamefont {Xia},\ and\ \citenamefont {del Campo}}]{Zeng:2022hut}%
  \BibitemOpen
  \bibfield  {author} {\bibinfo {author} {\bibfnamefont {H.-B.}\ \bibnamefont
  {Zeng}}, \bibinfo {author} {\bibfnamefont {C.-Y.}\ \bibnamefont {Xia}},\ and\
  \bibinfo {author} {\bibfnamefont {A.}~\bibnamefont {del Campo}},\ }\bibfield
  {title} {\bibinfo {title} {{Universal Breakdown of Kibble-Zurek Scaling in
  Fast Quenches across a Phase Transition}},\ }\href
  {https://doi.org/10.1103/PhysRevLett.130.060402} {\bibfield  {journal}
  {\bibinfo  {journal} {Phys. Rev. Lett.}\ }\textbf {\bibinfo {volume} {130}},\
  \bibinfo {pages} {060402} (\bibinfo {year} {2023})},\ \Eprint
  {https://arxiv.org/abs/2204.13529} {arXiv:2204.13529 [cond-mat.stat-mech]}
  \BibitemShut {NoStop}%
\bibitem [{\citenamefont {Yang}\ \emph {et~al.}(2026)\citenamefont {Yang},
  \citenamefont {Xia}, \citenamefont {Grieninger}, \citenamefont {Zeng},\ and\
  \citenamefont {Baggioli}}]{Yang:2025bsw}%
  \BibitemOpen
  \bibfield  {author} {\bibinfo {author} {\bibfnamefont {P.}~\bibnamefont
  {Yang}}, \bibinfo {author} {\bibfnamefont {C.-Y.}\ \bibnamefont {Xia}},
  \bibinfo {author} {\bibfnamefont {S.}~\bibnamefont {Grieninger}}, \bibinfo
  {author} {\bibfnamefont {H.-B.}\ \bibnamefont {Zeng}},\ and\ \bibinfo
  {author} {\bibfnamefont {M.}~\bibnamefont {Baggioli}},\ }\bibfield  {title}
  {\bibinfo {title} {{Topological Defect Formation beyond the Kibble-Zurek
  Mechanism in Crossover Transitions with Approximate Symmetries}},\ }\href
  {https://doi.org/10.1103/clvs-yk7v} {\bibfield  {journal} {\bibinfo
  {journal} {Phys. Rev. Lett.}\ }\textbf {\bibinfo {volume} {136}},\ \bibinfo
  {pages} {051602} (\bibinfo {year} {2026})},\ \Eprint
  {https://arxiv.org/abs/2508.05964} {arXiv:2508.05964 [cond-mat.stat-mech]}
  \BibitemShut {NoStop}%
\bibitem [{\citenamefont {Xia}\ \emph {et~al.}(2026)\citenamefont {Xia},
  \citenamefont {Grabarits}, \citenamefont {Zeng},\ and\ \citenamefont {del
  Campo}}]{Xia:2026yrj}%
  \BibitemOpen
  \bibfield  {author} {\bibinfo {author} {\bibfnamefont {C.-Y.}\ \bibnamefont
  {Xia}}, \bibinfo {author} {\bibfnamefont {A.}~\bibnamefont {Grabarits}},
  \bibinfo {author} {\bibfnamefont {H.-B.}\ \bibnamefont {Zeng}},\ and\
  \bibinfo {author} {\bibfnamefont {A.}~\bibnamefont {del Campo}},\ }\bibfield
  {title} {\bibinfo {title} {{Evolution of Vortex Strings after a Thermal
  Quench in a Holographic Superfluid}},\ }\href@noop {} {\  (\bibinfo {year}
  {2026})},\ \Eprint {https://arxiv.org/abs/2601.14328} {arXiv:2601.14328
  [hep-th]} \BibitemShut {NoStop}%
\bibitem [{\citenamefont {Xia}\ \emph {et~al.}(2023)\citenamefont {Xia},
  \citenamefont {Zeng}, \citenamefont {Chen},\ and\ \citenamefont {del
  Campo}}]{Xia:2023pom}%
  \BibitemOpen
  \bibfield  {author} {\bibinfo {author} {\bibfnamefont {C.-Y.}\ \bibnamefont
  {Xia}}, \bibinfo {author} {\bibfnamefont {H.-B.}\ \bibnamefont {Zeng}},
  \bibinfo {author} {\bibfnamefont {C.-M.}\ \bibnamefont {Chen}},\ and\
  \bibinfo {author} {\bibfnamefont {A.}~\bibnamefont {del Campo}},\ }\bibfield
  {title} {\bibinfo {title} {{Structural phase transition and its critical
  dynamics from holography}},\ }\href
  {https://doi.org/10.1103/PhysRevD.108.026017} {\bibfield  {journal} {\bibinfo
   {journal} {Phys. Rev. D}\ }\textbf {\bibinfo {volume} {108}},\ \bibinfo
  {pages} {026017} (\bibinfo {year} {2023})},\ \Eprint
  {https://arxiv.org/abs/2302.11597} {arXiv:2302.11597 [hep-th]} \BibitemShut
  {NoStop}%
\bibitem [{\citenamefont {Su}\ \emph {et~al.}(2024)\citenamefont {Su},
  \citenamefont {Xia}, \citenamefont {Yang},\ and\ \citenamefont
  {Zeng}}]{Su:2023vqa}%
  \BibitemOpen
  \bibfield  {author} {\bibinfo {author} {\bibfnamefont {J.-H.}\ \bibnamefont
  {Su}}, \bibinfo {author} {\bibfnamefont {C.-Y.}\ \bibnamefont {Xia}},
  \bibinfo {author} {\bibfnamefont {W.-C.}\ \bibnamefont {Yang}},\ and\
  \bibinfo {author} {\bibfnamefont {H.-B.}\ \bibnamefont {Zeng}},\ }\bibfield
  {title} {\bibinfo {title} {{Vortex-antivortex lattices in a holographic
  superconductor}},\ }\href {https://doi.org/10.1103/PhysRevD.109.046019}
  {\bibfield  {journal} {\bibinfo  {journal} {Phys. Rev. D}\ }\textbf {\bibinfo
  {volume} {109}},\ \bibinfo {pages} {046019} (\bibinfo {year} {2024})},\
  \Eprint {https://arxiv.org/abs/2311.05856} {arXiv:2311.05856 [hep-th]}
  \BibitemShut {NoStop}%
\bibitem [{\citenamefont {An}\ \emph {et~al.}(2024)\citenamefont {An},
  \citenamefont {Li}, \citenamefont {Xia},\ and\ \citenamefont
  {Zeng}}]{An:2024ebg}%
  \BibitemOpen
  \bibfield  {author} {\bibinfo {author} {\bibfnamefont {Y.-P.}\ \bibnamefont
  {An}}, \bibinfo {author} {\bibfnamefont {L.}~\bibnamefont {Li}}, \bibinfo
  {author} {\bibfnamefont {C.-Y.}\ \bibnamefont {Xia}},\ and\ \bibinfo {author}
  {\bibfnamefont {H.-B.}\ \bibnamefont {Zeng}},\ }\bibfield  {title} {\bibinfo
  {title} {{Interface dynamics of strongly interacting binary superfluids}},\
  }\href {https://doi.org/10.1103/PhysRevD.109.106022} {\bibfield  {journal}
  {\bibinfo  {journal} {Phys. Rev. D}\ }\textbf {\bibinfo {volume} {109}},\
  \bibinfo {pages} {106022} (\bibinfo {year} {2024})},\ \Eprint
  {https://arxiv.org/abs/2401.09189} {arXiv:2401.09189 [cond-mat.quant-gas]}
  \BibitemShut {NoStop}%
\bibitem [{\citenamefont {Yang}\ \emph {et~al.}(2024)\citenamefont {Yang},
  \citenamefont {Xia}, \citenamefont {Tian}, \citenamefont {Tsubota},\ and\
  \citenamefont {Zeng}}]{Yang:2024hom}%
  \BibitemOpen
  \bibfield  {author} {\bibinfo {author} {\bibfnamefont {W.-C.}\ \bibnamefont
  {Yang}}, \bibinfo {author} {\bibfnamefont {C.-Y.}\ \bibnamefont {Xia}},
  \bibinfo {author} {\bibfnamefont {Y.}~\bibnamefont {Tian}}, \bibinfo {author}
  {\bibfnamefont {M.}~\bibnamefont {Tsubota}},\ and\ \bibinfo {author}
  {\bibfnamefont {H.-B.}\ \bibnamefont {Zeng}},\ }\bibfield  {title} {\bibinfo
  {title} {{Emergence of Large-Scale Structures in Holographic Superfluid
  Turbulence}},\ }\href@noop {} {\  (\bibinfo {year} {2024})},\ \Eprint
  {https://arxiv.org/abs/2402.17980} {arXiv:2402.17980 [hep-th]} \BibitemShut
  {NoStop}%
\bibitem [{\citenamefont {Xia}\ and\ \citenamefont {Zeng}(2025)}]{Xia:2024ton}%
  \BibitemOpen
  \bibfield  {author} {\bibinfo {author} {\bibfnamefont {C.-Y.}\ \bibnamefont
  {Xia}}\ and\ \bibinfo {author} {\bibfnamefont {H.-B.}\ \bibnamefont {Zeng}},\
  }\bibfield  {title} {\bibinfo {title} {{Pattern formation from gauge/gravity
  duality}},\ }\href {https://doi.org/10.1103/PhysRevD.111.026004} {\bibfield
  {journal} {\bibinfo  {journal} {Phys. Rev. D}\ }\textbf {\bibinfo {volume}
  {111}},\ \bibinfo {pages} {026004} (\bibinfo {year} {2025})},\ \Eprint
  {https://arxiv.org/abs/2404.04274} {arXiv:2404.04274 [hep-th]} \BibitemShut
  {NoStop}%
\bibitem [{\citenamefont {Zeng}\ \emph {et~al.}(2025)\citenamefont {Zeng},
  \citenamefont {Xia}, \citenamefont {Yang}, \citenamefont {Tian},\ and\
  \citenamefont {Tsubota}}]{Zeng:2024rwn}%
  \BibitemOpen
  \bibfield  {author} {\bibinfo {author} {\bibfnamefont {H.-B.}\ \bibnamefont
  {Zeng}}, \bibinfo {author} {\bibfnamefont {C.-Y.}\ \bibnamefont {Xia}},
  \bibinfo {author} {\bibfnamefont {W.-C.}\ \bibnamefont {Yang}}, \bibinfo
  {author} {\bibfnamefont {Y.}~\bibnamefont {Tian}},\ and\ \bibinfo {author}
  {\bibfnamefont {M.}~\bibnamefont {Tsubota}},\ }\bibfield  {title} {\bibinfo
  {title} {{Dissipation and Decay of Three-Dimensional Holographic Quantum
  Turbulence}},\ }\href {https://doi.org/10.1103/PhysRevLett.134.091603}
  {\bibfield  {journal} {\bibinfo  {journal} {Phys. Rev. Lett.}\ }\textbf
  {\bibinfo {volume} {134}},\ \bibinfo {pages} {091603} (\bibinfo {year}
  {2025})},\ \Eprint {https://arxiv.org/abs/2408.13620} {arXiv:2408.13620
  [hep-th]} \BibitemShut {NoStop}%
\bibitem [{\citenamefont {Janik}\ \emph
  {et~al.}(2016{\natexlab{a}})\citenamefont {Janik}, \citenamefont
  {Jankowski},\ and\ \citenamefont {Soltanpanahi}}]{Janik:2015iry}%
  \BibitemOpen
  \bibfield  {author} {\bibinfo {author} {\bibfnamefont {R.~A.}\ \bibnamefont
  {Janik}}, \bibinfo {author} {\bibfnamefont {J.}~\bibnamefont {Jankowski}},\
  and\ \bibinfo {author} {\bibfnamefont {H.}~\bibnamefont {Soltanpanahi}},\
  }\bibfield  {title} {\bibinfo {title} {{Nonequilibrium Dynamics and Phase
  Transitions in Holographic Models}},\ }\href
  {https://doi.org/10.1103/PhysRevLett.117.091603} {\bibfield  {journal}
  {\bibinfo  {journal} {Phys. Rev. Lett.}\ }\textbf {\bibinfo {volume} {117}},\
  \bibinfo {pages} {091603} (\bibinfo {year} {2016}{\natexlab{a}})},\ \Eprint
  {https://arxiv.org/abs/1512.06871} {arXiv:1512.06871 [hep-th]} \BibitemShut
  {NoStop}%
\bibitem [{\citenamefont {Janik}\ \emph
  {et~al.}(2016{\natexlab{b}})\citenamefont {Janik}, \citenamefont
  {Jankowski},\ and\ \citenamefont {Soltanpanahi}}]{Janik:2016btb}%
  \BibitemOpen
  \bibfield  {author} {\bibinfo {author} {\bibfnamefont {R.~A.}\ \bibnamefont
  {Janik}}, \bibinfo {author} {\bibfnamefont {J.}~\bibnamefont {Jankowski}},\
  and\ \bibinfo {author} {\bibfnamefont {H.}~\bibnamefont {Soltanpanahi}},\
  }\bibfield  {title} {\bibinfo {title} {{Quasinormal modes and the phase
  structure of strongly coupled matter}},\ }\href
  {https://doi.org/10.1007/JHEP06(2016)047} {\bibfield  {journal} {\bibinfo
  {journal} {JHEP}\ }\textbf {\bibinfo {volume} {06}},\ \bibinfo {pages}
  {047}},\ \Eprint {https://arxiv.org/abs/1603.05950} {arXiv:1603.05950
  [hep-th]} \BibitemShut {NoStop}%
\bibitem [{\citenamefont {Janik}\ \emph {et~al.}(2017)\citenamefont {Janik},
  \citenamefont {Jankowski},\ and\ \citenamefont
  {Soltanpanahi}}]{Janik:2017ykj}%
  \BibitemOpen
  \bibfield  {author} {\bibinfo {author} {\bibfnamefont {R.~A.}\ \bibnamefont
  {Janik}}, \bibinfo {author} {\bibfnamefont {J.}~\bibnamefont {Jankowski}},\
  and\ \bibinfo {author} {\bibfnamefont {H.}~\bibnamefont {Soltanpanahi}},\
  }\bibfield  {title} {\bibinfo {title} {{Real-Time dynamics and phase
  separation in a holographic first order phase transition}},\ }\href
  {https://doi.org/10.1103/PhysRevLett.119.261601} {\bibfield  {journal}
  {\bibinfo  {journal} {Phys. Rev. Lett.}\ }\textbf {\bibinfo {volume} {119}},\
  \bibinfo {pages} {261601} (\bibinfo {year} {2017})},\ \Eprint
  {https://arxiv.org/abs/1704.05387} {arXiv:1704.05387 [hep-th]} \BibitemShut
  {NoStop}%
\bibitem [{\citenamefont {Attems}\ \emph {et~al.}(2017)\citenamefont {Attems},
  \citenamefont {Bea}, \citenamefont {Casalderrey-Solana}, \citenamefont
  {Mateos}, \citenamefont {Triana},\ and\ \citenamefont
  {Zilhao}}]{Attems:2017ezz}%
  \BibitemOpen
  \bibfield  {author} {\bibinfo {author} {\bibfnamefont {M.}~\bibnamefont
  {Attems}}, \bibinfo {author} {\bibfnamefont {Y.}~\bibnamefont {Bea}},
  \bibinfo {author} {\bibfnamefont {J.}~\bibnamefont {Casalderrey-Solana}},
  \bibinfo {author} {\bibfnamefont {D.}~\bibnamefont {Mateos}}, \bibinfo
  {author} {\bibfnamefont {M.}~\bibnamefont {Triana}},\ and\ \bibinfo {author}
  {\bibfnamefont {M.}~\bibnamefont {Zilhao}},\ }\bibfield  {title} {\bibinfo
  {title} {{Phase Transitions, Inhomogeneous Horizons and Second-Order
  Hydrodynamics}},\ }\href {https://doi.org/10.1007/JHEP06(2017)129} {\bibfield
   {journal} {\bibinfo  {journal} {JHEP}\ }\textbf {\bibinfo {volume} {06}},\
  \bibinfo {pages} {129}},\ \Eprint {https://arxiv.org/abs/1703.02948}
  {arXiv:1703.02948 [hep-th]} \BibitemShut {NoStop}%
\bibitem [{\citenamefont {Attems}\ \emph {et~al.}(2020)\citenamefont {Attems},
  \citenamefont {Bea}, \citenamefont {Casalderrey-Solana}, \citenamefont
  {Mateos},\ and\ \citenamefont {Zilh{\~a}o}}]{Attems:2019yqn}%
  \BibitemOpen
  \bibfield  {author} {\bibinfo {author} {\bibfnamefont {M.}~\bibnamefont
  {Attems}}, \bibinfo {author} {\bibfnamefont {Y.}~\bibnamefont {Bea}},
  \bibinfo {author} {\bibfnamefont {J.}~\bibnamefont {Casalderrey-Solana}},
  \bibinfo {author} {\bibfnamefont {D.}~\bibnamefont {Mateos}},\ and\ \bibinfo
  {author} {\bibfnamefont {M.}~\bibnamefont {Zilh{\~a}o}},\ }\bibfield  {title}
  {\bibinfo {title} {{Dynamics of Phase Separation from Holography}},\ }\href
  {https://doi.org/10.1007/JHEP01(2020)106} {\bibfield  {journal} {\bibinfo
  {journal} {JHEP}\ }\textbf {\bibinfo {volume} {01}},\ \bibinfo {pages}
  {106}},\ \Eprint {https://arxiv.org/abs/1905.12544} {arXiv:1905.12544
  [hep-th]} \BibitemShut {NoStop}%
\bibitem [{\citenamefont {Bellantuono}\ \emph {et~al.}(2019)\citenamefont
  {Bellantuono}, \citenamefont {Janik}, \citenamefont {Jankowski},\ and\
  \citenamefont {Soltanpanahi}}]{Bellantuono:2019wbn}%
  \BibitemOpen
  \bibfield  {author} {\bibinfo {author} {\bibfnamefont {L.}~\bibnamefont
  {Bellantuono}}, \bibinfo {author} {\bibfnamefont {R.~A.}\ \bibnamefont
  {Janik}}, \bibinfo {author} {\bibfnamefont {J.}~\bibnamefont {Jankowski}},\
  and\ \bibinfo {author} {\bibfnamefont {H.}~\bibnamefont {Soltanpanahi}},\
  }\bibfield  {title} {\bibinfo {title} {{Dynamics near a first order phase
  transition}},\ }\href {https://doi.org/10.1007/JHEP10(2019)146} {\bibfield
  {journal} {\bibinfo  {journal} {JHEP}\ }\textbf {\bibinfo {volume} {10}},\
  \bibinfo {pages} {146}},\ \Eprint {https://arxiv.org/abs/1906.00061}
  {arXiv:1906.00061 [hep-th]} \BibitemShut {NoStop}%
\bibitem [{\citenamefont {Attems}(2021)}]{Attems:2020qkg}%
  \BibitemOpen
  \bibfield  {author} {\bibinfo {author} {\bibfnamefont {M.}~\bibnamefont
  {Attems}},\ }\bibfield  {title} {\bibinfo {title} {{Holographic approach of
  the spinodal instability to criticality}},\ }\href
  {https://doi.org/10.1007/JHEP08(2021)155} {\bibfield  {journal} {\bibinfo
  {journal} {JHEP}\ }\textbf {\bibinfo {volume} {08}},\ \bibinfo {pages}
  {155}},\ \Eprint {https://arxiv.org/abs/2012.15687} {arXiv:2012.15687
  [hep-th]} \BibitemShut {NoStop}%
\bibitem [{\citenamefont {Ning}\ \emph {et~al.}(2024)\citenamefont {Ning},
  \citenamefont {Chen}, \citenamefont {Tian}, \citenamefont {Wu},\ and\
  \citenamefont {Zhang}}]{Ning:2023edr}%
  \BibitemOpen
  \bibfield  {author} {\bibinfo {author} {\bibfnamefont {Z.}~\bibnamefont
  {Ning}}, \bibinfo {author} {\bibfnamefont {Q.}~\bibnamefont {Chen}}, \bibinfo
  {author} {\bibfnamefont {Y.}~\bibnamefont {Tian}}, \bibinfo {author}
  {\bibfnamefont {X.}~\bibnamefont {Wu}},\ and\ \bibinfo {author}
  {\bibfnamefont {H.}~\bibnamefont {Zhang}},\ }\bibfield  {title} {\bibinfo
  {title} {{Spontaneous deformation of an AdS spherical black hole}},\ }\href
  {https://doi.org/10.1103/PhysRevD.109.064082} {\bibfield  {journal} {\bibinfo
   {journal} {Phys. Rev. D}\ }\textbf {\bibinfo {volume} {109}},\ \bibinfo
  {pages} {064082} (\bibinfo {year} {2024})},\ \Eprint
  {https://arxiv.org/abs/2307.14156} {arXiv:2307.14156 [gr-qc]} \BibitemShut
  {NoStop}%
\bibitem [{\citenamefont {Yoon}\ \emph {et~al.}(2018)\citenamefont {Yoon},
  \citenamefont {Ha}, \citenamefont {Lee},\ and\ \citenamefont
  {Lee}}]{Yoon_2018}%
  \BibitemOpen
  \bibfield  {author} {\bibinfo {author} {\bibfnamefont {T.~J.}\ \bibnamefont
  {Yoon}}, \bibinfo {author} {\bibfnamefont {M.~Y.}\ \bibnamefont {Ha}},
  \bibinfo {author} {\bibfnamefont {W.~B.}\ \bibnamefont {Lee}},\ and\ \bibinfo
  {author} {\bibfnamefont {Y.-W.}\ \bibnamefont {Lee}},\ }\bibfield  {title}
  {\bibinfo {title} {“two-phase” thermodynamics of the frenkel line},\
  }\href {https://doi.org/10.1021/acs.jpclett.8b01955} {\bibfield  {journal}
  {\bibinfo  {journal} {The Journal of Physical Chemistry Letters}\ }\textbf
  {\bibinfo {volume} {9}},\ \bibinfo {pages} {4550–4554} (\bibinfo {year}
  {2018})}\BibitemShut {NoStop}%
\bibitem [{\citenamefont {Brazhkin}\ \emph {et~al.}(2013)\citenamefont
  {Brazhkin}, \citenamefont {Fomin}, \citenamefont {Lyapin}, \citenamefont
  {Ryzhov}, \citenamefont {Tsiok},\ and\ \citenamefont
  {Trachenko}}]{PhysRevLett.111.145901}%
  \BibitemOpen
  \bibfield  {author} {\bibinfo {author} {\bibfnamefont {V.~V.}\ \bibnamefont
  {Brazhkin}}, \bibinfo {author} {\bibfnamefont {Y.~D.}\ \bibnamefont {Fomin}},
  \bibinfo {author} {\bibfnamefont {A.~G.}\ \bibnamefont {Lyapin}}, \bibinfo
  {author} {\bibfnamefont {V.~N.}\ \bibnamefont {Ryzhov}}, \bibinfo {author}
  {\bibfnamefont {E.~N.}\ \bibnamefont {Tsiok}},\ and\ \bibinfo {author}
  {\bibfnamefont {K.}~\bibnamefont {Trachenko}},\ }\bibfield  {title} {\bibinfo
  {title} {``liquid-gas'' transition in the supercritical region: Fundamental
  changes in the particle dynamics},\ }\href
  {https://doi.org/10.1103/PhysRevLett.111.145901} {\bibfield  {journal}
  {\bibinfo  {journal} {Phys. Rev. Lett.}\ }\textbf {\bibinfo {volume} {111}},\
  \bibinfo {pages} {145901} (\bibinfo {year} {2013})}\BibitemShut {NoStop}%
\bibitem [{\citenamefont {Bolmatov}\ \emph {et~al.}(2013)\citenamefont
  {Bolmatov}, \citenamefont {Brazhkin},\ and\ \citenamefont
  {Trachenko}}]{Bolmatov2013}%
  \BibitemOpen
  \bibfield  {author} {\bibinfo {author} {\bibfnamefont {D.}~\bibnamefont
  {Bolmatov}}, \bibinfo {author} {\bibfnamefont {V.~V.}\ \bibnamefont
  {Brazhkin}},\ and\ \bibinfo {author} {\bibfnamefont {K.}~\bibnamefont
  {Trachenko}},\ }\bibfield  {title} {\bibinfo {title} {Thermodynamic behaviour
  of supercritical matter},\ }\href {https://doi.org/10.1038/ncomms3331}
  {\bibfield  {journal} {\bibinfo  {journal} {Nature Communications}\ }\textbf
  {\bibinfo {volume} {4}},\ \bibinfo {pages} {2331} (\bibinfo {year}
  {2013})}\BibitemShut {NoStop}%
\bibitem [{\citenamefont {Prescher}\ \emph {et~al.}(2017)\citenamefont
  {Prescher}, \citenamefont {Fomin}, \citenamefont {Prakapenka}, \citenamefont
  {Stefanski}, \citenamefont {Trachenko},\ and\ \citenamefont
  {Brazhkin}}]{Prescher_2017}%
  \BibitemOpen
  \bibfield  {author} {\bibinfo {author} {\bibfnamefont {C.}~\bibnamefont
  {Prescher}}, \bibinfo {author} {\bibfnamefont {Y.~D.}\ \bibnamefont {Fomin}},
  \bibinfo {author} {\bibfnamefont {V.~B.}\ \bibnamefont {Prakapenka}},
  \bibinfo {author} {\bibfnamefont {J.}~\bibnamefont {Stefanski}}, \bibinfo
  {author} {\bibfnamefont {K.}~\bibnamefont {Trachenko}},\ and\ \bibinfo
  {author} {\bibfnamefont {V.~V.}\ \bibnamefont {Brazhkin}},\ }\bibfield
  {title} {\bibinfo {title} {Experimental evidence of the frenkel line in
  supercritical neon},\ }\bibfield  {journal} {\bibinfo  {journal} {Physical
  Review B}\ }\textbf {\bibinfo {volume} {95}},\ \href
  {https://doi.org/10.1103/physrevb.95.134114} {10.1103/physrevb.95.134114}
  (\bibinfo {year} {2017})\BibitemShut {NoStop}%
\bibitem [{\citenamefont {Bolmatov}\ \emph {et~al.}(2015)\citenamefont
  {Bolmatov}, \citenamefont {Zhernenkov}, \citenamefont {Zav’yalov},
  \citenamefont {Tkachev}, \citenamefont {Cunsolo},\ and\ \citenamefont
  {Cai}}]{Bolmatov_2015}%
  \BibitemOpen
  \bibfield  {author} {\bibinfo {author} {\bibfnamefont {D.}~\bibnamefont
  {Bolmatov}}, \bibinfo {author} {\bibfnamefont {M.}~\bibnamefont
  {Zhernenkov}}, \bibinfo {author} {\bibfnamefont {D.}~\bibnamefont
  {Zav’yalov}}, \bibinfo {author} {\bibfnamefont {S.~N.}\ \bibnamefont
  {Tkachev}}, \bibinfo {author} {\bibfnamefont {A.}~\bibnamefont {Cunsolo}},\
  and\ \bibinfo {author} {\bibfnamefont {Y.~Q.}\ \bibnamefont {Cai}},\
  }\bibfield  {title} {\bibinfo {title} {The frenkel line: a direct
  experimental evidence for the new thermodynamic boundary},\ }\bibfield
  {journal} {\bibinfo  {journal} {Scientific Reports}\ }\textbf {\bibinfo
  {volume} {5}},\ \href {https://doi.org/10.1038/srep15850} {10.1038/srep15850}
  (\bibinfo {year} {2015})\BibitemShut {NoStop}%
\bibitem [{\citenamefont {Fomin}\ \emph {et~al.}(2018)\citenamefont {Fomin},
  \citenamefont {Ryzhov}, \citenamefont {Tsiok}, \citenamefont {Proctor},
  \citenamefont {Prescher}, \citenamefont {Prakapenka}, \citenamefont
  {Trachenko},\ and\ \citenamefont {Brazhkin}}]{Fomin_2018}%
  \BibitemOpen
  \bibfield  {author} {\bibinfo {author} {\bibfnamefont {Y.~D.}\ \bibnamefont
  {Fomin}}, \bibinfo {author} {\bibfnamefont {V.~N.}\ \bibnamefont {Ryzhov}},
  \bibinfo {author} {\bibfnamefont {E.~N.}\ \bibnamefont {Tsiok}}, \bibinfo
  {author} {\bibfnamefont {J.~E.}\ \bibnamefont {Proctor}}, \bibinfo {author}
  {\bibfnamefont {C.}~\bibnamefont {Prescher}}, \bibinfo {author}
  {\bibfnamefont {V.~B.}\ \bibnamefont {Prakapenka}}, \bibinfo {author}
  {\bibfnamefont {K.}~\bibnamefont {Trachenko}},\ and\ \bibinfo {author}
  {\bibfnamefont {V.~V.}\ \bibnamefont {Brazhkin}},\ }\bibfield  {title}
  {\bibinfo {title} {Dynamics, thermodynamics and structure of liquids and
  supercritical fluids: crossover at the frenkel line},\ }\href
  {https://doi.org/10.1088/1361-648X/aaaf39} {\bibfield  {journal} {\bibinfo
  {journal} {Journal of Physics: Condensed Matter}\ }\textbf {\bibinfo {volume}
  {30}},\ \bibinfo {pages} {134003} (\bibinfo {year} {2018})}\BibitemShut
  {NoStop}%
\bibitem [{\citenamefont {Fomin}\ \emph {et~al.}(2015)\citenamefont {Fomin},
  \citenamefont {Ryzhov}, \citenamefont {Tsiok},\ and\ \citenamefont
  {Brazhkin}}]{Fomin2015}%
  \BibitemOpen
  \bibfield  {author} {\bibinfo {author} {\bibfnamefont {Y.~D.}\ \bibnamefont
  {Fomin}}, \bibinfo {author} {\bibfnamefont {V.~N.}\ \bibnamefont {Ryzhov}},
  \bibinfo {author} {\bibfnamefont {E.~N.}\ \bibnamefont {Tsiok}},\ and\
  \bibinfo {author} {\bibfnamefont {V.~V.}\ \bibnamefont {Brazhkin}},\
  }\bibfield  {title} {\bibinfo {title} {Dynamical crossover line in
  supercritical water},\ }\href {https://doi.org/10.1038/srep14234} {\bibfield
  {journal} {\bibinfo  {journal} {Scientific Reports}\ }\textbf {\bibinfo
  {volume} {5}},\ \bibinfo {pages} {14234} (\bibinfo {year}
  {2015})}\BibitemShut {NoStop}%
\bibitem [{\citenamefont {Brazhkin}\ \emph {et~al.}(2012)\citenamefont
  {Brazhkin}, \citenamefont {Fomin}, \citenamefont {Lyapin}, \citenamefont
  {Ryzhov},\ and\ \citenamefont {Trachenko}}]{PhysRevE.85.031203}%
  \BibitemOpen
  \bibfield  {author} {\bibinfo {author} {\bibfnamefont {V.~V.}\ \bibnamefont
  {Brazhkin}}, \bibinfo {author} {\bibfnamefont {Y.~D.}\ \bibnamefont {Fomin}},
  \bibinfo {author} {\bibfnamefont {A.~G.}\ \bibnamefont {Lyapin}}, \bibinfo
  {author} {\bibfnamefont {V.~N.}\ \bibnamefont {Ryzhov}},\ and\ \bibinfo
  {author} {\bibfnamefont {K.}~\bibnamefont {Trachenko}},\ }\bibfield  {title}
  {\bibinfo {title} {Two liquid states of matter: A dynamic line on a phase
  diagram},\ }\href {https://doi.org/10.1103/PhysRevE.85.031203} {\bibfield
  {journal} {\bibinfo  {journal} {Phys. Rev. E}\ }\textbf {\bibinfo {volume}
  {85}},\ \bibinfo {pages} {031203} (\bibinfo {year} {2012})}\BibitemShut
  {NoStop}%
\bibitem [{\citenamefont {{Huang}}\ \emph {et~al.}(2023)\citenamefont
  {{Huang}}, \citenamefont {{Baggioli}}, \citenamefont {{Lu}}, \citenamefont
  {{Ma}},\ and\ \citenamefont {{Feng}}}]{2023PhRvR...5a3149H}%
  \BibitemOpen
  \bibfield  {author} {\bibinfo {author} {\bibfnamefont {D.}~\bibnamefont
  {{Huang}}}, \bibinfo {author} {\bibfnamefont {M.}~\bibnamefont {{Baggioli}}},
  \bibinfo {author} {\bibfnamefont {S.}~\bibnamefont {{Lu}}}, \bibinfo {author}
  {\bibfnamefont {Z.}~\bibnamefont {{Ma}}},\ and\ \bibinfo {author}
  {\bibfnamefont {Y.}~\bibnamefont {{Feng}}},\ }\bibfield  {title} {\bibinfo
  {title} {{Revealing the supercritical dynamics of dusty plasmas and their
  liquidlike to gaslike dynamical crossover}},\ }\href
  {https://doi.org/10.1103/PhysRevResearch.5.013149} {\bibfield  {journal}
  {\bibinfo  {journal} {Physical Review Research}\ }\textbf {\bibinfo {volume}
  {5}},\ \bibinfo {eid} {013149} (\bibinfo {year} {2023})},\ \Eprint
  {https://arxiv.org/abs/2301.08449} {arXiv:2301.08449 [physics.plasm-ph]}
  \BibitemShut {NoStop}%
\bibitem [{\citenamefont {Jiang}\ \emph {et~al.}(2025)\citenamefont {Jiang},
  \citenamefont {Zheng}, \citenamefont {Chen}, \citenamefont {Baggioli},\ and\
  \citenamefont {Zhang}}]{Jiang_2025}%
  \BibitemOpen
  \bibfield  {author} {\bibinfo {author} {\bibfnamefont {C.}~\bibnamefont
  {Jiang}}, \bibinfo {author} {\bibfnamefont {Z.}~\bibnamefont {Zheng}},
  \bibinfo {author} {\bibfnamefont {Y.}~\bibnamefont {Chen}}, \bibinfo {author}
  {\bibfnamefont {M.}~\bibnamefont {Baggioli}},\ and\ \bibinfo {author}
  {\bibfnamefont {J.}~\bibnamefont {Zhang}},\ }\bibfield  {title} {\bibinfo
  {title} {Experimental observation of gapped shear waves and liquid-like to
  gas-like dynamical crossover in active granular matter},\ }\bibfield
  {journal} {\bibinfo  {journal} {Communications Physics}\ }\textbf {\bibinfo
  {volume} {8}},\ \href {https://doi.org/10.1038/s42005-025-02008-1}
  {10.1038/s42005-025-02008-1} (\bibinfo {year} {2025})\BibitemShut {NoStop}%
\bibitem [{\citenamefont {Xu}\ \emph {et~al.}(2005)\citenamefont {Xu},
  \citenamefont {Kumar}, \citenamefont {Buldyrev}, \citenamefont {Chen},
  \citenamefont {Poole}, \citenamefont {Sciortino},\ and\ \citenamefont
  {Stanley}}]{Xu_2005}%
  \BibitemOpen
  \bibfield  {author} {\bibinfo {author} {\bibfnamefont {L.}~\bibnamefont
  {Xu}}, \bibinfo {author} {\bibfnamefont {P.}~\bibnamefont {Kumar}}, \bibinfo
  {author} {\bibfnamefont {S.~V.}\ \bibnamefont {Buldyrev}}, \bibinfo {author}
  {\bibfnamefont {S.-H.}\ \bibnamefont {Chen}}, \bibinfo {author}
  {\bibfnamefont {P.~H.}\ \bibnamefont {Poole}}, \bibinfo {author}
  {\bibfnamefont {F.}~\bibnamefont {Sciortino}},\ and\ \bibinfo {author}
  {\bibfnamefont {H.~E.}\ \bibnamefont {Stanley}},\ }\bibfield  {title}
  {\bibinfo {title} {Relation between the widom line and the dynamic crossover
  in systems with a liquid–liquid phase transition},\ }\href
  {https://doi.org/10.1073/pnas.0507870102} {\bibfield  {journal} {\bibinfo
  {journal} {Proceedings of the National Academy of Sciences}\ }\textbf
  {\bibinfo {volume} {102}},\ \bibinfo {pages} {16558–16562} (\bibinfo {year}
  {2005})}\BibitemShut {NoStop}%
\bibitem [{\citenamefont {Ruppeiner}\ \emph {et~al.}(2012)\citenamefont
  {Ruppeiner}, \citenamefont {Sahay}, \citenamefont {Sarkar},\ and\
  \citenamefont {Sengupta}}]{Ruppeiner_2012}%
  \BibitemOpen
  \bibfield  {author} {\bibinfo {author} {\bibfnamefont {G.}~\bibnamefont
  {Ruppeiner}}, \bibinfo {author} {\bibfnamefont {A.}~\bibnamefont {Sahay}},
  \bibinfo {author} {\bibfnamefont {T.}~\bibnamefont {Sarkar}},\ and\ \bibinfo
  {author} {\bibfnamefont {G.}~\bibnamefont {Sengupta}},\ }\bibfield  {title}
  {\bibinfo {title} {Thermodynamic geometry, phase transitions, and the widom
  line},\ }\bibfield  {journal} {\bibinfo  {journal} {Physical Review E}\
  }\textbf {\bibinfo {volume} {86}},\ \href
  {https://doi.org/10.1103/physreve.86.052103} {10.1103/physreve.86.052103}
  (\bibinfo {year} {2012})\BibitemShut {NoStop}%
\bibitem [{\citenamefont {Luo}\ \emph {et~al.}(2014)\citenamefont {Luo},
  \citenamefont {Xu}, \citenamefont {Lascaris}, \citenamefont {Stanley},\ and\
  \citenamefont {Buldyrev}}]{PhysRevLett.112.135701}%
  \BibitemOpen
  \bibfield  {author} {\bibinfo {author} {\bibfnamefont {J.}~\bibnamefont
  {Luo}}, \bibinfo {author} {\bibfnamefont {L.}~\bibnamefont {Xu}}, \bibinfo
  {author} {\bibfnamefont {E.}~\bibnamefont {Lascaris}}, \bibinfo {author}
  {\bibfnamefont {H.~E.}\ \bibnamefont {Stanley}},\ and\ \bibinfo {author}
  {\bibfnamefont {S.~V.}\ \bibnamefont {Buldyrev}},\ }\bibfield  {title}
  {\bibinfo {title} {Behavior of the widom line in critical phenomena},\ }\href
  {https://doi.org/10.1103/PhysRevLett.112.135701} {\bibfield  {journal}
  {\bibinfo  {journal} {Phys. Rev. Lett.}\ }\textbf {\bibinfo {volume} {112}},\
  \bibinfo {pages} {135701} (\bibinfo {year} {2014})}\BibitemShut {NoStop}%
\bibitem [{\citenamefont {Banuti}\ \emph {et~al.}(2017)\citenamefont {Banuti},
  \citenamefont {Raju},\ and\ \citenamefont {Ihme}}]{PhysRevE.95.052120}%
  \BibitemOpen
  \bibfield  {author} {\bibinfo {author} {\bibfnamefont {D.~T.}\ \bibnamefont
  {Banuti}}, \bibinfo {author} {\bibfnamefont {M.}~\bibnamefont {Raju}},\ and\
  \bibinfo {author} {\bibfnamefont {M.}~\bibnamefont {Ihme}},\ }\bibfield
  {title} {\bibinfo {title} {Similarity law for widom lines and coexistence
  lines},\ }\href {https://doi.org/10.1103/PhysRevE.95.052120} {\bibfield
  {journal} {\bibinfo  {journal} {Phys. Rev. E}\ }\textbf {\bibinfo {volume}
  {95}},\ \bibinfo {pages} {052120} (\bibinfo {year} {2017})}\BibitemShut
  {NoStop}%
\bibitem [{\citenamefont {Gallo}\ \emph {et~al.}(2014)\citenamefont {Gallo},
  \citenamefont {Corradini},\ and\ \citenamefont {Rovere}}]{Gallo2014}%
  \BibitemOpen
  \bibfield  {author} {\bibinfo {author} {\bibfnamefont {P.}~\bibnamefont
  {Gallo}}, \bibinfo {author} {\bibfnamefont {D.}~\bibnamefont {Corradini}},\
  and\ \bibinfo {author} {\bibfnamefont {M.}~\bibnamefont {Rovere}},\
  }\bibfield  {title} {\bibinfo {title} {Widom line and dynamical crossovers as
  routes to understand supercritical water},\ }\href
  {https://doi.org/10.1038/ncomms6806} {\bibfield  {journal} {\bibinfo
  {journal} {Nature Communications}\ }\textbf {\bibinfo {volume} {5}},\
  \bibinfo {pages} {5806} (\bibinfo {year} {2014})}\BibitemShut {NoStop}%
\bibitem [{\citenamefont {Zhao}\ \emph
  {et~al.}(2025{\natexlab{b}})\citenamefont {Zhao}, \citenamefont {Nie},
  \citenamefont {Zhang},\ and\ \citenamefont {Zhang}}]{Zhao:2025ecg}%
  \BibitemOpen
  \bibfield  {author} {\bibinfo {author} {\bibfnamefont {Z.-Q.}\ \bibnamefont
  {Zhao}}, \bibinfo {author} {\bibfnamefont {Z.-Y.}\ \bibnamefont {Nie}},
  \bibinfo {author} {\bibfnamefont {J.-F.}\ \bibnamefont {Zhang}},\ and\
  \bibinfo {author} {\bibfnamefont {X.}~\bibnamefont {Zhang}},\ }\bibfield
  {title} {\bibinfo {title} {{Characterized Behaviors of Black Hole
  Thermodynamics in the Supercritical Region}},\ }\href
  {https://doi.org/10.1088/0256-307X/42/10/101102} {\bibfield  {journal}
  {\bibinfo  {journal} {Chin. Phys. Lett.}\ }\textbf {\bibinfo {volume} {42}},\
  \bibinfo {pages} {101102} (\bibinfo {year} {2025}{\natexlab{b}})},\ \Eprint
  {https://arxiv.org/abs/2504.04995} {arXiv:2504.04995 [gr-qc]} \BibitemShut
  {NoStop}%
\bibitem [{\citenamefont {Xu}\ and\ \citenamefont {Mann}(2026)}]{Xu:2025jrk}%
  \BibitemOpen
  \bibfield  {author} {\bibinfo {author} {\bibfnamefont {Z.-M.}\ \bibnamefont
  {Xu}}\ and\ \bibinfo {author} {\bibfnamefont {R.~B.}\ \bibnamefont {Mann}},\
  }\bibfield  {title} {\bibinfo {title} {{Thermodynamic Supercriticality and
  Complex Phase Diagram for the AdS Black Hole}},\ }\href
  {https://doi.org/10.1103/c39y-zcz6} {\bibfield  {journal} {\bibinfo
  {journal} {Phys. Rev. Lett.}\ }\textbf {\bibinfo {volume} {136}},\ \bibinfo
  {pages} {041402} (\bibinfo {year} {2026})},\ \Eprint
  {https://arxiv.org/abs/2504.05708} {arXiv:2504.05708 [gr-qc]} \BibitemShut
  {NoStop}%
\bibitem [{\citenamefont {Li}\ \emph {et~al.}(2025)\citenamefont {Li},
  \citenamefont {Zhang}, \citenamefont {Yang}, \citenamefont {Mann},\ and\
  \citenamefont {Wang}}]{Li:2025tdd}%
  \BibitemOpen
  \bibfield  {author} {\bibinfo {author} {\bibfnamefont {R.}~\bibnamefont
  {Li}}, \bibinfo {author} {\bibfnamefont {K.}~\bibnamefont {Zhang}}, \bibinfo
  {author} {\bibfnamefont {J.}~\bibnamefont {Yang}}, \bibinfo {author}
  {\bibfnamefont {R.~B.}\ \bibnamefont {Mann}},\ and\ \bibinfo {author}
  {\bibfnamefont {J.}~\bibnamefont {Wang}},\ }\bibfield  {title} {\bibinfo
  {title} {{Critical slowing down of black hole phase transition and kinetic
  crossover in supercritical regime}},\ }\href
  {https://doi.org/10.1103/z8bq-kvq3} {\bibfield  {journal} {\bibinfo
  {journal} {Phys. Rev. D}\ }\textbf {\bibinfo {volume} {112}},\ \bibinfo
  {pages} {064004} (\bibinfo {year} {2025})},\ \Eprint
  {https://arxiv.org/abs/2505.24148} {arXiv:2505.24148 [gr-qc]} \BibitemShut
  {NoStop}%
\bibitem [{\citenamefont {Wang}\ \emph {et~al.}(2025)\citenamefont {Wang},
  \citenamefont {Li}, \citenamefont {Jin},\ and\ \citenamefont
  {Li}}]{Wang:2025ctk}%
  \BibitemOpen
  \bibfield  {author} {\bibinfo {author} {\bibfnamefont {S.}~\bibnamefont
  {Wang}}, \bibinfo {author} {\bibfnamefont {X.}~\bibnamefont {Li}}, \bibinfo
  {author} {\bibfnamefont {Y.}~\bibnamefont {Jin}},\ and\ \bibinfo {author}
  {\bibfnamefont {L.}~\bibnamefont {Li}},\ }\bibfield  {title} {\bibinfo
  {title} {{Analogous supercritical crossovers in black holes and water}},\
  }\href@noop {} {\  (\bibinfo {year} {2025})},\ \Eprint
  {https://arxiv.org/abs/2506.10808} {arXiv:2506.10808 [gr-qc]} \BibitemShut
  {NoStop}%
\bibitem [{\citenamefont {Li}\ \emph {et~al.}(2026)\citenamefont {Li},
  \citenamefont {Chen}, \citenamefont {Wu},\ and\ \citenamefont
  {Xu}}]{Li:2025lrq}%
  \BibitemOpen
  \bibfield  {author} {\bibinfo {author} {\bibfnamefont {Z.-Y.}\ \bibnamefont
  {Li}}, \bibinfo {author} {\bibfnamefont {X.-R.}\ \bibnamefont {Chen}},
  \bibinfo {author} {\bibfnamefont {B.}~\bibnamefont {Wu}},\ and\ \bibinfo
  {author} {\bibfnamefont {Z.-M.}\ \bibnamefont {Xu}},\ }\bibfield  {title}
  {\bibinfo {title} {{Thermodynamic supercriticality and complex phase diagram
  for charged Gauss-Bonnet AdS black holes}},\ }\href
  {https://doi.org/10.1016/j.physletb.2026.140438} {\bibfield  {journal}
  {\bibinfo  {journal} {Phys. Lett. B}\ }\textbf {\bibinfo {volume} {876}},\
  \bibinfo {pages} {140438} (\bibinfo {year} {2026})},\ \Eprint
  {https://arxiv.org/abs/2511.10357} {arXiv:2511.10357 [hep-th]} \BibitemShut
  {NoStop}%
\bibitem [{\citenamefont {Anand}\ and\ \citenamefont
  {Wang}(2026)}]{Anand:2025rzh}%
  \BibitemOpen
  \bibfield  {author} {\bibinfo {author} {\bibfnamefont {A.}~\bibnamefont
  {Anand}}\ and\ \bibinfo {author} {\bibfnamefont {S.}~\bibnamefont {Wang}},\
  }\bibfield  {title} {\bibinfo {title} {{Universal supercritical behavior in
  global monopole-charged AdS black holes}},\ }\href
  {https://doi.org/10.1007/JHEP05(2026)278} {\bibfield  {journal} {\bibinfo
  {journal} {JHEP}\ }\textbf {\bibinfo {volume} {05}},\ \bibinfo {pages}
  {278}},\ \Eprint {https://arxiv.org/abs/2512.12723} {arXiv:2512.12723
  [hep-th]} \BibitemShut {NoStop}%
\bibitem [{\citenamefont {Guo}\ and\ \citenamefont {Xu}(2026)}]{Guo:2026xlk}%
  \BibitemOpen
  \bibfield  {author} {\bibinfo {author} {\bibfnamefont {F.}~\bibnamefont
  {Guo}}\ and\ \bibinfo {author} {\bibfnamefont {Z.-M.}\ \bibnamefont {Xu}},\
  }\bibfield  {title} {\bibinfo {title} {{Complex plane phase diagram and Widom
  line for the Born{\textendash}Infeld black holes with reentrant phase
  transition}},\ }\href {https://doi.org/10.1140/epjc/s10052-026-15792-z}
  {\bibfield  {journal} {\bibinfo  {journal} {Eur. Phys. J. C}\ }\textbf
  {\bibinfo {volume} {86}},\ \bibinfo {pages} {524} (\bibinfo {year} {2026})},\
  \Eprint {https://arxiv.org/abs/2602.09770} {arXiv:2602.09770 [hep-th]}
  \BibitemShut {NoStop}%
\bibitem [{\citenamefont {Zhao}\ \emph
  {et~al.}(2026{\natexlab{c}})\citenamefont {Zhao}, \citenamefont {Nie},
  \citenamefont {Wei}, \citenamefont {Zhang},\ and\ \citenamefont
  {Zhang}}]{Zhao:2026mkx}%
  \BibitemOpen
  \bibfield  {author} {\bibinfo {author} {\bibfnamefont {Z.-Q.}\ \bibnamefont
  {Zhao}}, \bibinfo {author} {\bibfnamefont {Z.-Y.}\ \bibnamefont {Nie}},
  \bibinfo {author} {\bibfnamefont {S.-W.}\ \bibnamefont {Wei}}, \bibinfo
  {author} {\bibfnamefont {J.-F.}\ \bibnamefont {Zhang}},\ and\ \bibinfo
  {author} {\bibfnamefont {X.}~\bibnamefont {Zhang}},\ }\bibfield  {title}
  {\bibinfo {title} {{Interior geometry of black holes as a probe of
  first-order phase transition}},\ }\href
  {https://doi.org/10.1140/epjc/s10052-026-16149-2} {\bibfield  {journal}
  {\bibinfo  {journal} {Eur. Phys. J. C}\ }\textbf {\bibinfo {volume} {86}},\
  \bibinfo {pages} {937} (\bibinfo {year} {2026}{\natexlab{c}})},\ \Eprint
  {https://arxiv.org/abs/2604.01818} {arXiv:2604.01818 [gr-qc]} \BibitemShut
  {NoStop}%
\bibitem [{\citenamefont {Li}\ \emph {et~al.}(2020)\citenamefont {Li},
  \citenamefont {Nie},\ and\ \citenamefont {Tian}}]{Li:2020ayr}%
  \BibitemOpen
  \bibfield  {author} {\bibinfo {author} {\bibfnamefont {X.}~\bibnamefont
  {Li}}, \bibinfo {author} {\bibfnamefont {Z.-Y.}\ \bibnamefont {Nie}},\ and\
  \bibinfo {author} {\bibfnamefont {Y.}~\bibnamefont {Tian}},\ }\bibfield
  {title} {\bibinfo {title} {{Holographic boiling and generalized thermodynamic
  description beyond local equilibrium}},\ }\href
  {https://doi.org/10.1007/JHEP09(2020)063} {\bibfield  {journal} {\bibinfo
  {journal} {JHEP}\ }\textbf {\bibinfo {volume} {09}},\ \bibinfo {pages}
  {063}},\ \Eprint {https://arxiv.org/abs/2003.12987} {arXiv:2003.12987
  [hep-th]} \BibitemShut {NoStop}%
\bibitem [{\citenamefont {Chen}\ \emph {et~al.}(2025)\citenamefont {Chen},
  \citenamefont {Liu}, \citenamefont {Tian}, \citenamefont {Wu},\ and\
  \citenamefont {Zhang}}]{Chen:2024pyy}%
  \BibitemOpen
  \bibfield  {author} {\bibinfo {author} {\bibfnamefont {Q.}~\bibnamefont
  {Chen}}, \bibinfo {author} {\bibfnamefont {Y.}~\bibnamefont {Liu}}, \bibinfo
  {author} {\bibfnamefont {Y.}~\bibnamefont {Tian}}, \bibinfo {author}
  {\bibfnamefont {X.}~\bibnamefont {Wu}},\ and\ \bibinfo {author}
  {\bibfnamefont {H.}~\bibnamefont {Zhang}},\ }\bibfield  {title} {\bibinfo
  {title} {{Dynamics of holographic steady flows near a first-order phase
  transition}},\ }\href {https://doi.org/10.1007/s11433-025-2633-y} {\bibfield
  {journal} {\bibinfo  {journal} {Sci. China Phys. Mech. Astron.}\ }\textbf
  {\bibinfo {volume} {68}},\ \bibinfo {pages} {260414} (\bibinfo {year}
  {2025})},\ \Eprint {https://arxiv.org/abs/2408.09679} {arXiv:2408.09679
  [hep-th]} \BibitemShut {NoStop}%
\bibitem [{\citenamefont {Zhao}\ \emph
  {et~al.}(2026{\natexlab{d}})\citenamefont {Zhao}, \citenamefont {Nie},
  \citenamefont {Zhang},\ and\ \citenamefont {Zhang}}]{Zhao:2026eav}%
  \BibitemOpen
  \bibfield  {author} {\bibinfo {author} {\bibfnamefont {Z.-Q.}\ \bibnamefont
  {Zhao}}, \bibinfo {author} {\bibfnamefont {Z.-Y.}\ \bibnamefont {Nie}},
  \bibinfo {author} {\bibfnamefont {J.-F.}\ \bibnamefont {Zhang}},\ and\
  \bibinfo {author} {\bibfnamefont {X.}~\bibnamefont {Zhang}},\ }\bibfield
  {title} {\bibinfo {title} {{Topological defect induced phase separation in a
  holographic system}},\ }\href@noop {} {\  (\bibinfo {year}
  {2026}{\natexlab{d}})},\ \Eprint {https://arxiv.org/abs/2604.00690}
  {arXiv:2604.00690 [hep-th]} \BibitemShut {NoStop}%
\bibitem [{\citenamefont {Jin}\ \emph {et~al.}(2026)\citenamefont {Jin},
  \citenamefont {An},\ and\ \citenamefont {Li}}]{Jin:2026lzp}%
  \BibitemOpen
  \bibfield  {author} {\bibinfo {author} {\bibfnamefont {Z.-h.}\ \bibnamefont
  {Jin}}, \bibinfo {author} {\bibfnamefont {Y.-p.}\ \bibnamefont {An}},\ and\
  \bibinfo {author} {\bibfnamefont {L.}~\bibnamefont {Li}},\ }\bibfield
  {title} {\bibinfo {title} {{Bubble dynamics and vortex formation in
  holographic first-order superfluid phase transitions}},\ }\href@noop {} {\
  (\bibinfo {year} {2026})},\ \Eprint {https://arxiv.org/abs/2604.17216}
  {arXiv:2604.17216 [hep-th]} \BibitemShut {NoStop}%
\bibitem [{\citenamefont {Zhao}\ \emph
  {et~al.}(2026{\natexlab{e}})\citenamefont {Zhao}, \citenamefont {Nie},
  \citenamefont {Zhang},\ and\ \citenamefont {Zhang}}]{Zhao:2026jqp}%
  \BibitemOpen
  \bibfield  {author} {\bibinfo {author} {\bibfnamefont {Z.-Q.}\ \bibnamefont
  {Zhao}}, \bibinfo {author} {\bibfnamefont {Z.-Y.}\ \bibnamefont {Nie}},
  \bibinfo {author} {\bibfnamefont {J.-F.}\ \bibnamefont {Zhang}},\ and\
  \bibinfo {author} {\bibfnamefont {X.}~\bibnamefont {Zhang}},\ }\bibfield
  {title} {\bibinfo {title} {{Nonequilibrium crossover in the supercritical
  region from quench dynamics}},\ }\href
  {https://doi.org/10.1016/j.physletb.2026.140819} {\bibfield  {journal}
  {\bibinfo  {journal} {Phys. Lett. B}\ }\textbf {\bibinfo {volume} {880}},\
  \bibinfo {pages} {140819} (\bibinfo {year} {2026}{\natexlab{e}})},\ \Eprint
  {https://arxiv.org/abs/2604.13104} {arXiv:2604.13104 [cond-mat.stat-mech]}
  \BibitemShut {NoStop}%
\bibitem [{\citenamefont {Zhao}\ \emph
  {et~al.}(2026{\natexlab{f}})\citenamefont {Zhao}, \citenamefont {Nie},
  \citenamefont {Zhang},\ and\ \citenamefont {Zhang}}]{Zhao:2026als}%
  \BibitemOpen
  \bibfield  {author} {\bibinfo {author} {\bibfnamefont {Z.-Q.}\ \bibnamefont
  {Zhao}}, \bibinfo {author} {\bibfnamefont {Z.-Y.}\ \bibnamefont {Nie}},
  \bibinfo {author} {\bibfnamefont {J.-F.}\ \bibnamefont {Zhang}},\ and\
  \bibinfo {author} {\bibfnamefont {X.}~\bibnamefont {Zhang}},\ }\bibfield
  {title} {\bibinfo {title} {{Phase separation seeded by Z2 and U(1)
  topological defects from holography}},\ }\href@noop {} {\  (\bibinfo {year}
  {2026}{\natexlab{f}})},\ \Eprint {https://arxiv.org/abs/2606.00163}
  {arXiv:2606.00163 [hep-th]} \BibitemShut {NoStop}%
\end{thebibliography}%

\end{document}